%% file: main.tex
\PassOptionsToPackage{dvipsnames}{xcolor}
\documentclass[10pt]{article}
\usepackage{graphicx,amssymb, amstext, amsmath, epstopdf, booktabs, verbatim, geometry, appendix, natbib, lmodern}
\usepackage{sfmath}
\usepackage{ulem}
\usepackage[rightcaption]{sidecap}
\usepackage{tabularx}
\usepackage{tcolorbox}

\usepackage{comment}

\newtcolorbox{mybox}[2]{
    arc=0pt,
    boxrule=#2pt,
    colback=#1,
    width=10cm,
    halign=left,
}
\usepackage[font=small]{caption} 

\usepackage[dvipsnames]{xcolor}
\usepackage{titlesec}
\usepackage[colorlinks=true,citecolor=blue,urlcolor=blue,linkcolor=blue]{hyperref}
\newcommand*\Title{HRMOS}
\newcommand*\cpiType{Science Motivation v.1}
\newcommand*\Date{December 2023}

\title{White Paper}
\author{}
\date{December 2023}

\usepackage{cpistuff/cpi} 

\begin{document}

\begin{titlepage}
\maketitle
\end{titlepage}

\linespread{1.15} 

\begin{executive}

The High-Resolution Multi-Object Spectrograph (HRMOS) is a facility instrument that we plan to propose for the Very Large Telescope (VLT) of the European Southern Observatory (ESO), following the initial presentation at the VLT 2030 workshop held at ESO in June 2019. HRMOS provides a combination of capabilities that are essential to carry out breakthrough science across a broad range of active research areas from stellar astrophysics and exoplanet studies to Galactic and Local Group archaeology. HRMOS fills a gap in capabilities amongst the landscape of future instrumentation planned for the next decade. The key characteristics of HRMOS will be high spectral resolution ($R = 60\,000 - 80\,000$) combined with multi-object ($20-100$) capabilities and long term stability that will provide excellent radial velocity precision and accuracy ($\sim$10\,m\,s$^{-1}$). Initial designs predict that a  $SNR\approx 100$ will be achievable in about one hour for a star with $\rm mag(AB) = 15$, while with the same exposure time a $SNR\approx 30$  will be reached for a  star with $\rm mag(AB) = 17$. The combination of high resolution and multiplexing with wavelength coverage extending to relatively blue wavelengths (down to 380\,nm), makes HRMOS a spectrograph that will push the boundaries of our knowledge and that is envisioned as a workhorse instrument in the future. 

The science cases presented in this White Paper include topics and ideas developed by the Core Science Team with the contributions from the astronomical community, also through the wide participation in the first HRMOS Workshop\footnote{\url{https://indico.ict.inaf.it/event/1547/}} that took place in Firenze (Italy) in October 2021. 

\end{executive}

\begin{center}

{\bf HRMOS Core Science Team}:

Laura Magrini$^{1}$, Thomas Bensby$^{2}$, Anna Brucalassi$^{1}$,  
Sofia Randich$^{1}$,
Robin Jeffries$^{3}$,
Gayandhi de Silva$^{4}$,
Ása Skúladóttir$^{5, 1}$,
Rodolfo Smiljanic$^{6}$,
Oscar Gonzalez$^{7}$, Vanessa Hill$^{8}$,  Nad\`ege Lagarde$^{9}$,     Eline Tolstoy$^{10}$.

{\bf Contributors from the Science team: }

 José María Arroyo-Polonio$^{11}$, 
 Martina Baratella$^{12}$, 
 John R. Barnes$^{13}$, 
 Giuseppina Battaglia$^{11}$,  
 Holger Baumgardt$^{14}$, 
 Michele Bellazzini$^{15}$, 
 Katia Biazzo$^{16}$, 
 Angela Bragaglia$^{15}$, 
 Bradley Carter$^{17}$, 
 Giada Casali$^{18,15}$, 
 Gabriele Cescutti$^{19,20,21}$, 
 Camilla Danielski$^{1}$, 
 Elisa Delgado Mena$^{22}$, 
 Arnas Drazdauskas$^{23}$, 
 Mark Gieles $^{24, 25}$, 
 Riano Giribaldi$^{26}$, 
 Keith Hawkins$^{27}$, 
 H. Jens Hoeijmakers$^{2}$, 
 Pascale Jablonka$^{28}$, 
 Devika Kamath$^{29, 30}$, 
 Tom Louth$^{7}$, 
 Anna Fabiola Marino$^{1, 31}$, 
 Sarah Martell$^{32}$, 
 Thibault Merle$^{25}$, 
 Benjamin Montet$^{32}$, 
 Michael T.  Murphy$^{33}$, 
 Brunella Nisini$^{16}$, 
 Thomas Nordlander$^{34, 35}$,  
 Valentina D'Orazi$^{31, 36}$, 
 Lorenzo Pino$^{1}$, 
 Donatella Romano$^{15}$, 
 Germano Sacco$^{1}$,  
 Nathan R. Sandford$^{37}$, 
 Antonio Sollima$^{15}$,  
 Lorenzo Spina$^{1}$, 
 Gra{\v z}ina Tautvai{\v s}ien{\. e}$^{23}$, 
 Yuan-Sen Ting$^{4}$, 
 Andrea Tozzi$^{1}$, 
 Mathieu Van der Swaelmen$^{1}$, 
 Sophie Van Eck$^{25}$, 
 Stephen Watson$^{7}$, 
 C.Clare Worley$^{38}$, 
 Alice Zocchi$^{39}$
 
{\bf Affiliations: }

$^{1}$ INAF - Osservatorio Astrofisico di Arcetri, Firenze, Italy\\
$^{2}$ Lund Observatory, Division of Astrophysics, Department of Physics, Lund University, Lund, Sweden \\
$^{3}$ Keele University, Staffordshire, United Kingdom\\
$^{4}$ Australian Astronomical Observatory, Macquarie University, Australia\\
$^{5}$ Dipartimento di Fisica e Astronomia, Universitá degli Studi di Firenze, Sesto Fiorentino, Italy\\
$^{6}$ Nicolaus Copernicus Astronomical Center, Polish Academy of Sciences, Warsaw, Poland\\
$^{7}$ UK Astronomy Technology Ctr., Royal Observatory, Edinburgh, United Kingdom\\
$^{8}$ Observatoire de la Cote d’Azur, Nice, France\\
$^{9}$ Laboratoire d'Astrophysique de Bordeaux, Universit\'e Bordeaux, CNRS, B18N, All\'ee Geoffroy Saint-Hilaire, 33615 Pessac, France \\
$^{10}$ Kapteyn Astronomical Institute, University of Groningen, Groningen, The Netherlands\\
$^{11}$ Instituto de Astrofisica de Canarias, San Cristobal de La Laguna, Spain\\
$^{12}$ ESO, Santiago,  Chile\\
$^{13}$ Department of Physical Sciences, The Open University, Walton Hall, Milton Keynes, United Kingdom\\
$^{14}$ The University of Queensland, Australia\\
$^{15}$ INAF – Osservatorio di Astrofisica e Scienza dello Spazio di Bologna, Bologna, Italy\\
$^{16}$ INAF - Osservatorio Astronomico di Roma, Monte Porzio Catone, Italy\\
$^{17}$ Centre for Astrophysics, University of Southern Queensland, Toowoomba, Queensland, Australia\\
$^{18}$ Dipartimento di Fisica e Astronomia, Università di Bologna, Bologna, Italy \\
$^{19}$ Dipartimento di Fisica, Sezione di Astronomia, Università di Trieste, Trieste, Italy\\
$^{20}$ INAF – Osservatorio Astronomico di Trieste, Trieste, Italy\\
$^{21}$ INFN – Sezione di Trieste, Trieste, Italy\\
$^{22}$ Instituto de Astrofísica e Cie\^ncias do Espaco, Universidade do Porto, CAUP, Porto, Portugal\\
$^{23}$ Institute of Theoretical Physics and Astronomy, Vilnius University, Vilnius, Lithuania\\
$^{24}$ ICREA, Pg. Llu\'{i}s Companys 23, E08010 Barcelona, Spain,\\
$^{25}$ Institut de Ci\`{e}ncies del Cosmos (ICCUB), Universitat de Barcelona (IEEC-UB), Mart\'{i} i Franqu\`{e}s 1, E08028 Barcelona, Spain\\
$^{26}$ Institut d'Astronomie et d'Astrophysique, Université libre de Bruxelles, Belgium\\
$^{27}$ University of Texas, Austin, US\\
$^{28}$ Laboratoire d’astrophysique, EPFL, Observatoire, Versoix, Switzerland\\
$^{29}$ School of Mathematical and Physical Sciences, Macquarie University, Sydney, Australia\\
$^{30}$ Research Centre for Astronomy, Astrophysics and Astrophotonics, Macquarie Univ., Sydney, Australia\\
$^{31}$ INAF - Osservatorio Astronomico di Padova, Padova, Italy\\
$^{32}$ University of New South Wales, Australia\\
$^{33}$ Swinburne University, Australia\\
$^{34}$ Research School of Astronomy and Astrophysics, Australian National University, Canberra, ACT 2611, Australia\\
$^{35}$ ARC Centre of Excellence for Astrophysics in Three Dimensions (ASTRO-3D), Australia\\
$^{36}$ Dipartimento di Fisica, Universita di Roma Tor Vergata, Roma, Italy\\
$^{37}$ Department of Astronomy, University of California Berkeley, Berkeley, CA, USA\\
$^{38}$ University of Canterbury, Christchurch, New Zeland \\
$^{39}$ Department of Astrophysics, University of Vienna, Vienna, Austria\\
\end {center}

\clearpage

\tableofcontents

\clearpage

\listoftables

\clearpage
\input{intro.tex}

\clearpage

\input{instrument}

\clearpage

\input{young_stars}

\clearpage

\input{exoplanets}

\clearpage
\input{clusters}

\clearpage
 
\input{gal_archaeology}

\clearpage

\input{dwarf_gal}

\clearpage

\phantomsection
\addcontentsline{toc}{section}{List of acronyms}
\input{acronyms}

\clearpage

\phantomsection
\addcontentsline{toc}{section}{References}
\bibliographystyle{aa}
\bibliography{referenser}{}

\end{document}

%% file: intro.tex
\newcommand{\asacom}[1]{\color{teal}[ÁS: #1] \color{black}}
\newcommand{\magenta}{\textcolor{magenta}}
\newcommand{\asa}{\textcolor{teal}}

\section{Introduction} \label{sec:intro}
\subsection{The need for a multi-object spectrograph at very high resolution}
Recent years have witnessed a revolution in the investigation of the Milky Way and its stellar populations. On the one hand, the ESA {\sl Gaia} mission is providing high-precision parallaxes, proper motions, and photometry for almost two billion stars, down to a {\sl Gaia} magnitude $G\lesssim 20$; as well as radial velocities, stellar parameters, metallicity, and some elemental abundances for brighter stars \citep{gaia_prusti2016,gaiadr3_a, gaiadr3_b, gaiadr3_c}. On the other hand, ground-based stellar spectroscopic surveys, such as RAVE \citep{Steinmetz2020}, {\sl Gaia}-ESO \citep{Gilmore2022,randich2022}; APOGEE \citep{majewski2017}, GALAH \citep{desilva2015}, and LAMOST \citep{lamost2012}, are complementing the {\sl Gaia} data with more precise radial velocities, atmospheric parameters and elemental abundances, down to fainter magnitudes. These data have allowed significant progress in the understanding of a variety of topics, including: the formation of stars and star clusters; stellar physics and evolution; and the formation and evolution of the Milky Way. At the same time, new, critical questions have emerged, which need to be addressed with much larger statistical samples, and more detailed and precise information on key elemental abundances and isotopic abundance ratios. 

In the coming years, new generation multi-object spectroscopic (MOS) facilities and surveys at low, medium, and relatively high spectral resolution will obtain spectroscopic data for millions of stars, for example the 4MOST survey \citep{dejong2019}, WEAVE \citep{dalton2014, jin2022},  MOONS \citep{cirasuolo2020,gonzalez2020}, the Milky Way Mapper (MWM) survey of SDSS-V \citep{mwm23}, the DESI survey \citep{desi23}, and, possibly, the Wide Field Spectroscopic Telescope -WST- \citep{bacon2023}. These up-coming surveys will allow further excellent {\sl Gaia} follow-up studies of the chemo-dynamical substructures in the Milky Way discs, halo, and bulge regions with samples of unprecedented size. Complementary to that, from the observation of radial and non-radial solar-like oscillations, asteroseismology also opens a new window to constrain stellar and Galactic physics. It delivers crucial parameters such as precise stellar surface gravities and more accurate (model-dependent) age estimates of giant stars, even several kiloparsecs away from the Sun. This incredible source of new constraints is expected to grow with the future PLATO space mission starting in 2026 \citep{Rauer2014}. 

Similarly, thanks to the very high-resolution spectrographs with R$>$80\,000, such as HARPS and ESPRESSO at the ESO VLT, and HARPS-N at the TNG, huge progress has also been achieved in the detection and characterisation of extra-solar planets and their atmospheres, as well as the properties of their host stars. Within the next decade, new facilities, such as the ESA Ariel space mission \citep{tinetti18, tinetti21} and ANDES on the E-ELT \citep{ANDES,ANDES22}, will provide novel information on the composition of their atmospheres. 
Nevertheless, some critical areas are, and will remain, poorly covered by observations, such as planets in clusters of different ages, or planets in crowded fields like the Galatic bulge. To make progress, efficient, high-precision and stable radial velocity measurements are needed of statistically significant samples of cluster members, as well as Galactic bulge stars.

\begin{figure}
\centering
\resizebox{0.999\hsize}{!}{
\includegraphics{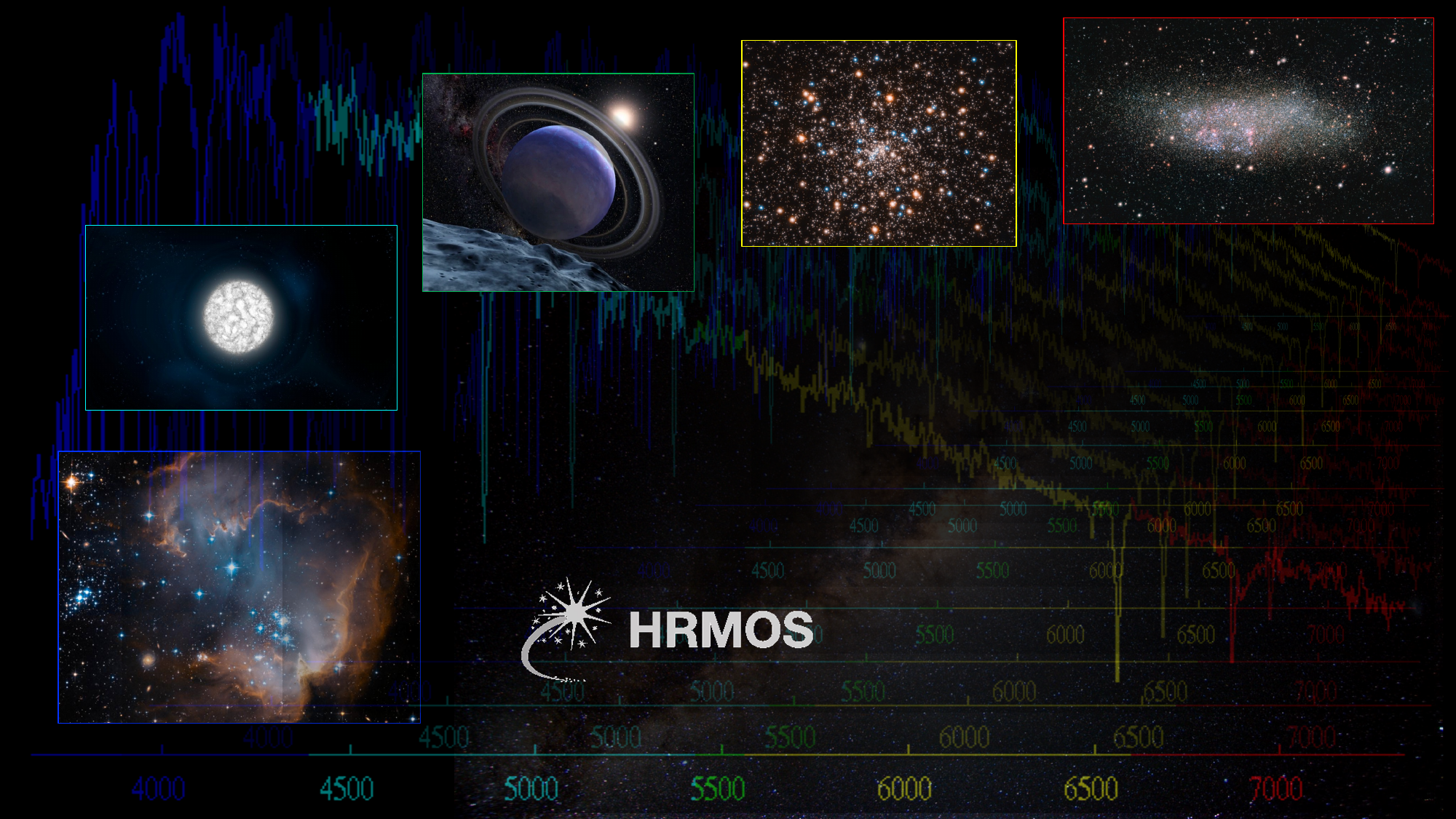}}
\caption{
\label{fig:sci_cases}
HRMOS will have a revolutionary impact in several fields of astrophysics: ranging from the discovery of exoplanets in crowded environments,  to the detection and characterisation of activity and magnetic fields in stars, to detailed studies of the stellar populations in star clusters and dwarf galaxies, to the most precise determination of stellar ages with nucleocosmochronology.}
\end{figure}

\begin{figure}
\centering
\resizebox{0.599\hsize}{!}{
\includegraphics{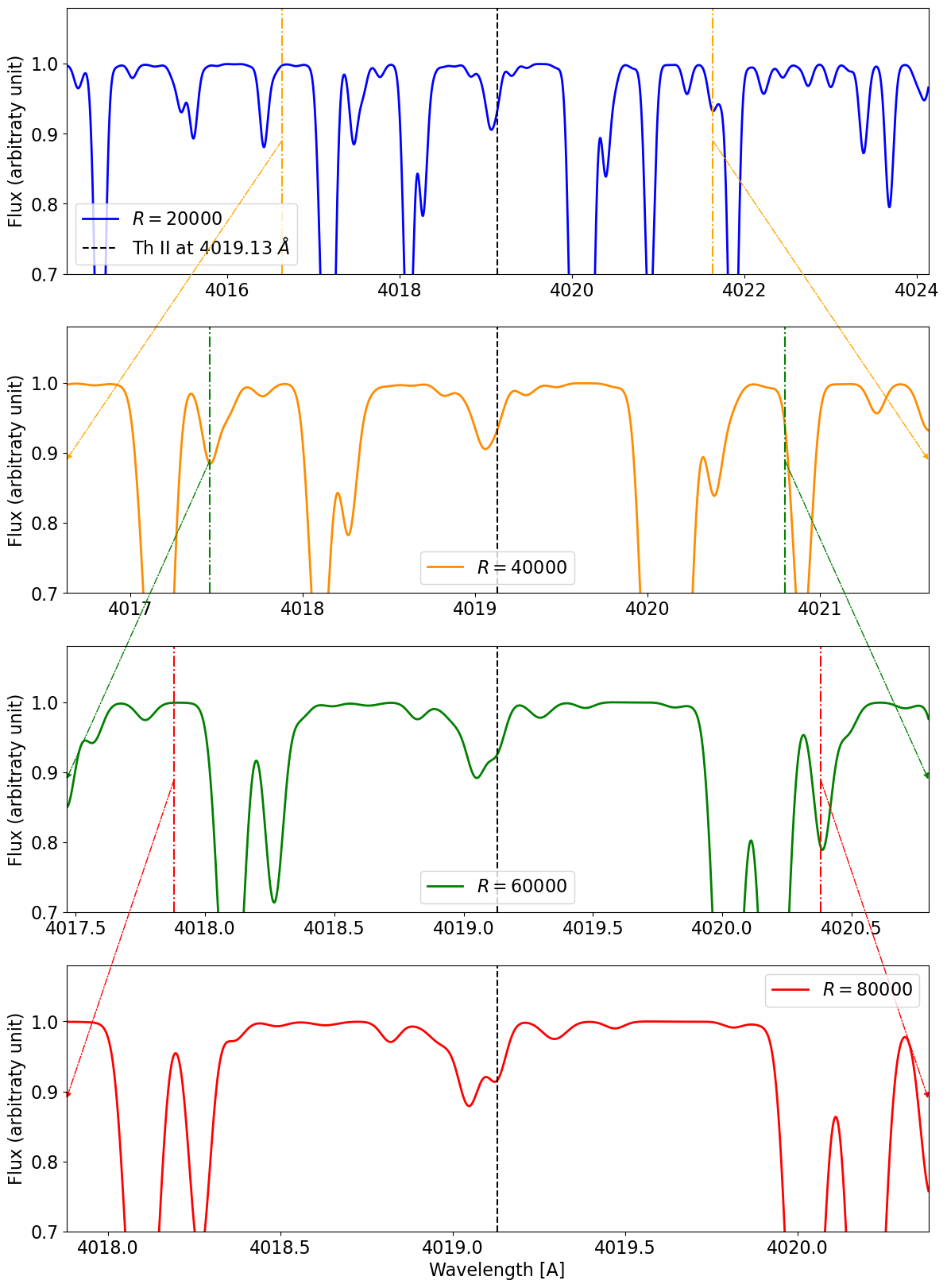}}
\caption{
\label{fig:syn_4res}
Synthetic spectrum of a giant star (4200 K, 2.5, -2.5) at four different resolutions. In each panel, a zoomed-in portion of the wavelength range of the corresponding upper panel is shown. The vertical dashed line indicates the Th~{\sc ii} line at 4019.13 \AA, which can be separated  at $R>60\,0000$. }
\end{figure}

In summary, an exponential growth is expected in the next decade, with photometric, astrometric, asteroseismic, kinematics, and chemical information being sourced for millions to billions of targets located in various Galactic environments. This is providing (and will soon provide to a much larger extent) new, exquisite, and complementary observational constraints,
improving our understanding of the Milky Way with its stellar and planetary populations, as well as of nearby galaxies. Whilst the discovery space of those facilities is certainly significant, there are important open questions that still require observations of sizeable samples with very high-resolution spectroscopy (Fig.~\ref{fig:sci_cases}). These data will enable us to determine challenging abundances from weak and/or crowded spectral lines of key elements (for example, those produced by neutron-capture, see Fig.~\ref{fig:syn_4res}), to measure isotopic abundance ratios, to detect and study exoplanets in crowded environments, and to measure line profile variations to map the presence of magnetic activity on stellar surfaces.

Therefore, in the framework of the large variety of spectroscopic surveys and multi-object instrumentation that have mapped, are mapping, or are planned to map, the stellar populations of our Galaxy and nearby galaxies (e.g. {\sl Gaia}-ESO, APOGEE, GALAH, LAMOST, 4MOST, WEAVE, MOONS, MWM, DESI, WST), we discuss the opportunity to have a new instrument for the ESO-VLT in the context of the VLT2030 roadmap. This is planned to be a very-high resolution, multi-object spectrograph (HRMOS) to replace FLAMES on UT2 \citep{pasquini2002}, inheriting the same field-of-view (FoV) of about 25\,arcmin in diameter. HRMOS is proposed to have a much higher spectral resolution ($R>60\,000$) than all current, and upcoming, spectrographs dedicated to spectroscopic surveys of Galactic stellar populations, with a relatively high multiplexing capability (20-100).
In this respect, it will have a significant advantage over very high-spectral resolution instruments such as ANDES \citep{ANDES} which will be mounted on the E-ELT, PEPSI \citep{PEPSI} at LBT, and ESPRESSO \citep{ESPRESSO} at VLT,  all of which are single-object spectrographs. 

The most salient features of HRMOS will be:
\frame{
\begin{itemize}
\item High-precision determination of elemental abundances with uncertainties down to 0.01\,dex;
\item Detection of weak and blended lines; and of isotopic features in stellar spectra; 
\item Precision and stability (also in the long term) in radial velocity as low as $10$\,m\,s$^{-1}$;
\item Simultaneous observations of $20-100$ objects in a $\sim25$\,arcmin diameter field.
\end{itemize}
}

\begin{figure}
\centering
\resizebox{0.999\hsize}{!}{
\includegraphics{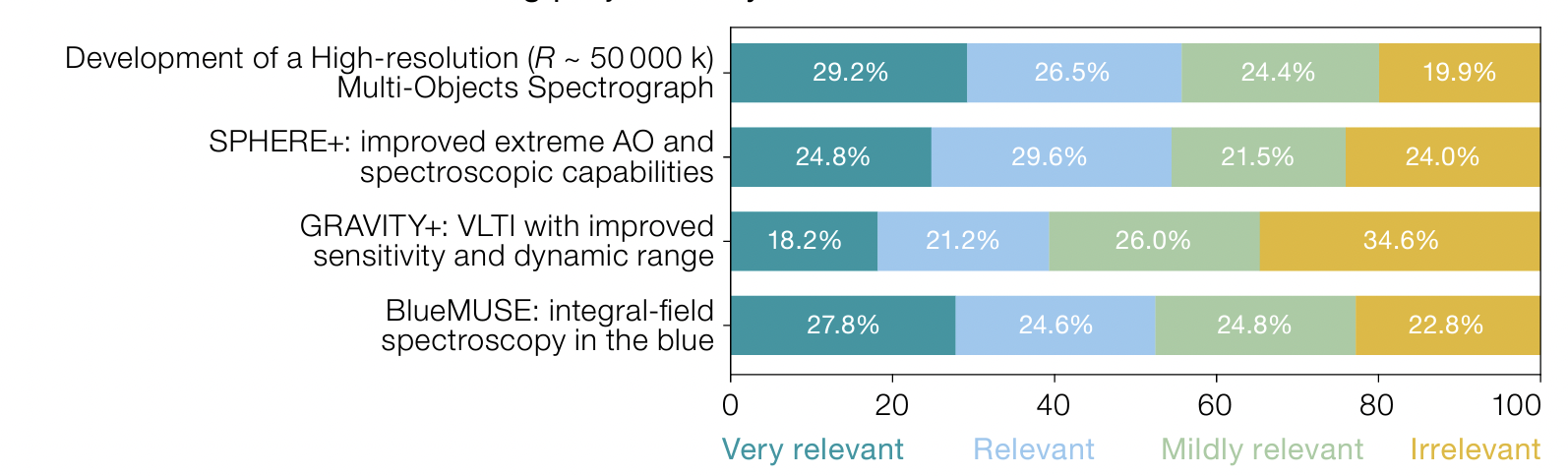}}
\caption{
\label{fig:eso_fut}
Answers to the question \textit{``How relevant are the following projects for your research?''} that ESO posed in a Scientific Prioritisation Community Poll in 2020, following the workshop VLT2030. Several projects were discussed, and four projects were selected by the Science and Technical Committee (STC) for further review (Figure adapted from Fig.~8 in \citealt{merand21}).}
\end{figure}

We stress that the need for such an instrument was highlighted by the results of the latest ``Report on the Scientific Prioritisation Community Poll'' conducted by ESO \citep{merand21}, as shown in Fig.~\ref{fig:eso_fut}. 

\begin{figure*}[ht]
\centering
\resizebox{\hsize}{!}{
\includegraphics[height=5cm]{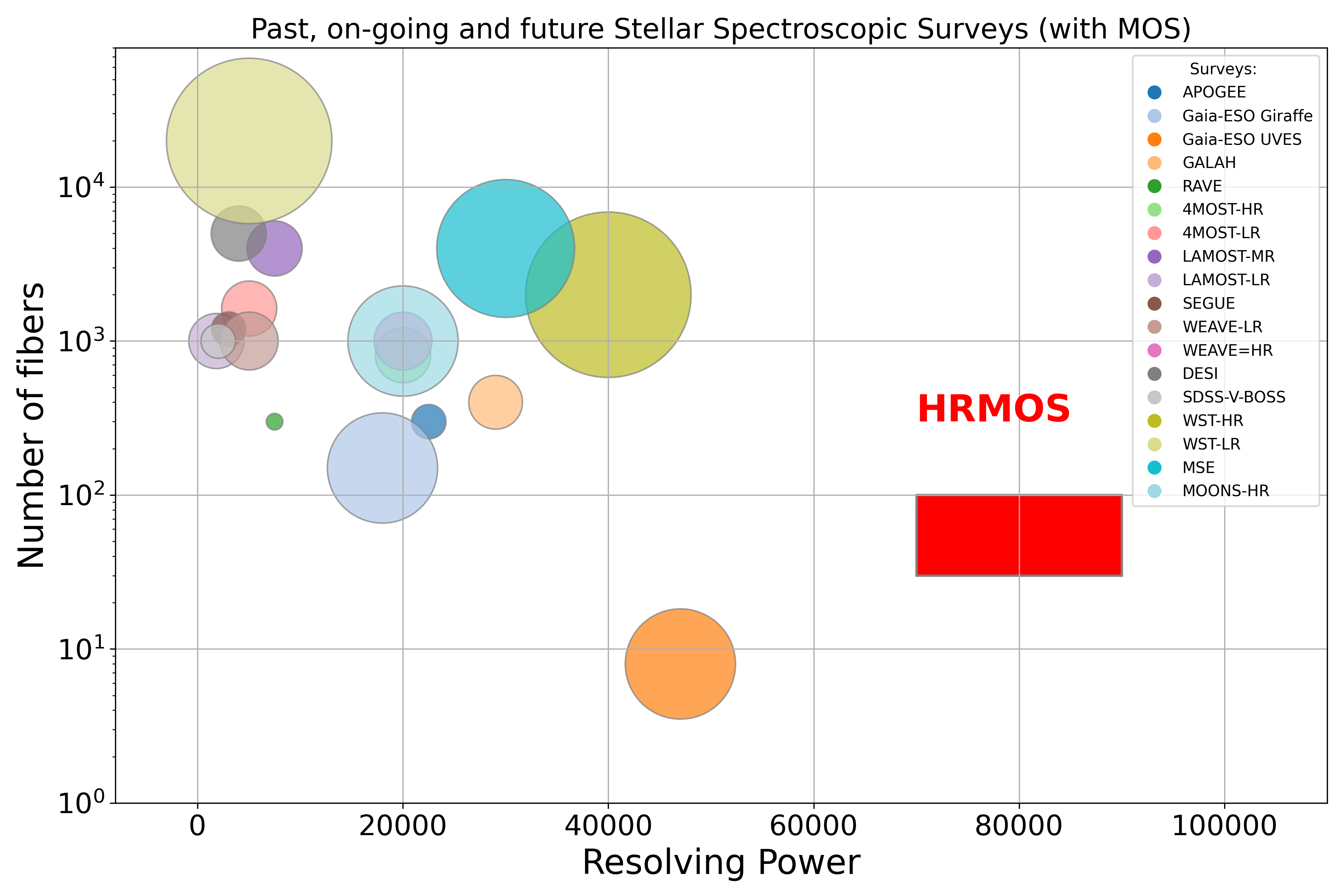}
}
\caption{
HRMOS in the landscape of spectroscopic surveys with completed, ongoing, and future MOS instrumentation: Number of fibres as a function of the resolving power ($R$). The size of the symbol is proportional to the collecting area of the telescope. HRMOS is represented with a red rectangle, with $R$ centred on $80\,000$, and the number of fibres from $20$ to $100$. 
\label{fig:fibers} 
}
\end{figure*}

The main characteristics of HRMOS (resolving power, fibre number, and collecting area of the telescope) compared to the major spectroscopic surveys/facilities completed, in progress, or planned for the future are shown in Fig.~~\ref{fig:fibers}. It is clear that HRMOS stands out as it provides a unique combination of high resolution, and a large collecting area, with considerable multiplexing.

In this document, we aim to discuss highly relevant scientific cases, for which the characteristics of an instrument such as HRMOS are mandatory and that cannot be addressed with current and planned facilities. The main science drivers of HRMOS are shown in Fig.~\ref{fig:sci_cases}.

These include the detailed study of the chemical compositions of stars traced by a wide range of elemental abundances, from the lightest to the heaviest, as well as their isotopic abundance ratios, whenever possible. In addition, HRMOS will allow us to study the properties of young stars and the prominent, although often overlooked, role of magnetic fields. The multi-object aspect of HRMOS is crucial to the analysis of the stellar populations in clusters and dwarf galaxies in the Local Group, and the very high resolution will provide a level of precision never before obtained on a systematic basis on large samples. Finally, HRMOS will be revolutionary to the detection of extrasolar planets and their characterisation in environments that have been very little explored so far.

\subsection{Instrument overview}

HRMOS is planned as an instrument that aims to take advantage of the large collecting area of the VLT. It will provide an upgraded version of the existing FLAMES instrument. In the next sections, we describe several science cases that, through the analysis of a set of simulated spectra, have driven the definition of the main features and characteristics of HRMOS. A short summary of the requirements based on the scientific cases is presented in Table~\ref{tab:sci_request} and described in each of the following sections. The common requirements are: 
\begin{enumerate}
    \item a very high spectral resolution, expected to range from $R=60\,000$ to $R=80\,000$
    \item a relatively high number of fibres ($>20$) in the $25-30$\,arcmin diameter field-of-view of the VLT;
    \item a wide spectral range (possibly divided in four windows),  including, the following approximate  wavelengths:
    \begin{itemize}
      \item $\sim770-800$\,nm: O\,{\sc i} triplet,   $^{12}$C$/^{13}$C isotopic ratio
        \item $\sim630-670$\,nm: H$\alpha$,  [O\,{\sc i}], CN molecular bands  [$647-649$\,nm] 
        \item $\sim510-570$\,nm:  C$_2$ band heads at 516.5 and 563.5\,nm
        \item $\sim380-420$\,nm:  neutron-capture elements, including Pb and Th
    \end{itemize}
\item high stability, also on the long term, to measure precise radial velocities, with expected uncertainties of at least 10\,m\,s$^{-1}$.
\end{enumerate}

To date, two main concept designs and a hybrid solution have been developed \citep{Brucalassi2022}:
\begin{enumerate}
    \item 4-Arms Design: based on a HERMES-like instrument, with a very large simultaneous wavelength coverage, but with challenging optics and dimensions.
    \item 1-Arm Design: based on a compact instrument, with interchangeable dispersing elements and feasible optics, but with less simultaneous wavelength coverage.
    \item Hybrid Solution: based on an intermediate solution, with a simultaneous wavelength coverage on three wavelength bands but not independent and with a challenging optical layout.
\end{enumerate}

In Sect.~\ref{sec:InstrumentBaseline} more detailed descriptions of the three proposed designs are given, highlighting the strengths and weaknesses of each configuration to address the scientific requirements.


\begin{center}
\begin{table*}[ht]%
\centering
\caption{Instrument requirement summary based on the main scientific cases}%
\footnotesize
\begin{tabularx}{\textwidth}{l|X|X|X|X|X}
\hline
\textbf{Parameter} & \textbf{Magnetic Fields}  & \textbf{Exoplanets} & \textbf{Star Clusters} & \textbf{Galactic Archaeology} & \textbf{MW Satellites}  \\
\hline
Resolving Power ($R$) &  $> 60\,000$ & $80\,000$ &  $>60\,000$ &  $80\,000$ &  $60\,000$ \\
Spectral range and lines & $380-850$\,nm; at least 75\,nm simultaneous coverage & As large as possible. It should allow simultaneous monitoring of Ca\,{\sc ii} H\&K lines as well as a window at $500$\,nm & Various windows between $380-800$\,nm & Various windows between $380-800.0$\,nm & 402\,nm (Th), 406\,nm (Pb), 481\,nm (Zn). Other key elements have some flexibility (of various degree)\\  
\hline
Multiplexing   &  50 & 100 & $>20$ & $20-100$ & 100   \\
\hline
Stability      &  $<$1 km s$^{-1}$ &  10 m s$^{-1}$ & 50 m s$^{-1}$ & 100 m s$^{-1}$ & $<200$ m s$^{-1}$  \\
\hline
\end{tabularx}
\label{tab:sci_request}
\end{table*}
\end{center}

\subsection{Structure and aim of this document}

After providing a summary of current concept designs, this document presents an overview of those science drivers, dividing them into five thematic chapters dedicated to: 
\begin{itemize}
\item Young stars, magnetic activity and magnetospheric accretion;
\item Exoplanets in crowded environments: in star clusters, in the bulge and in other galaxies; 
\item Very high-resolution spectroscopy of star clusters; 
\item Galactic Archaeology and Cosmological ages;
\item The Milky Way satellite galaxies. 
\end{itemize}
These scientific cases show the need for an instrument like HRMOS to achieve progress that would otherwise be unattainable with current or planned instrumentation in the near future. 
In each chapter, a summary of the technical requirements driven by the particular scientific case is included, highlighting the uniqueness of HRMOS in the global landscape.  The main scientific drivers that HRMOS will allow us to address are visually sketched in Fig.~\ref{fig:sci_cases}.

This document is not intended to be exhaustive but aims to introduce a number of relevant driving cases as well as to identify the technical requirements associated with them. At the moment, there are no other planned spectrographs with the combination of multiplexing, resolving power, wavelength coverage, and efficiency when mounted on an 8-m-class telescope. This makes it likely that the science cases we are proposing will still be unsolved in the early 2030s.

%% file: instrument.tex
\clearpage
\section{ Instrument Baseline Designs}
\label{sec:InstrumentBaseline}

This section presents the main concept designs currently developed for the new HRMOS instrument \citep[see also][]{Brucalassi2022}, according to the scientific requirements discussed in the following sections of this document. 

\subsection{4-Arms concept design}
\label{sec:4Arms_solution}

\begin{figure*}[h]
\centering
\resizebox{\hsize}{!}{
\includegraphics[height=5cm]{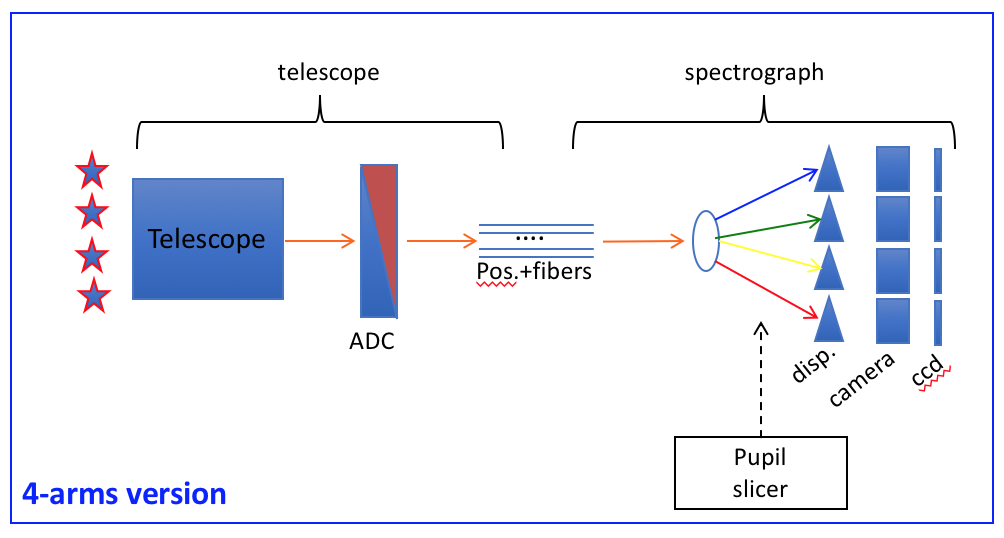}
\includegraphics[height=5cm]{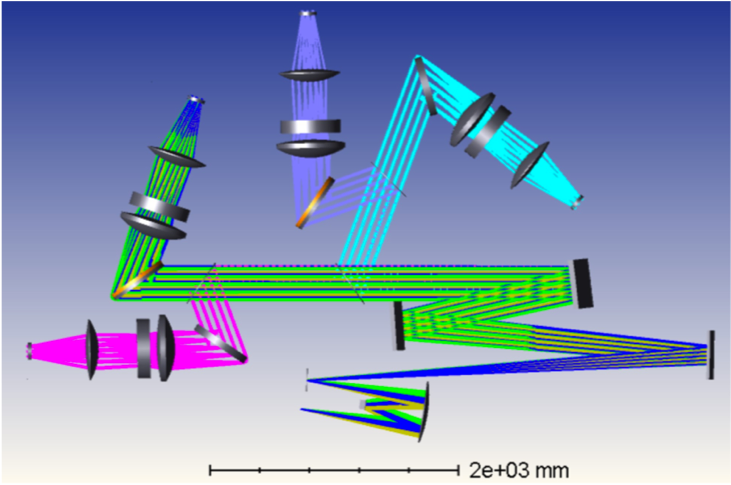}
}
\caption{
Instrument topology (left) and optical layout (right) of the 4-Arms concept design for HRMOS.
\label{fig:4Arms} 
}
\end{figure*}

A first proposed solution for HRMOS is based on a 4-Arms design, with a configuration similar to that of the HERMES spectrograph \citep{HERMES}, the four-channel fibre-fed spectrograph with high resolution and multi-object capability on the Anglo-Australian Telescope (AAT).
Figure~\ref{fig:4Arms} shows the instrument topology and optical layout of the HRMOS 4-Arms concept design. Following the light path, after an Atmospheric Dispersion Corrector (ADC) and a pupil slicer, the beam passes inside the instrument from the fibre-fed slit, through a folding mirror and a collimator, to a sequence of dichroic beamsplitters where the light is divided into four beams. Each beam then passes to a grating and finally to a camera and a detector, providing simultaneous observations in four fixed optical bands. The main optical specifications are summarised in Table~\ref{tab:DesignSpec}. 

This solution allows for very large simultaneous wavelength coverage ($\sim110$ nm) and thus, for the same field, shorter observing times to cover the full wavelength range. However, the large dimensions, the presence of four dichroics, four dispersing elements, four cameras, and four detectors, makes the realisation of this design demanding in terms of costs. Moreover, such a configuration would need the use of a huge ADC, placed very close to the VLT focal plane: from a technical point of view, this is an extremely interesting and challenging opportunity, but, on the other hand, it would have significant impact on the other instruments attached to the telescope focal plane.

\begin{center}
\begin{table*}[t]
\centering
\caption{Baseline design specification}%
\label{tab:DesignSpec}
\footnotesize
\begin{tabularx}{\textwidth}{|l|X|l|X|}
\noalign{\smallskip}
\hline
\hline
\textbf{Parameter} & \textbf{4-Arms Design} & \textbf{1-Arm Design} &  \textbf{Hybrid Design}   \\
\hline
Resolving power $R$     & 80\,000   & 80\,000 & 80\,000  \\
Multiplexing   & 137    & 50 & 132 \\
Slicer & Pupil & Fibre & Fibre \\
N of slicing & 4 & 31 or 37 & 19 \\
Anomorphism & 4 & & \\
Fibre Size on Sky  & 1 arcsec & 1.2 arcsec & 1.2 arcsec \\
Fibre Size on the spectrograph &    & 30 microns & 53 microns \\
Slit length & 175.1 mm & 200 mm & 294 mm \\
Collimated beam size & 305x303 mm & 400 mm (Diam.) & 280x280 mm        \\
Collimator F/\# & 4.9 & 3.5 & 10x10  \\
Collimator Focal length & 1500 mm & 1400 mm & 2800 mm \\
N of dispersing elements & 4 & 4 exchangeable & 1           \\
 Blaze angle & 54.6 deg & 37.5 deg & 48.7 deg \\
Disperser size & challenging & commercial & feasible \\
Spectral bandwidth (in the blue) & 15.5 nm & 36 nm & 24 nm \\
Spectral sampling & 2.5px & 2.7px & 2.3px \\
N Camera lens & 4 simple & 1 feasible & 3 simple \\
Camera lens F/\# & 2.64x2.64 & 3 & 2.2 \\
Camera lens focal length & 789.47 mm & $\sim$1200 mm & 606 mm \\
Dichroics & 4 & No & 3 \\
N Detectors & 4 (separated) & 1 (mosaic) & 3 (separated) \\
Detector size & 9K x 9K & 10K x 10K  & 9K x 9K \\
Pixel Size  & 10 microns  & 9 microns & 10 microns \\
ADC & yes (challenging) & No & TBD \\

\hline
\end{tabularx}
\end{table*}
\end{center}

\subsection{1-Arm concept design}
\label{sec:1Arm_solution}

\begin{figure} 
   \begin{center}
   \begin{tabular}{c} 
   \includegraphics[height=5cm]{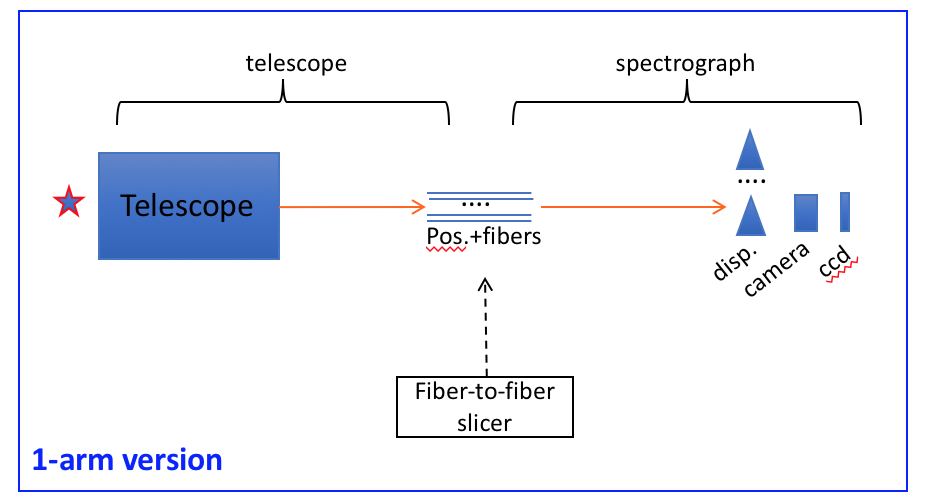}\\
   \includegraphics[height=5cm]{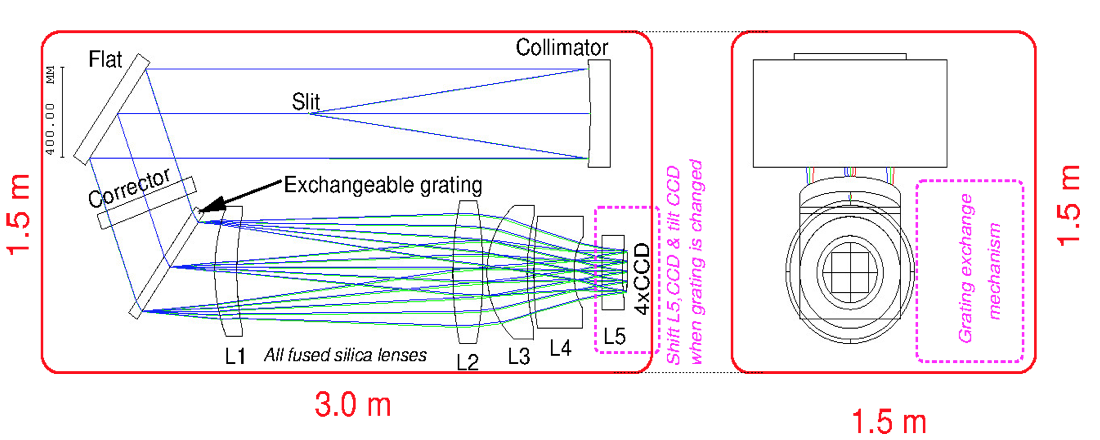}\\
\includegraphics[height=5cm]{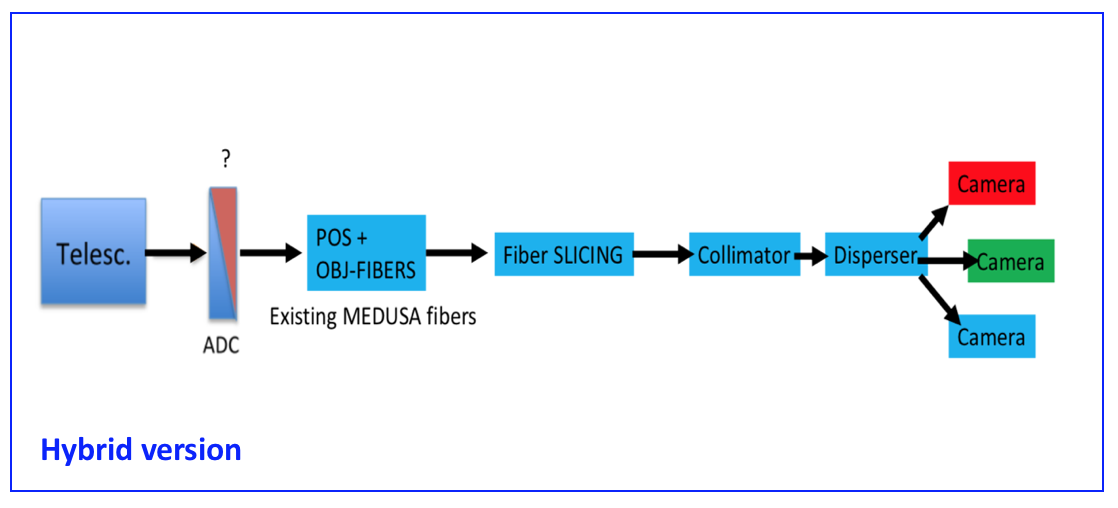}\\
   \includegraphics[height=5cm]{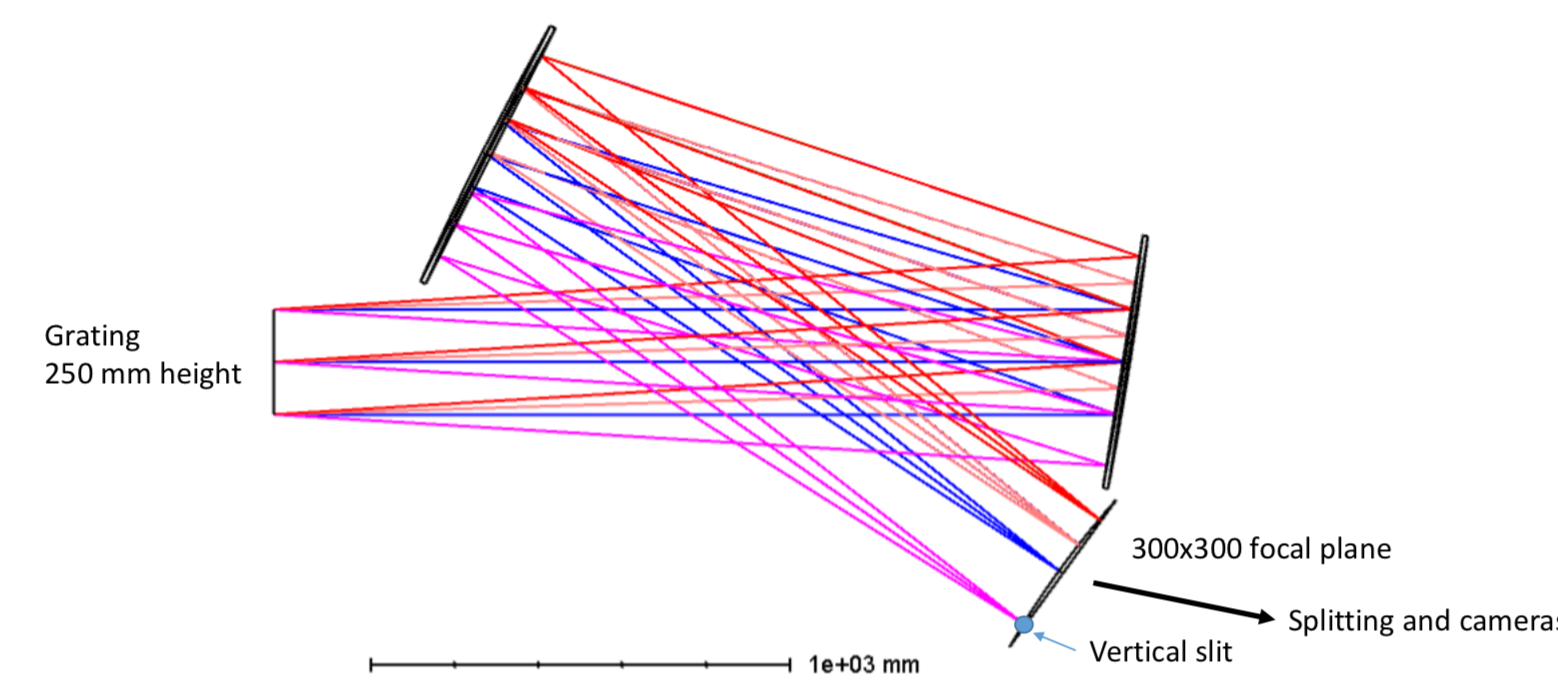}  
   \end{tabular}
   \end{center}
   \caption[1Arm] 
   { \label{fig:1ArmHybrid} 
{\sl Upper panels:} Instrument topology and optical layout of the 1-Arm concept design for HRMOS. {\sl Lower panels:} Instrument topology and optical layout for the collimator of the Hybrid concept design.}
   \end{figure}

Figure~\ref{fig:1ArmHybrid} shows in the two upper panels the instrument topology and optical layout for the second solution proposed for HRMOS, based on a 1-Arm design. The main optical specifications are summarised in Table~\ref{tab:DesignSpec} for a configuration at $R=80\,000$. The concept design for this solution foresees a collimator and a corrector system, followed by up to a maximum of four exchangeable dispersing elements without the use of an ADC or dichroics, a camera based on five lenses (with the last one movable accordingly with the grating selection), and a mosaic of four CCDs 10K$\times$10K (9 micron of pixel size). The introduction of a plane-folding mirror ensures a compact layout for the system. 
The simultaneous wavelength coverage for the central wavelengths $\lambda_{c}=390$ nm, $520$ nm, $660$ nm, and $860$ nm will be $\Delta\lambda= 36$ nm, $48$ nm, $61$ nm, and $80$ nm, respectively. The crucial optical element for this design are the camera lenses. However, because of the relatively slow working f-number, the overall design complexity is relaxed and the performance is good for all the selected wavelength bands and for a huge focal plane dimension. On the other hand, the absence of an ADC makes simultaneous spectral acquisition from 390 nm to 860 nm (as central wavelengths) impossible, increasing the observation times.

\subsection{Hybrid solution}
\label{sec:Hybrid_solution}
In order to solve the dichotomy between the 4-Arms and 1-Arm solutions, an intermediate concept design is also under study. Figure~\ref{fig:1ArmHybrid} (two lower panels) shows the instrument topology and a possible collimator layout for this hybrid concept. Following the optical path, the beam after fibre slicing is redirect to a collimator system and a single disperser element (Lithographic/Holographic Etched Grating). Then, the light is divided into three wavelength bands using dichroics and imaged in three different CCDs by the three camera lens. Although this layout can ensure in total a large simultaneous wavelength coverage, a critical point is that the three wavelength bands are not independent. Moreover, the location of the dichroics after the disperser makes the optical design more difficult, forcing the camera lens to be quite far away.

Finally, Table~\ref{tab:feasib_instr} summarises the instrument requirements and contains a note on the feasibility of each baseline design.

\begin{center}
\begin{table*}
\centering
\caption{Instrument requirement summary based on the main scientific cases with included a feasibility note for each baseline design solution : 4-Ar: 4-Arms solution; 1-Ar: 1-Arm solution; Hyb: Hybrid solution. Green colour: feasible goal; orange colour: challenging; red colour: impossible.}%
\footnotesize
\begin{tabularx}{\textwidth}{|l|X|X|X|X|X|}
\noalign{\smallskip}
\hline
\hline
\textbf{Parameter} & \textbf{Young stars} & \textbf{Exoplanets} &  \textbf{Star clusters} & \textbf{Galactic Archaeology}    & \textbf{Satellites}  \\
\hline
Resolving power $R$     &   80\,000 & $>$60\,000 & 80\,000 &  80\,000 &  60\,000  \\
\hline
            & 
            \textcolor{teal}{\textbf{4-Ar/1-Ar/Hyb}}   &
            \textcolor{teal}{\textbf{4-Ar/1-Ar/Hyb}}   & 
            \textcolor{teal}{\textbf{4-Ar/1-Ar/Hyb}} & 
            \textcolor{teal}{\textbf{4-Ar/1-Ar/Hyb}} &
            \textcolor{teal}{\textbf{4-Ar/1-Ar/Hyb}}   \\
\hline
Spectral range & as in Table \ref{tab:sci_request} & as in Table \ref{tab:sci_request}& as in Table \ref{tab:sci_request}& as in Table \ref{tab:sci_request}&as in Table \ref{tab:sci_request} \\
\hline
            & 
            \textcolor{teal}{\textbf{4-Ar}}/\textcolor{red}{\textbf{1-Ar}}/\textcolor{teal}{\textbf{Hyb}} &
            \textcolor{teal}{\textbf{4-Ar}}/\textcolor{red}{\textbf{1-Ar}}/\textcolor{red}{\textbf{Hyb}}  & 
             \textcolor{teal}{\textbf{4-Ar/1-Ar/Hyb}}   & 
           \textcolor{teal}{\textbf{4-Ar}}/\textcolor{orange}{\textbf{1-Ar}}/\textcolor{orange}{\textbf{Hyb}} &
            \textcolor{teal}{\textbf{4-Ar/1-Ar/Hyb}}     \\
\hline
Multiplexing   &  20-100 & $>$20  & 50 & 100 & 100 \\
\hline
            & 
            \textcolor{teal}{\textbf{4-Ar}}/\textcolor{teal}{\textbf{1-Ar}}/\textcolor{teal}{\textbf{Hyb}}&
            \textcolor{teal}{\textbf{4-Ar}}/\textcolor{red}{\textbf{1-Ar}}/\textcolor{teal}{\textbf{Hyb}}&     
             \textcolor{teal}{\textbf{4-Ar}}/\textcolor{teal}{\textbf{1-Ar}}/\textcolor{teal}{\textbf{Hyb}}&
            \textcolor{teal}{\textbf{4-Ar}}/\textcolor{orange}{\textbf{1-Ar}}/\textcolor{teal}{\textbf{Hyb}}&
            \textcolor{teal}{\textbf{4-Ar}}/\textcolor{red}{\textbf{1-Ar}}/\textcolor{teal}{\textbf{Hyb}}\\
\hline
Stability      &  100 m s$^{-1}$ & 50 m s$^{-1}$   & $<$1 km s$^{-1}$ & 10 m s$^{-1}$ & <200 m s$^{-1}$ \\
\hline
            & 
            \textcolor{teal}{\textbf{4-Ar}}/\textcolor{teal}{\textbf{1-Ar}}/\textcolor{teal}{\textbf{Hyb}}&
            \textcolor{red}{\textbf{4-Ar}}/\textcolor{orange}{\textbf{1-Ar}}/\textcolor{orange}{\textbf{Hyb}}&
            \textcolor{teal}{\textbf{4-Ar}}/\textcolor{teal}{\textbf{1-Ar}}/\textcolor{teal}{\textbf{Hyb}}&
           \textcolor{orange}{\textbf{4-Ar}}/\textcolor{teal}{\textbf{1-Ar}}/\textcolor{teal}{\textbf{Hyb}}&
            \textcolor{orange}{\textbf{4-Ar}}/\textcolor{teal}{\textbf{1-Ar}}/\textcolor{teal}{\textbf{Hyb}}\\
\hline
\end{tabularx}
\label{tab:feasib_instr}
\end{table*}
\end{center}

\subsection{Fibre positioning}

The HRMOS fibre positioning system is a crucial aspect for meeting the various scientific needs of the instrument. Several MOS positioner architectures have been designed for the VLT; of particular interest are OzPoz on FLAMES \citep[]{pasquini2002} and the MOONS fibre positioning module \citep[]{Watson2022}. The former uses a pick-and-place robot to position fibre buttons on one of two plates, and the latter uses individual fibre positioners in an overlapping theta-phi configuration. Both instruments reflect established positioner architectures with similar examples in operation at other telescopes and in active development.

OzPoz has the potential to be directly upgraded to meet the needs of HRMOS. The reduced cost and risk of such an upgrade would need to be traded off against any limitations imposed by any retained hardware; in particular in terms of instrument lifetime and reconfigure/acquisition time. A new state-of-the-art pick-and-place system, such as WEAVE \citep[]{dalton2016}, would significantly reduce reconfigure times, especially for the number of fibres considered in HRMOS. However, one disadvantage of this architecture is that the pick-and-place robot represents a single-point failure for the entire instrument.

An alternative design using independent positioners based on the MOONS design but with reduced density of scaled-up mechanisms could offer a balance of capability with minimised design effort and risk. However, bringing many fibres together requires considerable overlap of the patrol fields. For example, MOONS achieves 200\% coverage across the focal plane (two fibres can reach any point), but, with a theta-phi configuration, this results in challenging path-planning to reach the observation positions. For this reason, a novel positioner concept based on MOONS but employing a theta-r architecture is being investigated. This retains the primary rotation axis, but the second degree of freedom is achieved using a linkage to extend an arm radially outward from the axis. This presents the opportunity to significantly increase the reach of each positioner, and so increase the overall coverage, while maintaining simple path routing. This proposal could retain significant commonality with the overall MOONS front end in terms of field correction, acquisition cameras, meteorology system, control architecture, etc., therefore offering a significant simplification for the design stages of HRMOS. This presents an opportunity to address key aspects of the HRMOS science cases such as fast reconfiguration, including positioning fibres close together to observe clusters, and allowing for active correction for differential atmospheric dispersion during an observation. 

\begin{figure}[ht]
   \begin{center}
   \includegraphics[height=10cm]{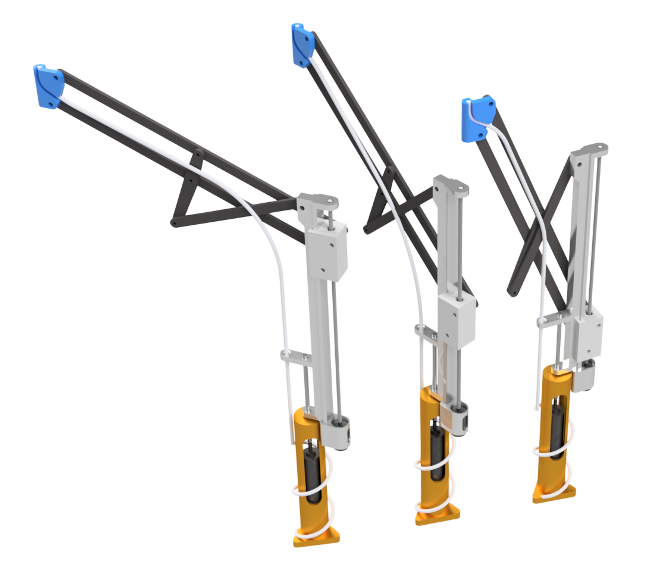} 
   \end{center}
   \caption[FPOS] 
   {Concept of a potential theta-r fibre positioner embodiment, incorporating Scott-Russel and parallelogram linkages to provide the linear movement. This Fibre positioner system could be integrated to a focal plate and rotating front end structure replicating the MOONS instrument at the VLT.}
   \end{figure}

\begin{table} 
    \centering
    \caption{Physical parameters of the new r-theta fibre positioner concept, providing up to 150 positioners in the field of view with 1000\% field coverage (10 fibres to any point in the field).
    \label{tab:my_label}}
    \begin{tabular}{|c|c|c|}
    \hline
        \textbf{Parameter} & \textbf{Optical Dimension} & \textbf{Physical Dimension}\\
    \hline    
        \textbf{Overall FoV} & 25 arcminute & 880 mm \\
    \hline
        \textbf{Positioner Spacing} & 1.8 arcminute & 65 mm \\
    \hline
        \textbf{Positioner Reach} & 3.1 arcminute & 110 mm \\
    \hline
        \textbf{Min Fibre Spacing} & <10 arcsec & <6 mm\\
    \hline
    \end{tabular}
\end{table}

\subsection{ETC development}
\label{sec:Preliminary ETC devolopment}

\begin{figure} 
   \begin{center}
   \begin{tabular}{c} 
   \includegraphics[height=10cm]{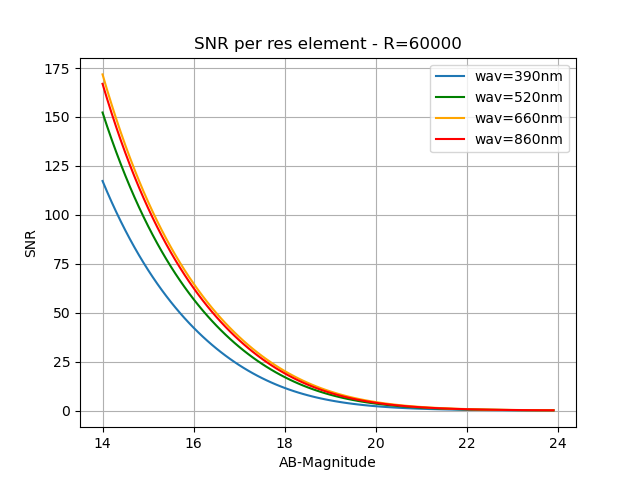}\\
   \includegraphics[height=10cm]{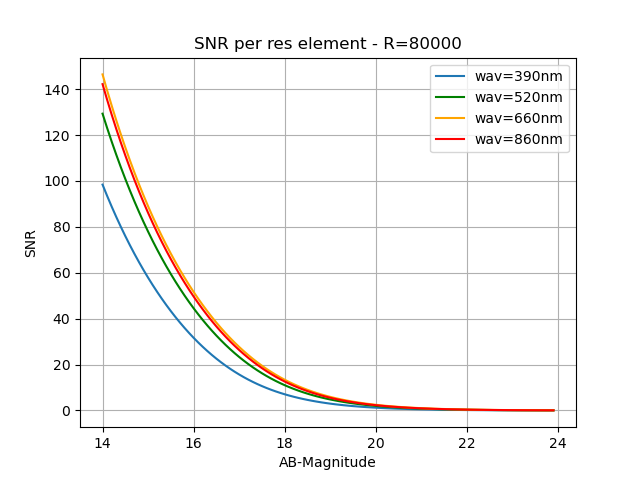}
   \end{tabular}
   \end{center}
   \vspace{-5mm}
   \caption[SNR6000080000] 
   { \label{fig:SNR6000080000} SNR per resolution element in the AB magnitude system, with central wavelengths at 390\,nm, 520\,nm, 660\,nm, 860\,nm,  and an exposure time of 3600\,s and $R=80\,000$ (upper panel) and $R=60\,000$ (lower panel). Telescope efficiency (including detector): $0.48$ (at $390$ nm), $0.55$ (at $520$ nm), $0.55$ (at $660$ nm) and $0.44$ (at $860$ nm); instrument efficiency: $0.15$, $0.17$, $0.19$, $0.21$ and slit efficiency: $0.64$, $0.69$, $0.72$, $0.75$ respectively.}
   \end{figure}
A preliminary version of Exposure Time Calculator (ETC) has been developed to predict the performance of the spectrometer for different parameters and environmental conditions in order to better define the science case drivers presented in this work. The theoretical equations used are based on the Andes ETC baseline (\url{https://drive.google.com/file/d/18dM5OzTxdvshHY86XPMlfDsbm0hN6juZ/view}).
Figure~\ref{fig:SNR6000080000} shows the SNR per resolution element in the AB magnitude system for the same four central wavelengths considered in the optical designs and a $R=80\,000$ or $R=60\,000$. Telescope and instrument parameters are taken from Table~\ref{tab:DesignSpec} (1-Arm solution). Default values for the efficiency of the telescope outside the atmosphere are obtained from the ESO ETC website (\url{https://www.eso.org/observing/etc/)}.
Throughout the text, the term SNR will be referred to as ``SNR per resolution element".

%% file: young_stars.tex
\clearpage

\section{Magnetic Fields: Shaping the Early Lives of Stars and Planetary Systems}

\label{sect:ystars}

Dynamo-generated magnetic fields play a prominent role in shaping the evolution of stars and planetary systems, particularly at young ages. Magnetic fields strongly influence the atmospheres and spectra of young stars through non-radiative heating processes, control their accretion of matter, the geometry of their winds and outflows and rates of angular momentum loss, and potentially determine the habitability of their planets. The high spectral resolution of HRMOS will allow us to spatially resolve magnetic structures with Doppler tomography and map the dynamos, starspots, prominences, magnetospheres, jets and winds of young stars.
Whilst such studies have been possible before for individual examples, a multiplexing instrument like HRMOS on the VLT permits the simultaneous synoptic monitoring of unprecedentedly large stellar samples in many young star clusters of known age and composition.

During their early lives, young, low-mass stars ($0.1-1.5$\,M$_{\odot}$) are rapidly rotating, magnetically active, and may be accreting gas from a circumstellar disc. As they get older, discs disperse leaving fully-formed planetary systems, and stars spin down, becoming less magnetically active. The magnetic field, generated by a rotation-driven dynamo process, is thought to play a leading role in all of these events: the stressing and twisting of buoyant magnetic fields leads to manifestations such as starspots or chromospheric and coronal heating; the magnetic field can also couple to the stellar wind or to any surrounding accretion disc providing a means to regulate and lose angular momentum, and this leads to slower rotation which, in turn, feeds back into the dynamo mechanism \citep[e.g.][]{weber1967,skumanich1972,durney1978,noyes1984,koenigl1991}.

Recent observations at millimetre wavelengths, in the near infrared (NIR) with adaptive optics and interferometric observations have started to spatially resolve structure and planets on scales of $1-100$\,au around young stars \citep[e.g.][]{lagrange2010,andrews2018,garcialopez2020, Benisty2022a}. However, there is no prospect of spatially resolving or imaging inner discs, magnetospheres, close-in planets or stellar surfaces using similar techniques. Instead, we must use indirect imaging techniques, of which the most important is Doppler tomography \citep[e.g.][]{marsh2001, strassmeier2009}. This relies on high-resolution, time-resolved spectroscopy to track the velocity evolution of emission or absorption features due to co-rotating structures associated with magnetic fields and accretion.

Led by observations of magnetic activity, rotation rates and accretion discs, phenomenological models have been built which can reasonably explain the evolution of spin and magnetic activity in ensembles of coeval stars in groups and clusters \citep[e.g.][]{kawaler1988,Denissenkov10,reiners2012,gallet2013,gallet2015}. However, a full understanding of the components of these models and how they interact with each other is missing. There are clear signs that the dynamos in fast-rotating young stars are not just scaled-up versions of the solar dynamo - young stars are extensively covered by cool spots, sometimes at high or even polar latitudes \citep[e.g.][]{barnes2000,barnes2001}; the evidence for magnetic cycles in the most active stars is weak; the dynamo appears to operate even when stars are fully convective and may “saturate” or even “supersaturate” at high rotation rates \citep{jeffries2011,wright2011,wright2018}. Young stellar chromospheres and coronae are orders of magnitude more powerful than those of the Sun; mass loss rates may be much higher and the role of extensive prominence systems inferred around some young stars in regulating angular momentum loss or triggering giant flares is unclear \citep[e.g.][]{colliercameron1990,wood2005,jardine2019,jardine2020}. The large-scale geometry of magnetic fields, which is crucial to angular momentum loss \citep[e.g.][]{matt2012,vidotto2014,reville2015} has begun to be explored and appears to vary with age, rotation rate and spectral type \citep[e.g.][]{donati2008b,donati2009,donati2011}. Whilst the role of coupling to an accretion disc in regulating angular momentum loss is also vital \citep[e.g.][]{koenigl1991,colliercameron1993}, a detailed understanding of how this works or what role magnetic activity may play in dispersing discs is missing \citep{gallet2019}.

\subsection{Doppler tomography  of magnetic structures and dynamo activity}

Making observations of low-mass stars that give spatial information about the relevant structures associated with their magnetic fields, investigating how these change on a variety of timescales and studying their evolution over tens and hundreds of million years, is crucial in developing a full understanding of these processes and how they combine to control the early evolution of rotation, magnetism, coronal and chromospheric activity, planet formation and migration, and possible star-planet interactions.

\frame{
\titlecol{Key Questions:}
\begin{itemize}
\item
How do the strengths and geometries of magnetic fields at the surface and in the coronae of stars change with age, spectral type and rotation rate?
\item
Does the nature of the magnetic dynamo evolve with time and do manifestations of magnetic activity change in just a quantitative way or more fundamentally?
\item
How is angular momentum lost from contracting pre-main-sequence stars and subsequently after they have reached the zero-age main sequence (ZAMS)? What is the role of stellar prominences and how do large-scale magnetic geometries and the distribution of energy density between global versus small-scale magnetic structures change with age?
\end{itemize}
}

To some extent, observational progress has stalled in these areas, whilst theoretical work has marched on. This is because although there are nearby examples of young field stars that can be studied (one-at-a-time), their ages are uncertain and the large-scale synoptic monitoring programmes that would be required to track the inherently dynamic and changing magnetic structures would require prohibitively large amounts of time on major telescopes.

HRMOS would represent a transformational step forward in this field. The majority of known young stars are found in clusters and star-forming regions at distances $>100$\,pc, where ages and masses can be more confidently assigned by considering the HR diagram of an ensemble. A multiplexing spectrograph offers the opportunity to gather data efficiently on statistically significant numbers of stars in young clusters and associations and to follow their evolution over days, weeks, and years. Such a dataset could broadly sample the mass/age plane and provide a comprehensive and detailed picture of both the mean and range of possible behaviours. High resolution is essential since Doppler tomography relies on resolving the rotational profile into many discrete elements in order to map the features of interest. The higher the resolution, the broader the range of rotation rates that can be sampled, increasing the number of viable targets and extending studies to stars of median rotation level in young clusters, rather than being restricted to “ultra-fast rotators”.

\paragraph{\bf Key projects would include:}
Investigating the distribution and extent of magnetic starspots as a function of age and mass between star-forming regions and ZAMS clusters. Magnetic dynamo models make predictions of these parameters \citep{schuessler1992,solanki2004}. Synoptic monitoring over months and years can reveal the presence (or not) of magnetic activity cycles and the migration of spot patterns. Monitoring on timescales of days and weeks will reveal the details of spot lifetimes and surface differential rotation with important implications for interpreting the one-dimensional light curves provided by Kepler, TESS, CHEOPS and other forthcoming missions like PLATO, and for ongoing radial velocity searches for planets around young stars \citep[e.g.][and Sect.~\ref{sect:exoplanets}]{colliercameron1995,jeffers2007,petit2015,barnes2017}. Note that the investigation of these topics would have also an important impact on understanding how metallicity and abundances in young stars are affected by stellar activity (see Sect.~\ref{sect:clusters}).

Circumstellar prominences map the large-scale structure and dynamics of the extended magnetic fields around young stars. These play a crucial role in angular momentum loss, wind and stellar flare dynamics, star-disc interactions, and possible interactions between stars, close-in planets, and the stellar corona \citep[e.g.][]{dunstone2006,hussain2007,lanza2018,jardine2019,villarealdangelo2019}. Monitoring programmes can locate prominences, determine their physical properties and lifetimes and follow their evolution as they fall back towards the star or escape from beyond the Keplerian co-rotation radius.

Doppler tomography works by tracking contrasts between emission from the immaculate stellar surface and emission/absorption from something on, or above, the stellar surface as it moves across the stellar surface. Analysis of Doppler tomograms can: provide us with contrast maps of the stellar surface, betraying the location and extent of magnetic starspots on young, cool stars \citep[e.g.][]{strassmeier2002,strassmeier2009}; map the magnetic fields and abundance anomalies in magnetic and chemically peculiar early-type stars; trace the heights above the surface and physical conditions in co-rotating prominences \citep{colliercameron1989,jeffries1993} ascertain the location and extent of accretion hotspots and accretion flows \citep{donati2013}; confirm the existence and orbital orientation of transiting planets, even around hot stars \citep[e.g.][]{temple2019}, and many more.

The requirements of Doppler tomography are that a star has a rotationally broadened spectrum that allows smaller and perhaps unresolved features to be identified as they move with respect to an underlying spectral line profile. For example, a starspot has a different spectrum to the immaculate photosphere. As the spot rotates with the star, it produces a bump in the spectral line that moves in a prograde fashion across the line profile. The spotted surface can be forward-modelled to match a time-series of observed line profiles \citep[e.g.][]{vogt1987,kuerster1994,colliercameron1994}. Another example would be the discovery that young stars host prominences, held in co-rotation above the stellar surface by extended magnetic field structures. The prominences absorb H$\alpha$ photons from the star when they are in front of it, producing dark features in a tomogram of the H$\alpha$ line \citep[e.g.][]{colliercameron1989}. The dark features move across the profile at a rate that tells us how high above the stellar surface they are. Observing the features in a number of line diagnostics tells us about the physical conditions in the plasma \citep{colliercameron1990}.


\frame{\vspace{-7mm}
\paragraph{Uniqueness}
Doppler tomography can and has been done on many 4-m$+$ class telescopes around the world. However, these have been for bright, nearby targets and at lower resolving power ($R\sim50\,000$) than proposed here, and targets have been observed one at a time. An 8-m$+$ telescope is required to fully unlock the potential of multiplexing spectroscopy in clusters, by providing sufficient viable targets in single instrument fields. The multiplexing spectrographs WEAVE and 4MOST are on 4-m telescopes and, like MOONS, which is on the VLT, operate at resolving powers of at most $R\sim20\,000$. This gives poor tomographic/spatial resolution in all but the fastest rotating stars of any cluster (see below), severely reducing the number of viable tomographic targets. The Mauna Kea Explorer will be an 11-m telescope and will have a wide field of view and many fibres. However, again, the maximum resolving power will be $R\sim40\,000$ meaning that its tomographic resolution will be inferior and therefore the range of targets it can observe and the effective resolution of its tomography is more restricted. Similar arguments apply for the Wide-field Spectroscopic Telescope \citep[WST,][]{bacon2023}.
}

\subsubsection{Requirements for Doppler tomography}

\begin{figure}
\centering
\resizebox{0.65\hsize}{!}{
\includegraphics{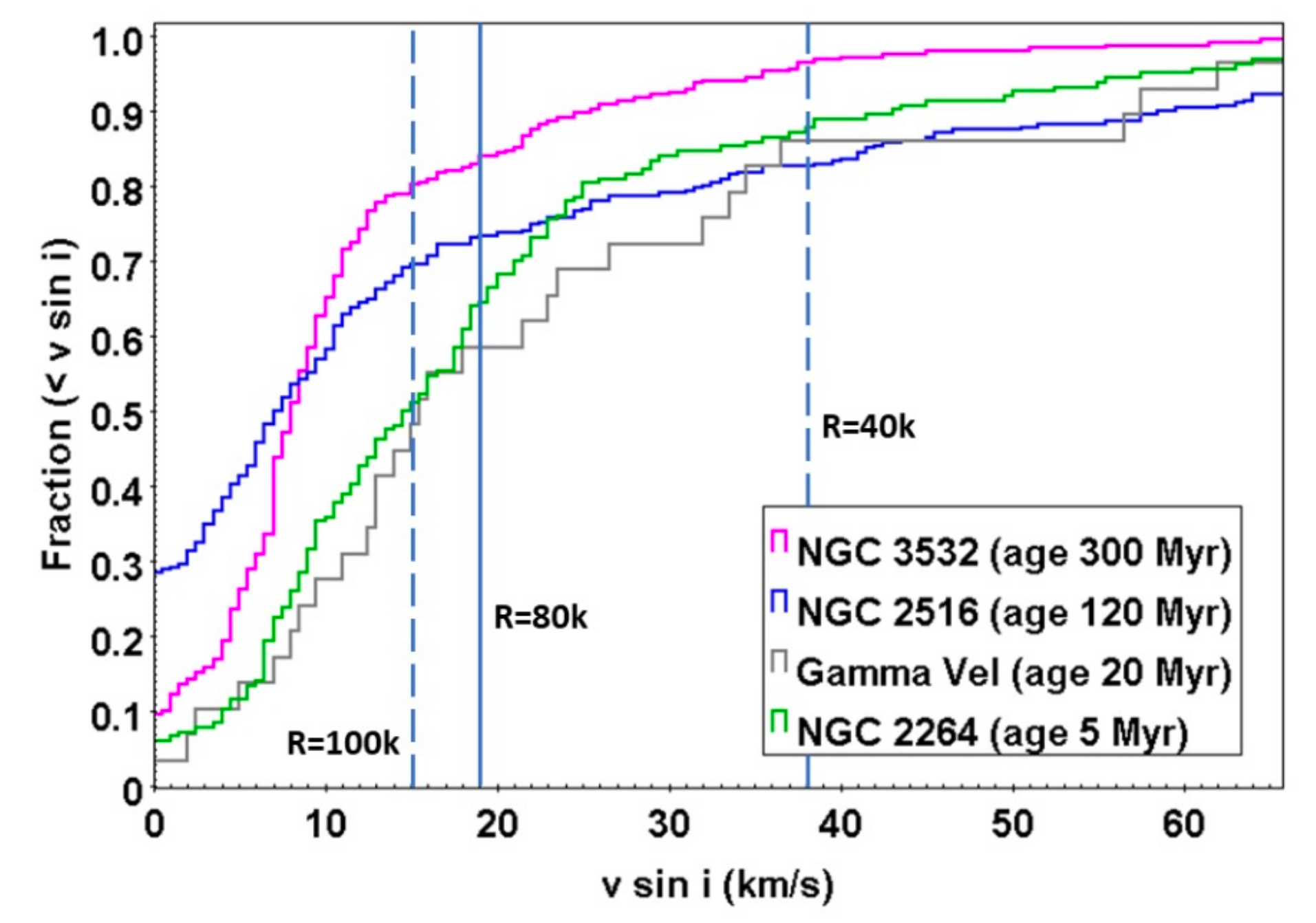}}
\caption{
A normalised cumulative histogram of $v\,\sin i$ for solar type stars 
($4000<T_{\rm eff}/{\rm K}<6500$) in clusters with ages from $5-300$\,Myr \citep[data from the {\it Gaia}-ESO Survey,][]{randich2022}. The vertical lines mark the thresholds where the rotation profile of a star can be resolved into 10 elements for different values of the instrument resolving power. With $R=80\,000$, approximately $20-40$\,\% of stars are viable targets for Doppler imaging in clusters of age $5-300$\,Myr. Increasing $R$ to $100\,000$ would increase these numbers by $\sim 10$\,\%. $R=40\,000$ (similar to the Maunakea Spectroscopic Explorer) would only reach $5-15$\,\%. 
\label{fig:ystars1}
}
\end{figure}

Tomographic methods demand a high enough resolution to divide the stellar rotation profile into many spectral elements; that is, $2 v\, \sin i/{\rm FWHM} > 10$, where $v\sin i$ is the projected equatorial velocity of the star and FWHM is the resolution of the spectrograph, equal to $c/$resolving power ($R$). In practice, stars in young clusters have rotation periods from about 0.25 to a few days and radii of $\sim ~1\,R_\odot$ (depending on mass and age) and 
$v\,\sin i$ of $10-100$\,km\,s$^{-1}$. This leads to the requirement of a spectrograph with a resolving power of at least $R\sim60\,000$ and preferably $R\sim80\,000$ or higher. The higher the resolution, the better the tomographic resolving power and the better the spatial resolution of inferred structure on and around the target stars. A higher resolution also lowers the threshold $v\,\sin i$ for which targets will have resolvable spatial structure and hence increases the range of angular velocities accessible to experiment and increases the number of viable targets in each cluster. A resolving power of $80\,000$ would mean about $20-40$\% of the solar-type stars in a typical young cluster, with $v\,\sin i > 19$ km s$^{-1}$, would give well-resolved surface maps \citep[e.g.][and see Fig.~\ref{fig:ystars1}]{queloz1998}. For lower resolutions, the threshold $v\,\sin i$ at which tomography can be attempted becomes proportionately higher and the number of viable targets decreases rapidly – to the extent that the advantage of a highly multiplexing spectrograph is much reduced because only the very fastest $\sim 10$\% of rotators in a cluster could yield resolved Doppler images.

The magnitude limit for these investigations is set by a combination of the simultaneously
observed wavelength range $\lambda_r$, and the maximum possible exposure time $\tau$. The former is because (at least for spot mapping) each photospheric absorption feature contributes towards the tomographic reconstruction and the effective SNR achieved is $\propto \lambda^{0.5}_{r}$. The latter is because exposure times should be limited to avoid “smearing” unresolved features in the rotation profile across more than one resolution element as the star rotates. For a feature that co-rotates with the star (e.g. a cool starspot or a prominence) at a radius $r$ (where $r=R$ for a surface feature), with a period $P$ and rotation-axis line-of-sight inclination of $i$, then this requirement is:
\begin{equation}
\tau \left( \frac{4\pi^2}{P^2} \right) r\sin i < \Delta v\ ,
\end{equation}
where the left hand side is the maximum radial acceleration multiplied by the exposure time and $\Delta v$ is the velocity resolution element. For a resolving power of $80\,000$ and inclination of $90^\circ$, this leads to
\begin{equation}
\tau < 1000 \left( \frac{P}{1\,{\rm day}} \right)^2 \left( \frac{r}{R_\odot}\right)^{-1}\ {\rm s}\ .
\end{equation}
If our aim is to exploit the full spectral resolution available then exposure times should not be longer than this and given that many targets will have periods of about 0.5\,days and radii $\sim 1 R_\odot$, then exposure times can be no longer than about 250\,s.

The likely reach of such observations can be judged by comparison with previous investigations of similar stars. \cite{barnes2001} were able to construct Doppler images of the spot distribution for two rapidly rotating ($V \simeq 12$, $P = 0.5$\,d, $v\sin i \simeq 60$\,km $s^{-1}$) early M-dwarfs in the solar vicinity using echelle spectrographs (UCLES and UES, resolving power $\simeq 50$k) on 4-m telescopes (the AAT and WHT). The tomograms were created using a series of 300\,s exposures, with individual SNR of $\sim 10$ (per 2.5\,km\,s$^{-1}$ pixel) and combining the signal from several hundred line profiles over a $\sim 175$\,nm wavelength range (using a process called least squares deconvolution). HRMOS should have a better throughput (but possibly a factor of two smaller $\lambda_r$), so similar experiments might be done at $V \leq 15$ with HRMOS. The current exposure time calculator confirms that a SNR of 15 per 2.5\,km\,s$^{-1}$ bin at 520\,nm can be obtained for a solar-type star at $V = 15$ in 250\,s. 

It will be important that the chosen spectral range can be tuned for the spectral-types of the stars to be observed. Cooler stars will benefit from moving the spectral window redward. It will be necessary in any case to have access to multiple diagnostics of prominence activity, from H$\alpha$ down to the Ca\,{\sc ii} H\&K lines at $393 + 397$\,nm, preferably simultaneously.

\subsection{Star-disc interaction and tests of magnetospheric accretion}

The generally accepted picture of star formation is that matter in a circumstellar accretion disc is channelled onto the stellar surface by the stellar magnetic field, which truncates the disc at the corotation radius. This magnetospheric accretion model has received support from several pieces of indirect observational evidence \citep[e.g.][]{hartmann2016}. However, due to the spatial scales involved in the process (typical values for the co-rotation radii are $R_c<0.01$\,au), direct spatially resolved information on the accretion mechanism(s) are lacking. Only very recently has the first direct observation of the magnetosphere in the young pre-main sequence (PMS) star TW Hydrae been obtained with the VLT interferometer \citep{garcialopez2020}.

\frame{
\titlecol{Key Questions:}
\begin{itemize}
\item
Understanding the process of magnetospheric accretion in young stars is a major issue for star and planet formation theories. While the general picture is widely accepted, the way in which this process occurs is very poorly known due to the small spatial scales involved ($< 0.1$\,au). In particular, how are the properties and geometry of the accretion process  related with stellar and disc parameters?
\item
Disc winds and jets are believed to be the main ingredients in regulating angular momentum extraction from accreting discs and contributing to disc dissipation. However,  which mechanisms give rise to these outflows of matter and what is the role of photo-evaporation in driving winds at large disc radii? 
\end{itemize}
}

High-resolution spectroscopy can overcome the lack of spatial resolution through observations of features directly originating from hot gas in shocks created by the accreting material on the stellar surface (the hot spots), and from the magnetically funnelled accretion columns that link the inner disc to the star \citep[e.g.][]{bouvier2007b, alcala2017, gangi22}. In addition, information on the geometry of the magnetospheric accretion can be retrieved looking at radial velocity and intensity variations of emission and absorption features during stellar rotation, and adopting techniques like the Doppler tomography (or Doppler imaging). In particular, Doppler tomography has been successfully applied to spectro-polarimetric observations in order to reconstruct the large scale topology of magnetic fields in T Tauri stars and connect it with the distribution of accretion shocks \citep[e.g.][]{donati2008,donati2010}. Even without a polarimetric facility, with these techniques the location and size of accretion shocks and accretion funnel flows can be found, which can also give indirect information on properties of the stellar magnetic field such as the magnetic obliquity \citep[e.g.][]{bouvier2007a,mcginnis2020}.

These kinds of observations are however time consuming, requiring monitoring over the rotation period, and as such have been performed on only the few brightest, nearby sources. Consequently, crucial information on the evolution of accretion flows and how they are connected with the stellar (mass, spectral-type, radius, age), and accretion disc properties are still largely lacking.

Disc dispersal through winds, driven either magnetically or by photo-evaporation (X-rays and FUV), represent, in addition to accretion, the main mechanism for gas removal in proto-planetary discs; as such they may terminate gas giant planet formation, and constrain the available time for exoplanet orbital migration. Disc winds can extend from the inner (few au) to outer disc regions, up to 100\,au, thus contributing to the mass loss over a large disc surface. Models of disc evolution predict that photo-evaporation could significantly contribute to dissipation of the gas disc in a few Myr \citep[e.g.][]{armitage2013,pascucci2022a}. However, measurements of the disc mass and mass accretion rates in 10-20\,Myr old stars show that a significant fraction of them still have relatively high accretion rates \citep[e.g.][]{manara2020, manara23}. This demonstrates that we are still missing some important ingredients in disc dissipation theories. In addition, the historical paradigm of viscous accretion disc evolution, where the angular momentum is dissipated through turbulence, has recently shifted in favour of theories of accretion driven magnetised disc winds \citep[e.g.][]{bai2016,tabone2022a}. It is clear that further observational constraints are needed to follow these recent developments, in particular towards the direction of inferring physical properties and the kinematics of magneto-hydrodynamics (MHD) and photo-evaporative disc wind tracers in large samples of PMS sources at different evolutionary stages.

\subsubsection{Requirements for probing accretion and wind mechanisms}

High-resolution observations of samples of sources in young clusters at different ages are needed in order to answer the crucial question on how the accretion and wind mechanisms evolve together with the disc evolution and depends on stellar and disc properties. The key spectroscopic observations are summarised below:

\paragraph{\bf Magnetospheric accretion:}
Spectroscopic evidence for hot-spots derives from the presence of ‘veiling’ in the photospheric lines, i.e. of an excess of emission with respect to the pure photosphere whose intensity is modulated by the stellar rotation, as well as from radial velocity variations with the stellar period. Such modulations can be used, with Doppler imaging techniques, to reconstruct the location of hot spots on the stellar surface, providing that the veiling is not too high to significantly reduce the line EWs. Similarly, hot-spot location and geometry can be reconstructed using emission lines that originates from the post shock regions where the accretion columns hit the stellar surface. The main tracers are the narrow components (i.e. FWHM about $20-40$\,km\,s$^{-1}$) of a few permitted emission lines, the brightest being the He\,{\sc i} line at 587.6\,nm and the Ca\,{\sc ii} IRT at $\sim850$\,nm.

H~{\sc i} Balmer lines can instead be used to trace the accretion columns connecting the inner disc to the star. They sometimes exhibit direct or inverse P-Cygni profiles attributed to stellar winds and infalling matter. Studies of the modulation of the line profiles with rotational period give clues to the geometry of the accretion flows \citep[e.g.][]{bouvier2007b} and stellar winds.

\paragraph{\bf Disc winds:}
Disc winds and collimated jets are best probed by forbidden emission lines of neutral and weakly ionised species, the most relevant for a diagnostic analysis being [O\,{\sc i}] at  630\,nm and 557\,nm, [S\,{\sc ii}] at 406\,nm and  673\,nm, [N\,{\sc ii}] at 657\,nm. The profiles of these lines are composite, with components peaking at different velocities that trace physical distinct regions within the disc. The peak velocity and broadening of each component provide information on the emitting size and velocity of the different winds. Photo-evaporative winds are best probed by a blue-shifted narrow component peaking at about $1-2$\,km\,s$^{-1}$ with FWHM of $20-50$\,km\,s$^{-1}$. MHD winds are traced by a broader and often more blue-shifted component. Additionally, in the more active sources, a component tracing high velocity collimated jets is also present at velocities up to 200\,km\,s$^{-1}$. Instruments with high enough resolution (i.e. $R > 50\,000$) are needed to separate and individually study these components. Ratios between different lines can be used as a diagnostic of the physical conditions (e.g. temperature, density, ionisation fraction, e.g. \citealt{fang2018a,giannini2019, gangi23}), which, together with the kinematic information, give constraints on the mechanism driving such winds and on their mass loss rates.

\frame{\vspace{-7mm}
\paragraph{Uniqueness}
High resolution studies like those described here are time consuming and have been so far performed only on a few individual sources. However, the approach of observing large samples of young stellar objects in a given young cluster is the only way towards a full understanding of how the described processes relate with the central star and disc properties (e.g. stellar and disc mass, spectral type, age, mass accretion rate). Therefore, a high resolution ($R> 50\,000$), multiplexing instrument such as HRMOS is the only facility to have a transformational impact on the described science.
}

\subsection{Targeting strategy}

Catalogues of clusters and cluster members are available from  \citet{cantatgaudin2020}, which made use of {\it Gaia} DR2 parallaxes, proper motion and photometry for the membership analysis and to derive ages and distances. More recent catalogues have been released making use of {\it Gaia} DR3 data \citep[see, e.g.][]{hunt23, perren23, cavallo23}. 
They can be used to estimate the number of targets that can be observed in young clusters, and typical target densities.

\begin{figure*}
\centering
\resizebox{0.9\hsize}{!}{
\includegraphics{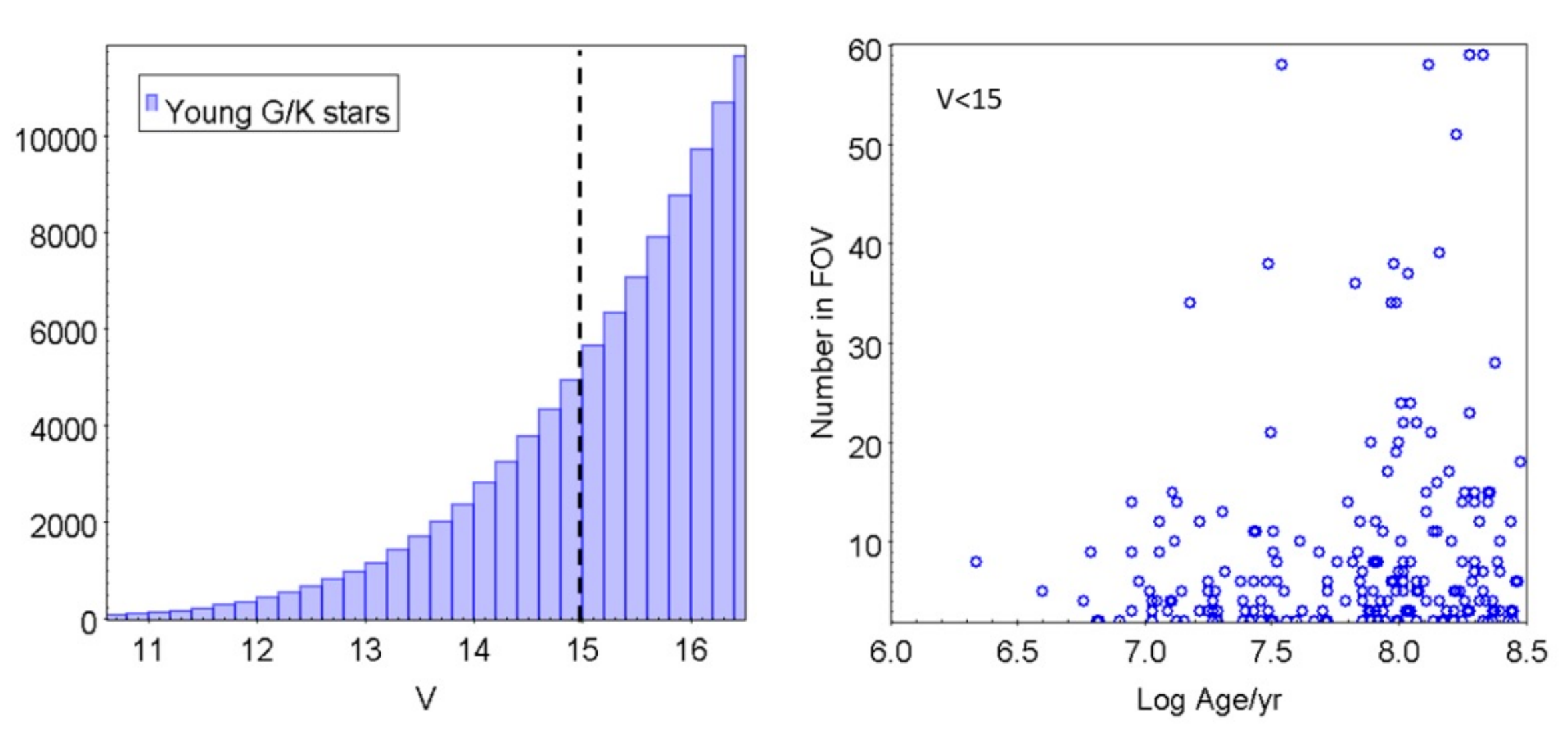}}
\vspace{-5mm}
\caption{
(a) The cumulative number of young ($\rm \log age/yr <8.3$) G/K-type targets observable from the VLT that are highly probable members of clusters. (b) The numbers of such targets within the 25\,arcmin diameter VLT FoV as a function of cluster age.
\label{fig:ystars2}
}
\end{figure*}

Figure~\ref{fig:ystars2} (left panel) shows a cumulative histogram of numbers of young ($\rm \log age/yr < 8.3$), G/K stars ($0.6 < (Bp-Rp)_0 <1.5$), with probability of cluster membership $\geq 0.9$ and at DEC$<20^{\circ}$ from \citet{cantatgaudin2020}. There are approximately 5000 young G/K targets in clusters with $V<15$. This number can increase or decrease by roughly a factor of two if we consider fainter  $V<16$ or brighter $V<14$ targets respectively. Figure~\ref{fig:ystars2} (right panel) shows the numbers of targets with $V<15$ and within 12.5\,arcmin of the cluster centre (the HRMOS field of view) as a function of the age of the cluster. There are about 1200 cluster members, belonging to 55 clusters, ranging in age from $<10$\,Myr to $300$\,Myr, that are in groups of $10-60$ that could be targeted by a multiplexing spectrograph. The peak target density in these clusters is low enough ($<1$/arcmin$^2$) that the minimum fibre spacing can be $>30$\,arcsec. 
These numbers are only preliminary, and will be improved from time to time with subsequent {\it Gaia} data releases. 

\begin{table*}[t]%
\centering
\caption{Instrument requirement summary (Young Stars)}%
\footnotesize
\begin{tabularx}{\textwidth}{lXX}
\noalign{\smallskip}
\hline
\hline
\textbf{Parameter} & \textbf{Value}   & \textbf{Justification}  \\
\hline
Resolving power ($R$)    & 80\,000     & To provide good tomographic resolution on stars with a broad range of $v\,\sin i$ in young clusters and to ensure a large enough number of viable targets to take advantage of multiplexing.    \\
Spectral range & H$\alpha$ to Ca\,{\sc ii} H\&K and higher Balmer lines. At least 75\,nm simultaneous coverage. Simultaneous observations of non-contiguous lines between 400 and 700\,nm: He\,{\sc i}~587~nm, [O\,{\sc i}]~630\,nm, 557\,nm, [S\,{\sc ii}]~406\,nm, Balmer lines, desirable Ca\,{\sc ii}\,IRT 850\,nm, [S\,{\sc ii}]~673\,nm  & Diagnostics of the physical conditions in prominence systems and chromospheres. Simultaneous wavelength coverage trades off with throughput in being able to do Doppler tomography. Reduced wavelength coverage would mean a lower threshold magnitude and lower viable target numbers. Simultaneous observations of a minimum set of non-contiguous lines located between 400 and 700\,nm are needed as diagnostics of the different phenomena, that is, accretion hot spots, accretion columns and disc winds.     \\
Multiplexing   & $\sim 50$ fibres      & Maximum number of suitable targets in young clusters within VLT FoV is $\leq 60$.   \\
Stability      & $<$ 0.2~km\,s$^{-1}$        & Needs to be stable or at least with the possibility of calibration to this precision during the night, since velocity bins for Doppler tomography are about 3.7~km\,s$^{-1}$  and photospheric line RV modulations are of order 1~km\,s$^{-1}$.  \\
Fibre spacing  & 30\,arcsec        & Target density does not exceed 1 object per square arcmin.     \\
\hline
\end{tabularx}
\end{table*}

%% file: exoplanets.tex


\clearpage

\section{Exoplanet populations in our Galaxy (and others)}

The search for, and characterisation of, planets around other stars has become one of the most compelling research areas in astrophysics since the discovery of the first exoplanets in the 1990s \citep{mayor95}. More than 5500 confirmed exoplanets are now known\footnote{See, for example, the NASA exoplanet archive \url{https://exoplanetarchive.ipac.caltech.edu/}}, and much work is now focused on detailed study of the most favourable individual examples using many facilities including ESPRESSO at the VLT, JWST and forthcoming E-ELT.

Yet an essential step towards an understanding of planetary formation and the development of the sheer diversity of observed planetary systems is the systematic investigation of exoplanet {\it populations} in samples of stars of known age and in differing environments \citep[see. e.g.][]{lovis2007, adi21, winter20, delgadomena2018, delgadomena23}. 

The inception of a very high resolution, stable, multiplexing spectrograph on an 8-m class telescope will allow for the first time, efficient systematic Doppler surveys for giant exoplanets in star clusters – detecting hundreds of exoplanets around pre main sequence and main sequence stars with precisely known age and composition. Such samples would give unique insights into the environmental factors that influence planetary formation and shape the development and evolution of planetary systems. Hundreds of exoplanets could also be detected around giants stars to great distances, probing planet formation at higher stellar masses and in environments of known age that can be radically different to the solar neighbourhood – e.g., star clusters over a wide range of Galactocentric radii, possibly establishing a Galactic habitable zone. The reach and efficiency of this spectrograph would also allow for the first exoplanet surveys in the bulge of our Galaxy and even for attempts to find exoplanets in the closest other galaxies. Studies of atmospheres of exoplanets detected in clusters can also be attempted.

\label{sect:exoplanets}

\begin{figure}
\centering
\resizebox{0.65\hsize}{!}{
\includegraphics{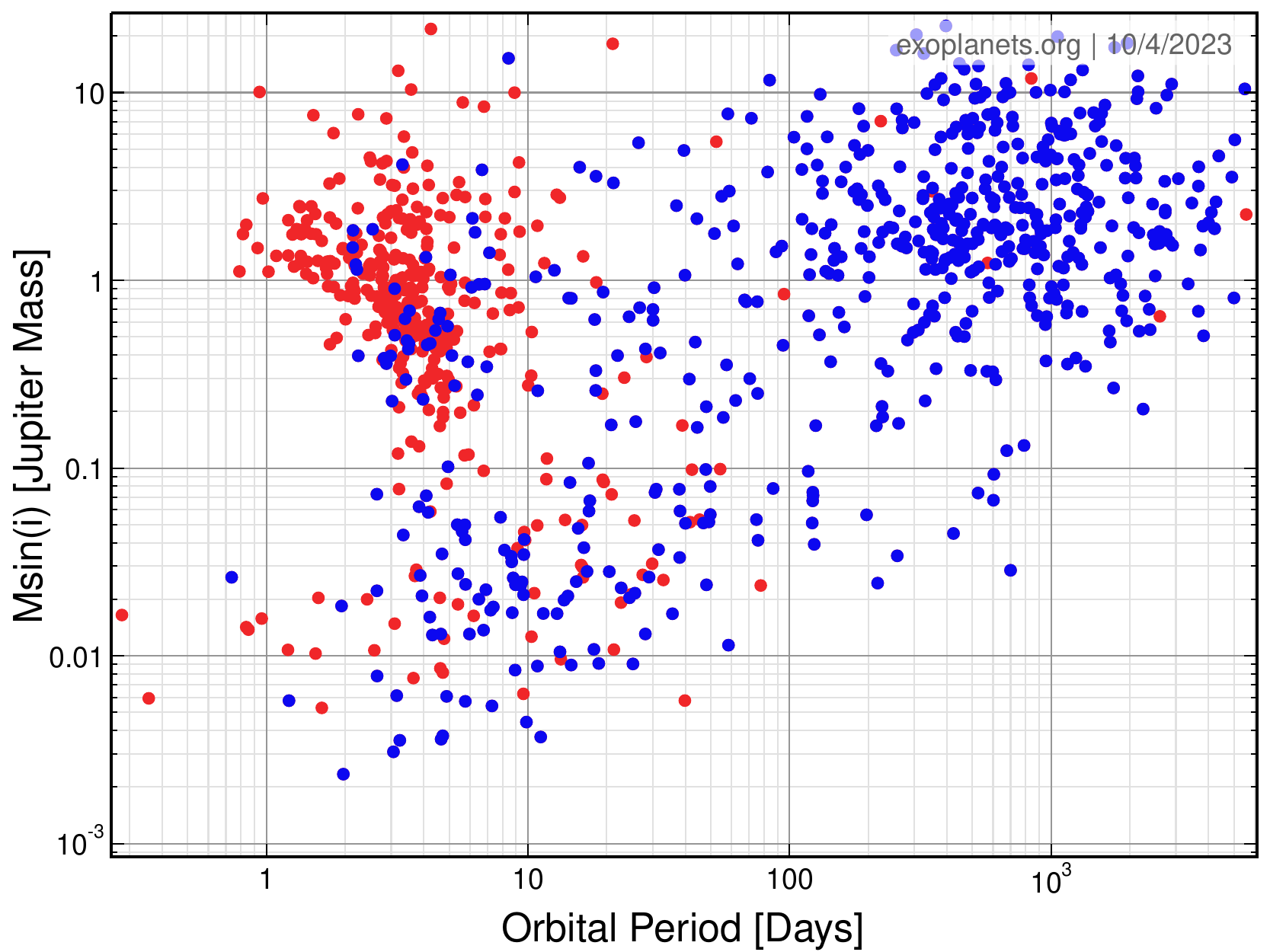}}
\caption{
The projected planetary mass ($M\,\sin\,i$) versus the orbital period for currently known exoplanets discovered either through the transit (red) or Doppler techniques (blue).
\label{fig:exoplanetdetections}
}
\end{figure}

Exoplanetary systems are expected to change significantly during their lives and their properties are almost certain to be strongly related to both their host stars and the environment in which those stars were born and reside. 
Taking our own Solar System as an example, we know that gas giants were born in a primordial disc, probably via the core accretion process, during the first 10\,Myr \citep[e.g.][]{pollack1996,ida2004}. Over the next $10-100$\,Myr, the terrestrial planets were built up from rocky debris through a process of accretion, agglomeration and punctuated by collisions \citep[e.g.][]{morbidelli2012}. Throughout this process and afterwards, it is likely that the orbital radii of the planets has changed and that the Solar System was strongly influenced by external radiation and close encounters with other stars \citep[e.g.][]{adams2010,walsh2011,nesvorny2012}. For many years now, it has been quite clear that our planetary system may not be so typical. The discovery of thousands of exoplanets through transit and Doppler surveys has uncovered a rich and diverse behaviour (see Fig.~\ref{fig:exoplanetdetections}) that demands explanation.

Dynamical changes are expected to occur during the birth and evolution of a planetary system – the presence of the so-called “hot Jupiters” and their range of eccentricities and obliquities, attest to multiple processes that cause the inward migration of giant planets, either though interactions with the disk in the early phase or through planet-planet scattering or perhaps from the influence of external third bodies \citep{lin1996,rasio1996,wu2003,ida2013}. Further changes in orbital radii, eccentricities and obliquities, or even ejection, can arise from interactions between planets or between planets and debris discs, Kozai-Lidov oscillations or more chaotic rearrangements \citep[e.g.,][]{davies2014}. Star-planet tidal interactions may ultimately result in the engulfment of hot Jupiters, however the theoretical timescales for this are uncertain by orders of magnitude and empirical evidence for the process is scarce \citep{levrard2009,patra2020}.

The exoplanets themselves will also experience physical changes even after their growth and accretion phases are ended. Gas giants should cool and contract over billions of years. Those that migrate close to their host stars may be subject to insolation and an increased radius, or their atmospheres may be photo-evaporated, depending on their mass and the proximity of their orbit to the star \citep{fortney2007,Sanz-Forcada2011,owen2013,Lampon2023,Zhang2023}; and whilst this latter process is likely accelerated by high levels of extreme and far ultraviolet radiation (EUV/FUV) when stars are young and fast-rotating, it may continue on Gyr timescales \citep{king2021}.

The influence of the wider environment is also likely to play an important role in the birth and evolution of planetary systems. The composition of the gas from which a star and its planets are born appears to be significant in determining the incidence of giant planets around main sequence stars, where a higher metallicity is correlated with a greater incidence of giant planets, particularly “hot Jupiters” \citep{fischer2005},  although the situation may be more ambiguous around giant (and therefore more massive) stars \citep[e.g.][]{jones2016}.  Planet formation is critically dependent on the properties of primordial discs and these in turn may be dissipated or truncated by the external radiation environment caused by proximate massive stars or by close encounters with third bodies \citep[e.g.][]{adams2004,adams2010}. Exoplanetary systems that are born, or remain, in dense stellar aggregates and clusters may have their orbits significantly altered, be subject to inward migration, or may even swap or eject planets due to close encounters with third bodies and tidal dissipation \citep[e.g.][]{davies2014,hamers2017, winter20}.

The exoplanets themselves may also influence the properties of the stars. Various differences in photospheric chemistry have been noted between stars with and without planets. For example, stars hosting hot Jupiters may be anomalously depleted in Li, but enhanced in C and N \citep[e.g.][]{israelian2004,israelian2009,delgadomena2015,ggonzalez2015,suarezandres2016,suarezandres2017}. Whilst the latter may simply be connected with the overall correlation between metallicity and planetary incidence noted above, the former may be a more fundamental observation, showing that exoplanets may influence the internal mixing or even rotation rates of their stars via tidal interactions or interactions with the primordial accretion disc during star and planet formation \citep[e.g.][]{bouvier2008}. There are suggestions of other more subtle anomalies in elements such as Al and Ba \citep{mishenina2016}. Others have found evidence that planet-hosting stars can be more metal-rich than a binary companion, particularly in refractory elements \citep[e.g.][]{saffe2017,liu2018}. This might indicate that the inward migration of planets can result in the accretion of rocky material or even the ingestion of planets during the star formation process. Star-planet interactions have also been suggested as a means of altering the rotation and enhancing or suppressing magnetic activity in some stars that host hot Jupiters, either through tides or the interactions of their magnetospheres \citep[e.g.][]{lanza2009,poppenhaeger2014,pillitteri2014}. A difficulty common to all these studies is firmly establishing the age of the star and obtaining a clear baseline level of chemical composition, rotation and magnetic activity for similar stars without planets.

On a grander scale, all that we know about exoplanet formation and evolution at present comes from a relatively small volume around the Sun. It is important for a broader view of planetary formation across the universe that we understand whether star and planet formation proceeds in the same way in different parts of the Galaxy that might have been subject to radically different initial or subsequent conditions. Likewise, exoplanets have only been studied in the Milky Way, but planet formation may have proceeded differently in low-metallicity or starburst environments that might be typified by local dwarf galaxies. The concept of a “Galactic habitable zone” \citep{lineweaver2004, spitoni17} is still one that has received little attention in the way of observational tests, although it is now being addressed from many perspectives \citep[see, e.g.][for the molecular content in the outer Galaxy]{bernal21, fontani22, fontani22a}.

\frame{
\titlecol{Key Questions:}
\begin{itemize}
\item How do the incidence, orbital and physical properties of giant exoplanets change with age, mass and composition of their host star?
\item Which are the characteristics that differentiate stars with planets and without planets when both are part of the same cluster? 
\item What is the influence of environmental effects on the frequency and properties of exoplanets? What are the properties of exoplanets formed in radically different environments?
\end{itemize}
}

Assembling large samples of objects with known age, mass and composition is key to testing theories for the formation and evolution of exoplanetary systems, discriminating the importance of various mechanisms and empirically calibrating the timescales where the physics is uncertain. Open clusters (OCs)  can play a leading role in this task. The fundamental questions that can be answered concern how the frequency, orbital parameters and mass distribution of exoplanets vary as a function of mass (within OCs at a common fixed age and composition) and also as a function of age and composition by looking at stars with similar mass across a broad range of OCs. In a similar way, observations of coeval giants in clusters at a range of Galactocentric radii and metallicity and of giants in the bulge and in local dwarf galaxies can answer wider questions about the environmental dependencies of exoplanetary formation and evolution.

The ongoing search for planets around other stars has taken many paths. Transit searches can target many stars but are only sensitive to a small fraction of exoplanets orbiting close to their parent (main sequence) stars ($a<1$\,au). Astrometric searches, especially with {\sl Gaia}, should reveal a larger population of massive planets in wider orbits, but are insensitive to close-in planets and will not find many planets beyond one kpc. Microlensing also probes wider orbits and can detect very low-mass exoplanets, but characterisation of the lensing star can often be problematic. 
Doppler searches have a broader sensitivity, both in mass and orbital parameters, but they are observationally expensive, requiring precise radial velocity measurements on the timescales of the orbital periods that are sought. Almost all current work relies on observing targets one-by-one and it is an ongoing challenge to accurately characterise these samples in terms of their ages and masses. 

A high-resolution, stable, multiplexing spectrograph on a large telescope will allow the survey of many thousands of stars (both main sequence and giants) in OCs, the Galactic bulge, GCs and perhaps even the nearest local galaxies. The key scientific advantage is that clusters have readily determined ages and composition, offering large coeval samples of stars with common chemistry across a broad range of well-estimated masses that can sample the full space of parameters that may influence the formation and evolution of planetary systems. The common technical advantage to these programs is that target samples can be found at densities that make a huge multiplex gain possible over the VLT FoV.

For many years, photometric and spectroscopic searches for planets in OCs gave no results, but the situation has dramatically changed over the last decade. More than 30 exoplanets and a similar number of strong exoplanet candidates have now been identified around stars in OCs (Doppler surveys: \citealt{sato2007,lovis2007,quinn2012,malavolta2016,brucalassi2014,brucalassi2016,brucalassi2017,delgadomena2018,leao2018} – Transit photometry: \citealt{meibom2013,mann2016,mann2017,mann2018,livingston2018,livingston2019,ciardi2018,obermeier2016,pepper2017,rizzuto2018,curtis2018,bouma2020, nardiello2021}). Finding exoplanets in GCs is more difficult – transit surveys have so far uncovered no clear-cut detections, few candidates, and hot Jupiters may be rare in these low metallicity environments \citep[e.g.][]{gilliland2000,weldrake2005,nardiello2019, wallace2020}. More success has been obtained in the Galactic bulge, where transit surveys have uncovered a number of planetary candidates around (presumably) quite metal-rich stars, most of which were too faint to be confirmed, with periods less than a few days \citep[][]{gaudi00, sahu2006, zhu17, penny16, penny20}. The large number of planet candidates in the bulge could be due to the choice of metal-rich star hosts and the higher incidence of giant planets in this type of star. 

The advent of the {\it Gaia} astrometric catalogue and of previous spectroscopic surveys (e.g. the {\it Gaia}-ESO survey) has made it comparatively easy to identify members of clusters based on their photometry, proper motions and radial velocities. Combining these target lists with a multiplexing HRMOS with 10\,m\,s$^{-1}$ radial velocity stability leads to at least four key exoplanet-related science programmes.

\subsection{Exoplanets in Open Clusters}

\label{exoplanets_ocs}

An instrument such as  HRMOS could survey stars in clusters of known age and composition for hot Jupiters and other close-in exoplanets – massive exoplanets with orbital periods less than $\sim100$\,days. This would fill a gap in our knowledge that will not be filled by {\it Gaia} astrometric searches. Kepler and TESS have surveyed a few, mostly nearby OCs using the transit technique, but the instrument proposed here would be able to efficiently reach many thousands of solar-type stars in clusters at a range of ages out to 1\,kpc or more. Since the transit technique is only sensitive to the $\sim10$\,\% of planets with fortuitous orbital plane alignment, HRMOS will detect about 10 times as many hot Jupiters in the clusters Kepler and TESS have surveyed. Given a hot Jupiter frequency of $\sim 1$\,\% and a high detection efficiency for planets with masses greater than 1\,M$_J$, then {\bf the potential is to discover of order 100 hot Jupiters around solar-type stars of precisely known age and composition.} This will be a unique dataset with robust statistics, to be used to explore how the incidence, orbital and physical characteristics of hot Jupiters depend on the age, mass and composition of the host star. Furthermore, the dataset will be an asset to test theories of how hot Jupiters are formed and evolve with time \citep[see, e.g.][]{wang22}. We can also track chemical composition differences between stars within the same cluster and and see whether this diversity is associated with (i) the current presence of planet(s) and the type of such planet(s),  (ii) the presence in the past of planet(s) that have been engulfed by the star, or (iii)  the influence of  close-in exoplanets on the rotation or magnetic activity of their hosts \citep[see.e.g.][for the detection of  such anomalies in M67]{church20}. The cluster population provides both a precise age and a large sample of comparison stars with which to establish a clear and precise baseline for chemistry, rotation and magnetic activity.

For example, work in the past 10 years has shown that M67, a solar-age solar metallicity OC, hosts an unusually high fraction of hot Jupiters, up to 5.7\,\%, compared to a fraction of about 1\,\% discovered among field stars of the same metallicity \citep{brucalassi2016,brucalassi2017}. M67 is a largely populated and old OC, and perhaps the high fraction of hot Jupiters has been caused by migration, induced by fly-by and interaction with other stars \citep{malmberg2011,shara2016,hamers2017}. This is an interesting hypothesis, but it is not clear yet whether this result is statistically robust (it differs from field stars by only 1\,sigma), whether M67 is an exception as a cluster, and when the planet migration occurred.  Recent  spectroscopic surveys of about 100 stars in the Hyades and Praesepe also find a high incidence of hot Jupiters \citep{quinn2012,quinn2014}, but these clusters are relatively metal rich ($\rm [Fe/H]\approx0.16$), so their high hot Jupiter frequency could be ascribed to their enhanced metallicity.  A major survey with HRMOS could settle these questions by looking at hot Jupiter frequencies in clusters with a variety of richness, whilst controlling for confounding parameters like chemical composition, and comparing with field populations.

This research becomes more difficult when targeting young stars, because of the increasing RV noise induced by physical phenomena that are stronger in this type of objects, such as stellar activity. Starspots may induce false RV signatures of $10-100$\,m\,s$^{-1}$ at the rotation period of the star and stellar activity can introduce jitter at levels of up to 50\,m\,s$^{-1}$ \citep[e.g.][]{wright2005,hojjatpanah2020}.  This is a challenge that can be solved, and it is linked to another HRMOS science case on mapping starspots and magnetically active regions on young stars (see Chapter 3). The ability to simultaneously monitor chromospheric activity using the Ca\,{\sc ii} H and K and Balmer lines would be a crucial aspect of such observations. It has already been shown that intensive observations that map the starspot distributions and/or monitor correlations between chromospheric activity, line bisector spans, and the RV are capable of reducing the effects of starspots and “jitter” by an order of magnitude \citep[e.g.][]{moulds2013,dumusque2014,petit2015,barnes2017}. Young clusters would benefit from a more frequent sampling strategy since spot patterns may evolve on timescales of weeks and months.

\subsection{Exoplanets around giant stars}

\label{exoplanets_giants}

Analysing the occurrence rate and the properties of exoplanets around giant stars allows us to study the effects of stellar evolution on the evolution of planetary systems. Furthermore, it also offers a means to study the properties of planetary systems around  more massive stars than those that we can observe when in the main sequence stage. Main sequence stars with $M>1.5$\,M$_{\odot}$ have relatively few spectral features and tend to be fast rotating, and their planets are difficult to find using the Doppler method. However, when they evolve to cooler G/K-giants, they become slow rotators and their spectra become much richer. Observing luminous giants also allows the investigation of planet evolution at a range of ages and over a larger range of Galactocentric radii than can be probed with main sequence stars and thus over a greater range of Galactic birth environments.

\begin{figure*}
\centering
\resizebox{\hsize}{!}{
\includegraphics{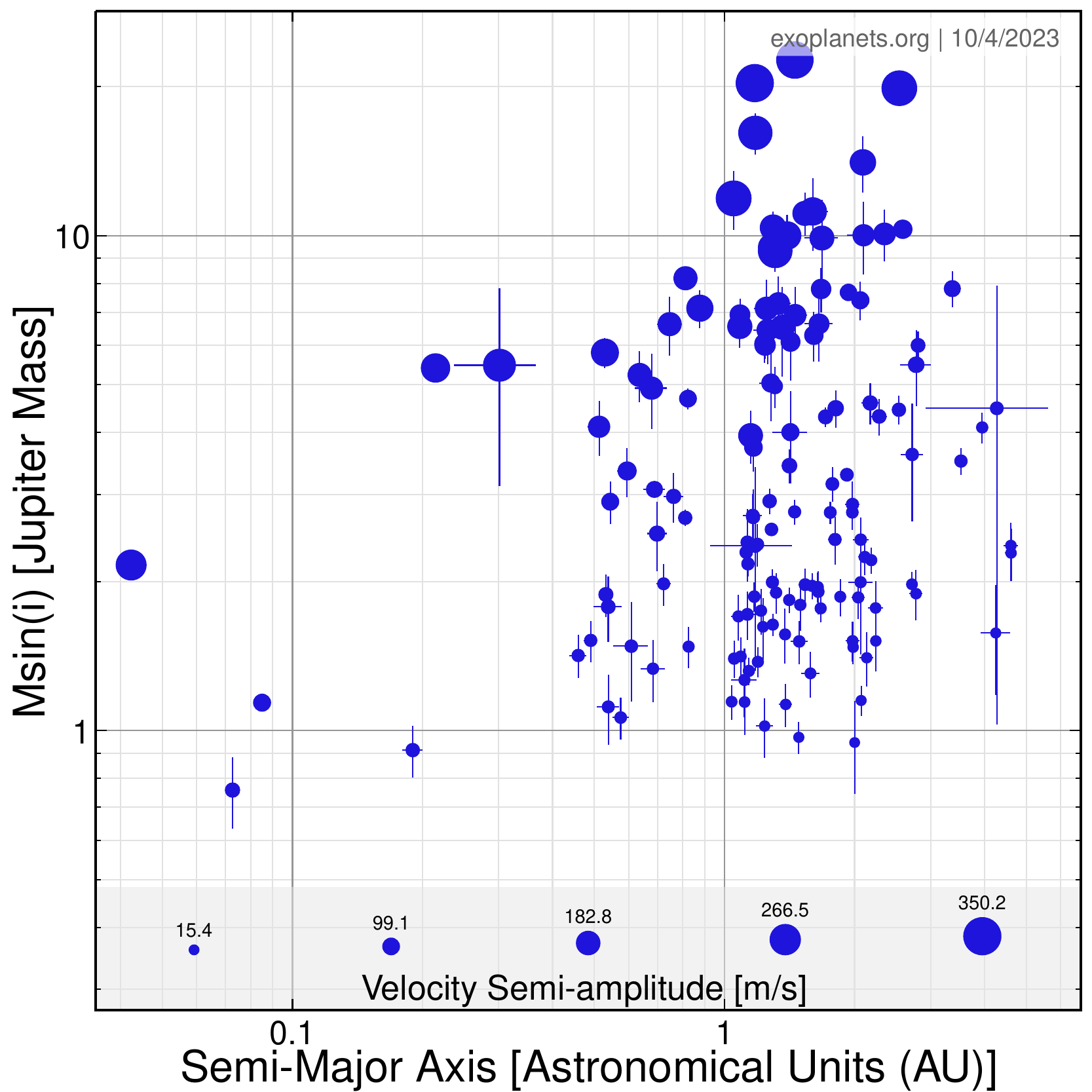}
\includegraphics{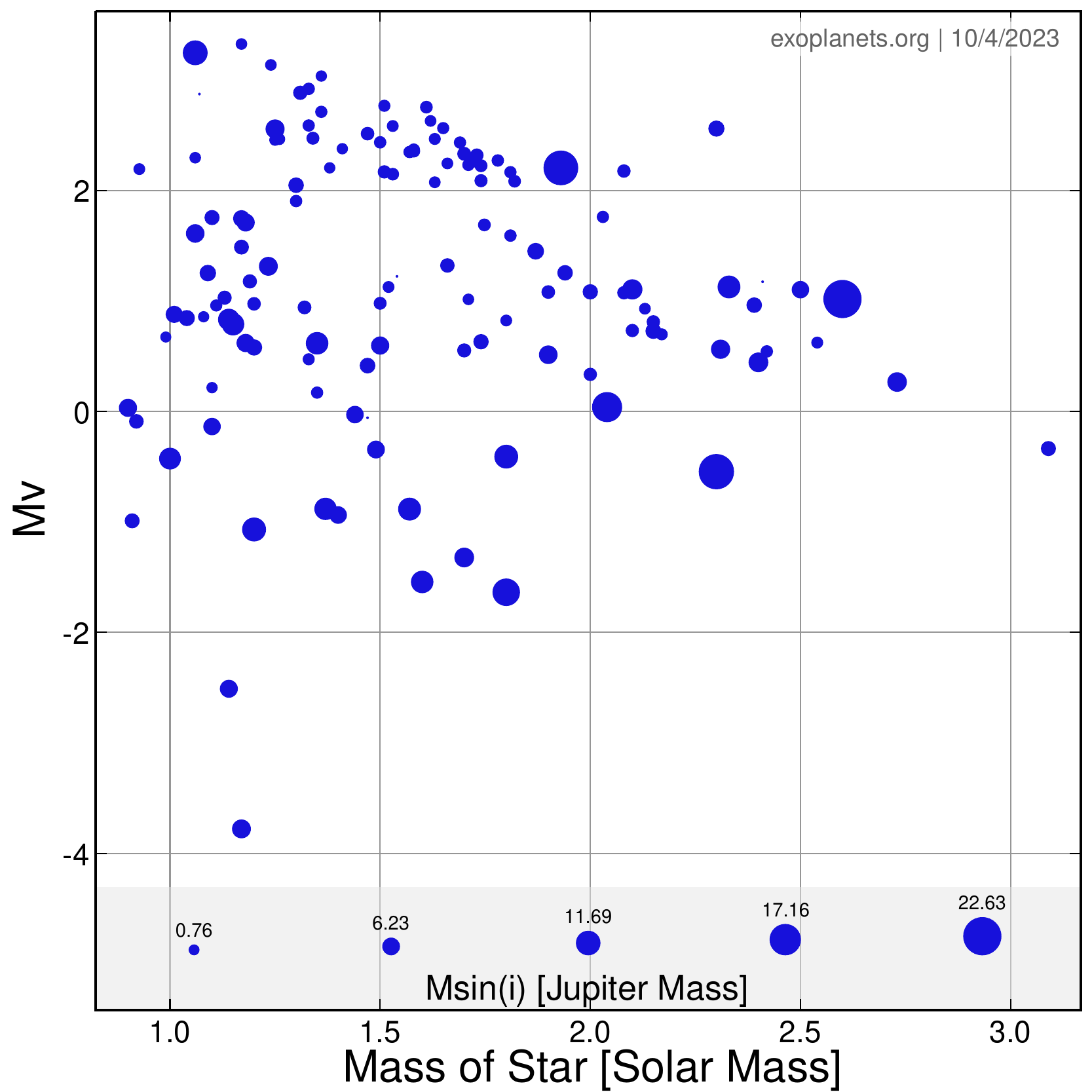}}
\caption{
Left panel: Exoplanet mass vs semi-major axis for currently known planets around giant stars with $\log g<3.5$. The symbol size encodes the observed RV semi-amplitude of the star. Right panel:  Absolute magnitude vs the stellar mass for the same objects. The symbol size encodes the planetary mass (data taken from the database  \url{exoplanets.org}).
\label{fig:exoplanetmass}
}
\end{figure*}

More than 100 exoplanets with $M_p >$M$_J$  have been found among evolved G/K giants using the Doppler method: with progenitor masses of $1-3$\,M$_{\odot}$, periods of about $100-2000$ days, semi-major axes of $0.5-5$\,au, RV semi-amplitudes of $10-300$\,m\,s$^{-1}$, and absolute magnitudes of  $-4<M_{\rm V}<2$   (see \citealt{jones2021} and references therein and Fig.~\ref{fig:exoplanetmass}). It has become apparent that long-term and careful monitoring of these RV signals is essential to attempt to distinguish planets from long term (magnetic?)  variability cycles, rotational modulation and shorter period oscillations \citep{hatzes2018,delgadomena2018}. 

Estimating the masses of field giants is difficult, but it is much easier in clusters of known age and metallicity. Observing large samples of giants in clusters, with the the efficiency of a multiplexing spectrograph such as HRMOS, can provide the much needed data to address issues such as: 
understanding the likelihood that both giant planets frequency and their mass increase with increasing stellar mass, 
and understanding whether exoplanets around giant planets have lower eccentricities, and whether they are overabundant at separations of $0.5-0.9$\,au but rarer in closer orbits compared to those around main sequence stars \citep[e.g.][]{sato2008,dollinger2009,johnson2010pasp,jones2021}.  
An HRMOS survey of giant stars, particularly in clusters can also shed light on the long-term evolution of planetary systems when the host star leaves the main sequence \citep[e.g.][]{veras16, vansluijs18,veras21}.

Constraining these phenomena with larger statistics and greater confidence in the robustness of host star parameters is pivotal for tracing planetary formation and migration scenarios. For instance, there is still considerable debate over the possibility that the correlation between giant exoplanet frequency and metallicity, that seems well established in main sequence stars, is either weaker or absent for exoplanets orbiting giant stars; this leads to some questions about whether these effects are a natural consequence of greater planet forming efficiency within metal-rich gas in the core-accretion model, or whether the observed higher stellar metallicities are a consequence of the ingestion of planetary material \citep{murray2002,pasquini2007}. 

Discovering more planetary systems around giant stars is difficult - the transit method is insensitive to planets around giants. Only a couple of systems have so far been discovered in this way. {\it Gaia} astrometry may find a few hundred in orbits $>1$\,au around field giants within $\sim600$\,pc \citep{perryman2014}.  HRMOS can efficiently survey thousands of G/K giants in open clusters down to $V \sim 16.5$, finding hundreds of exoplanets around giant stars of well-established age and composition.

\subsection{Exoplanets in the bulge and in nearby dwarf galaxies}

\label{exoplanets_bulge}

HRMOS offers an exciting opportunity to take a vital next step in our exploration of the exoplanetary universe – to systematically search for exoplanets beyond the Milky Way. By assessing the planetary frequency in the radically different environments of GCs, the Galactic bulge and perhaps even the nearest dwarf galaxies, we can start to understand how common the exoplanet formation process is in the universe and start to disentangle the influences of composition and density of birth environment on the frequency and properties of exoplanets. 

Surveys for transiting planets have been attempted in some of these environments before but with limited success \citep{sahu2006,wallace2020}.  An HRMOS Doppler survey of red giants in the clump and the upper part of the RGB offers a way to assess the exoplanetary statistics of Jovian planets in these environments, at orbital periods of $10-1000$ days through a long term monitoring program of fields containing more than 100 targets.

\subsection{Chemical characterisation of exoplanet atmospheres}

\label{exoplanets_chemistry}

In recent years, high-resolution spectroscopy has revealed that the absorption and emission spectra of hot-Jupiters are rich in metal lines at optical wavelengths. Besides the Na doublet, observed in intermediate-temperature hot Jupiters \citep{Snellen2008}, ultra-hot Jupiters have atmospheres in which many more metals lines are in the gas phase \citep[e.g.,][]{Hoeijmakers2018}. This aspect has greatly enhanced the use-case of stabilised optical high-resolution spectrographs for carrying out transit and even day-side observations \citep[e.g.,][]{Pino2020}.\\

Atmospheric studies of transiting exoplanets generally do not make use of multiplexing with many fibers. Instead, spectral time-series are self-calibrated by comparing in-transit and out-of-transit spectra. This leads to the effective removal of the continuum of the planetary spectrum, creating a degeneracy in the absolute abundance of absorbers in the transmission spectrum \citep{Heng2017,Gibson2020}. Such a loss of information does not affect space-based low-resolution observations, which instead keep providing absolute flux calibration. In this way ground-based high-resolution spectroscopy and space-based lower resolution spectroscopy are highly complementary \citep[e.g.,][]{Brogi2017}.\\

In crowded star clusters and bulge fields, HRMOS will have access to a large number of reference stars when performing time-series observations of a transiting exoplanet. This may allow the continuum of the exoplanet spectrum to be preserved, allowing flux-calibrated high-resolution spectroscopy of hot transiting exoplanets from the ground. In addition, HRMOS will provide simultaneous reference observations for active exoplanet host stars, when stellar activity may produce false-positive detections of an exoplanet atmosphere \citep[e.g.][]{Khalafinejad2017} -- which in the absence of reference stars can only be detrended via correlation with activity indicators.\\

Time-series observations of exoplanets are time-critical and time-intensive (typical durations greater than 4 hours). HRMOS will provide very deep spectroscopy of the reference star samples as a by-product, providing sensitivity to short-period spectroscopic binaries.\\

Spectral coverage is key to the extent of chemical elements and molecules that will be accessible to HRMOS. In a design with 4 narrow passbands, the main interest might lie in the monitoring of specific lines in the planet atmosphere that have the largest oscillator strengths, and can thus be individually detected (e.g., H$\alpha$, Na Fraunhofer doublet, K doublet; e.g. \citet{Wyttenbach2015}).
If the simultaneous spectral coverage will be sufficient, refractory elements such as iron or titanium might be accessible in the spectra of the hottest planets 
(T$_{\rm{eq}}\gtrsim2,000$~K) via cross-correlation spectroscopy in transmission  
\citep[e.g][]{Hoeijmakers2018}, or emission \citep[e.g.][]{Pino2020}.

\subsection{Requirements for the detection and characterisation of exoplanets}

\subsubsection{Main sequence stars}

\begin{figure*}
\centering
\resizebox{\hsize}{!}{
\includegraphics{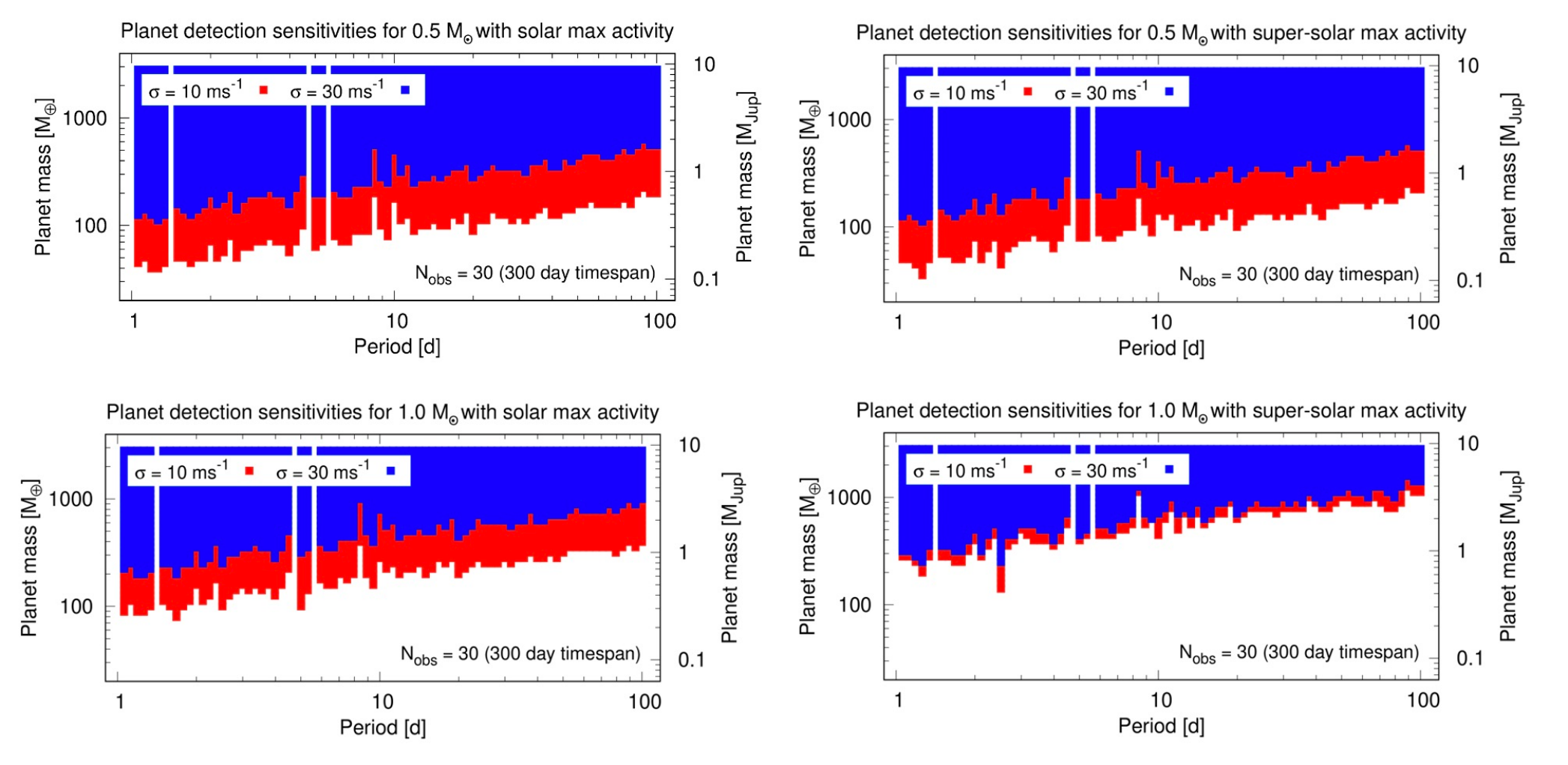}}
\caption{
Simulations illustrating the regions where exoplanets become detectable with a 0.1\,\% false alarm probability. The four simulations are labelled with the mass of star considered and the level of starspot activity assumed. In each panel, regions are shown corresponding to instrumental velocity noise of 10\,m\,s$^{-1}$ and 30\,m\,s$^{-1}$  respectively.
\label{fig:exoplanetdetectionsensitivity}
}
\end{figure*}

To illustrate the likely sensitivity of the observations, Fig.\ref{fig:exoplanetdetectionsensitivity} shows the results of simulations assuming that the velocity stability of the instrument is 10\,m\,s$^{-1}$ (or 30\,m\,s$^{-1}$) with a simple sampling strategy of 30 observations spread over 300 days, with some random offsets to reduce aliasing. In all cases it is assumed that the SNR of the observations is high enough that instrumental precision, or the effects of starspots on the RV signal, will dominate the uncertainties. 

Simulations were performed for cases of 1\,M$_{\odot}$ and 0.5\,M$_{\odot}$ cluster stars for two possible assumptions about the effects of spot activity on the RV precision – a solar maximum type spot coverage that might be appropriate for stars of $1-5$\,Gyr and a “super solar maximum” spot coverage that may be appropriate for younger stars (spot models 2 and 3 in \citealt{barnes2011}); the spot/photosphere temperature ratio is estimated from \cite{berdyugina2008}. The spots on the lower mass star were distributed over the entire surface, whereas for the solar mass star they were constrained to low and intermediate latitudes. The starspots add an extra “jitter” of $1.5-7$\,m\,s$^{-1}$ (for a 0.5\,M$_{\odot}$ star with modest spot coverage) to $7-44$\,m\,$s^{-1}$ (for 1\,M$_{\odot}$ stars with modest or high spot coverage). The simulations assume this jitter is added to the instrumental precision, but in practice the effects could be mitigated by monitoring activity and line bisectors as described in~\ref{exoplanets_ocs}. 

The plots illustrate the regions where exoplanet detections become significant for the two levels of instrument precision, with a false alarm probability of 0.1\,\% in each case.

An instrumental RV stability of 10\,m\,s$^{-1}$  will afford the sensitivity required to detect exoplanets of $>1$\,M$_J$  across the full range of periods 1-100 days and reach down to $\sim0.3$\,M$_J$ at the shortest periods. This will allow a full exploration of the separation range to which {\it Gaia} and WFIRST are insensitive and provide sensitivity down to the apparent local minimum in the density of currently detected short-period exoplanets in the mass-period plane. A lower precision would result in missing a substantial fraction of the giant planet population. There would be very little to be gained (in terms of sensitivity in the $1-100$ days range) by performing $>30$ observations.

The brightness limit to which data can be obtained that exploits this level of precision can be estimated using the work of \cite{bouchy2001}, that gives a simple expression for the photon-limited RV precision
\begin{equation}
\Delta {\rm RV}=  \frac{c}{Q\sqrt{N}}\ ,
\end{equation}
where $N$ is the number of detected photons over the whole spectral range observed and $Q$ is a “quality factor” which depends on the spectral type, wavelength, resolving power and degree of rotational broadening in the spectrum. This simple formula ignores everything except Poisson noise from the target. $Q$ is maximised for stars with many spectral lines, no rotational broadening and in high-resolution spectra.  The dependence on resolving power is linear up to the point where line profiles are resolved, and then flattens off. This break point is around $R=80\,000-100\,000$ in slow-rotating solar-type stars and giants. 

Using Bouchy’s simulations for $Q$, then the limiting magnitude in a 1-hour exposure to achieve a statistical RV precision of 5\,m\,s$^{-1}$ for a slowly rotating star ($v \sin i <  2$\,km\,s$^{-1}$) is given in Table~\ref{tab_bouchy}.  This threshold allows us to fully exploit an instrumental velocity stability of 10\,m\,s$^{-1}$ for HRMOS - the overall RV precision should then be dominated either by the spectrograph stability or intrinsic noise in the target. The calculation assumes that the star is observed near zenith in median seeing and that the sky signal is negligible (grey sky or darker). Estimates are given for resolving powers $R=80\,000$ and $R=60\,000$ and for corresponding wavelength ranges of 75\,nm and 100\,nm (the statistical RV precision would scale as the square root of the wavelength range). The central wavelength is chosen to optimise the trade-off between maximising $Q$ by maximising the number of spectral features (which increases towards the blue) and maximising the stellar flux (which decreases towards the blue for cooler stars).

\begin{table}[t]%
\centering
\caption{The limiting $V$ magnitude for slowly rotating targets that can yield a photon-limited velocity precision of $\sim5$\,m\,s$^{-1}$ in a 1-hour exposure with HRMOS. The central wavelength is a trade-off between density of spectral lines and spectral flux. The assumed $Q$ factor \citep[see][]{bouchy2001} is listed. }%
\footnotesize
\begin{tabular}{cccc}
\noalign{\smallskip}
\hline
\hline
Spectral type & Wavelength  & $Q/10^4$   & $V$     \\
                       & [nm]        &                     & [mag] \\
\hline
$R=80\,000$, $\Delta\lambda=75$\,nm &               &                     &                \\
F0 & 450 & 1.4 & 15.9 \\
G2 & 450 & 2.6 & 16.4 \\
K7 & 500 & 3.4 & 16.7 \\
$R=60\,000$, $\Delta\lambda=100$\,nm &               &                     &                \\
F0 & 450 & 1.1 & 15.6 \\
G2 & 450 & 2.0 & 16.1 \\
K7 & 500 & 2.4 & 16.4 \\
\hline
\end{tabular}
\label{tab_bouchy}
\end{table}

For faster rotating stars, the limiting magnitudes in Table~\ref{tab_bouchy} are reduced by $5 \log(v \sin i /2\,{\rm km\,s}^{-1} )$\,mag at  $R=80\,000$ because the line profiles become resolved and the value of $Q$ falls approximately as the reciprocal of $v \sin i$. Fainter targets become harder to access with even longer exposure times, because 1\,hour exposures at $V>16.5$ start to become limited by read noise, rather than just from Poisson noise in the target signal.
 
The calculations above suggest that targets should be sought with $V<16.5$ for solar-type targets at older ages and at brighter limits for potentially faster rotating stars in younger clusters.  The catalogues of clusters and cluster members compiled from \citet{cantatgaudin2020} are used to assess the number of targets that can be observed in clusters and typical target densities. For stars younger than 4\,Gyr, we assume that $v \sin i$ scales as $v \sin i \sim 2 {\rm (Age/4 Gyr)}^{0.5}$\,km\,s$^{-1}$ (which approximates a \cite{skumanich1972}-type magnetic wind braking law) and then subtract $5 \log(v \sin i /2\, {\rm km\,s}^{-1})$ from the $V=16.5$ threshold. 

\begin{figure*}
\centering
\resizebox{\hsize}{!}{
\includegraphics{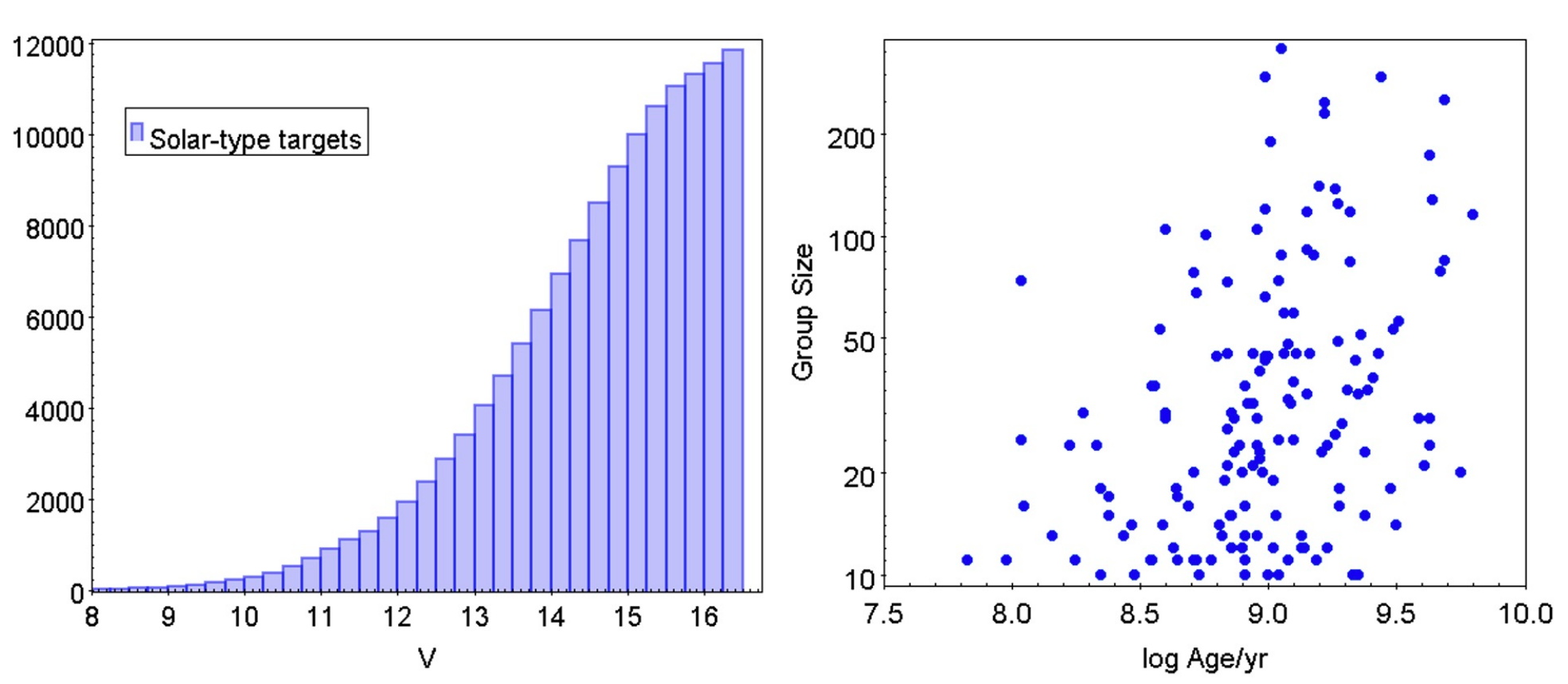}}
\caption{
{\sl Left:} The cumulative frequency of apparent magnitude for suitable solar-type cluster targets for HRMOS in the southern hemisphere. {\sl Right:} The number of targets within a single VLT FoV plotted versus the age of the cluster.
\label{fig:cumulativefrequency}
}
\end{figure*}

Figure~\ref{fig:cumulativefrequency} (left panel) shows a cumulative histogram of apparent $V$ magnitude for 11\,863 main sequence F/G/K stars ($0.3 < (Bp-Rp)_0 <1.5$), with probability of cluster membership $>0.9$ and at $\rm DEC <20$ degrees and which are bright enough, and likely to be slowly rotating enough, to act as targets. Figure~\ref{fig:cumulativefrequency} (right panel)  plots those clusters for which there are $>10$ targets within the VLT FoV of 25 arcmin diameter. This demonstrates that there are about 7000 cluster members, belonging to 140 clusters, ranging in age from 0.1 to 6.5\,Gyr in groups of 10-250.  HRMOS could most efficiently target the 5128 targets in 48 clusters with groups larger than 40. For best efficiency, the multiplexing factor would need to be $\geq100$ since this is the median number of targets in these clusters, and of course some fibers need to be reserved for sky subtraction. The highest target density in these clusters is low enough ($\sim4$/arcmin$^2$) that the minimum fibre spacing can be $\sim30$\,arcsec. 

The main programme would be a set of $\sim30$ monitoring observations on each cluster over the course of a couple of years to search for planets with periods out to $\sim100$ days. Beyond this, the massive planet population is likely to be picked up by {\it Gaia} astrometry. Such a large scale monitoring programme would use approximately 200 nights of telescope time and yield $\sim50-100$ hot Jupiters ($P<20$ days) and perhaps a similar number of exoplanets on wider orbits. 

It is worth noting that such a programme would also offer an exquisite and world-beating dataset for investigating the binary properties and frequencies of solar-type stars (sensitive to both stellar and substellar companions and in a separation range to which {\it Gaia} is insensitive) and for investigating the 3-dimensional kinematics of clusters in conjunction with {\it Gaia} astrometry. 
Needless to say, the data can be used for a complete characterisation of the planetary host star, and members without planets, providing very high quality data to study the correlation between stellar (particularly chemical) and planetary properties.

\subsubsection{Giant stars}

HRMOS can also perform a survey of thousands of G/K type giants in open clusters down to $V\sim16.5$. The RV precision of these observations will be similar (or perhaps slightly better, e.g.  \citealt{sozzetti2006}) to those for K dwarfs of similar brightness (see Table~\ref{tab_bouchy}), but the RV amplitudes for a given planet mass will be smaller because the host star is of higher mass. High resolution is required because the spectral features in most K giants will be narrow and the RV precision will increase up to the point where the line profiles become resolved.

\begin{figure}
\centering
\resizebox{0.75\hsize}{!}{
\includegraphics{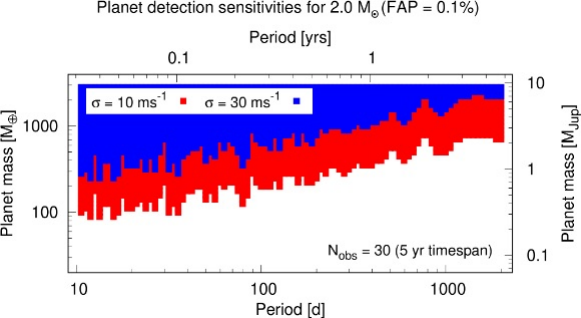}}
\caption{
Detection efficiency (with a 0.1\,\% false alarm probability) for exoplanets orbiting a 2\,M$_{\odot}$ giant star. The proposed 30 observations are distributed over a 5 year period as described in the text. Simulations where the instrumental velocity precision is 10\,m\,s$^{-1}$ or 30\,m\,s$^{-1}$ are shown.
\label{fig:detectionefficiency}
}
\end{figure}

Figure~\ref{fig:detectionefficiency} shows a simulation of the detection sensitivity on the basis of an instrumental precision of either 10\,m\,s$^{-1}$ or 30\,m\,s$^{-1}$ and assuming that the giant contributes no additional intrinsic noise. The adopted sampling is 14 observations spread uniformly over a year and then 4 observations per year spread over a further 4 years. (i.e. a 5 year monitoring programme). Comparing Figs.~\ref{fig:cumulativefrequency} and \ref{fig:detectionefficiency} shows that in order to detect a sizeable fraction of the exoplanet population found around field stars requires a precision of 10\,m\,s$^{-1}$, especially for orbits with periods longer than  100 days.

\begin{figure*}
\centering
\resizebox{\hsize}{!}{
\includegraphics{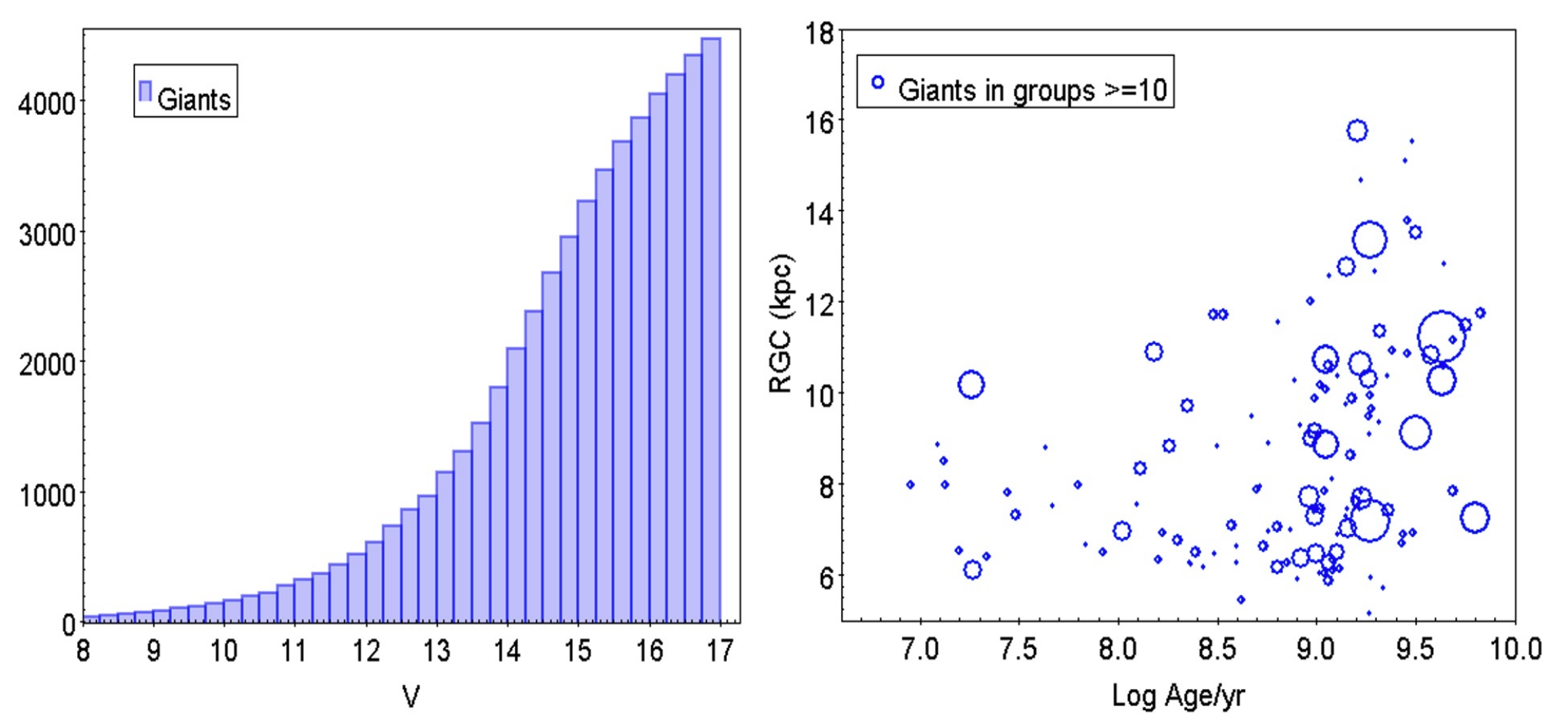}}
\caption{
{\sl Left:} Cumulative frequency of apparent magnitude for high probability cluster giants with $\rm DEC<20$ in the {\it Gaia} DR2 cluster catalogue of \cite{cantatgaudin2020}. {\sl Right:} Galactocentric radius as a function of  age for those clusters with more than 10 giants within a 25 arcmin FoV. Symbol size is linearly proportional to number of giants in the FoV.
\label{fig:cumulativefrequencygiants}
}
\end{figure*}

Figure~\ref{fig:cumulativefrequencygiants} (left panel) shows the cumulative histogram of $V$ magnitudes for high probability cluster giants at $\rm DEC<20$\,deg, selected in the {\it Gaia} absolute magnitude/intrinsic colour diagram from the catalogue of \cite{cantatgaudin2020}. There are approximately 4200 giants (RGB and red clump) with $V<16.5$ in 550 clusters. Figure~\ref{fig:cumulativefrequencygiants} (right panel) shows 198 clusters that have $10-121$ cluster giants within a 25\,arcmin diameter FoV (a total of 2778 stars), as a function of their age and distance from the Galactic centre. There are 1385 stars concentrated in groups with more than 30 stars, in 32 clusters which cover the full range of ages and Galactocentric radius. With a dedicated RV monitoring program of the richest clusters over a period of $\sim5$\,years, and assuming an occurrence rate of about $20-30$\,\% for Jovian planets within 3\,au \citep[e.g.][]{jones2021}, HRMOS should discover hundreds of $>1$\,M$_J$ exoplanets around giant stars in $\sim100$ nights of observing.

Table~\ref{tab_targets} then lists the $V$ magnitude of the red clump ($M_V \sim 0.7$) in several clusters, in the bulge and in dwarf galaxies. It provides the magnitude range for possible targets in an HRMOS survey in unexplored environments.

\begin{table*}[t]%
\centering
\caption{The $V$ magnitude of the red clump ($M_V \sim 0.7$). }%
\footnotesize
\begin{tabular}{ccccc}
\noalign{\smallskip}
\hline
\hline
Environment & Distance  & [Fe/H]   & Red clump $V$ & Target mag range    \\
            & [kpc]     &          & [mag]         & [mag]             \\
\hline
47 Tuc        &  4.0      &  $-0.8$             & 14.2    &   13.2-16  \\
Omega Cen     &  4.9      &  $-2$ to $-0.5$     & 15.0    &   14-16.5  \\
Bulge fields  &  8        &  $-0.8$ to $+0.6$   & 16.7    &   15.7-17  \\
Sgr dSph      &  20       &  $-1.5$ to $-0.2$   & 17.8    &   16.5-17.8  \\
LMC           &  50       &  $-0.3$             & 19.4    &   17-18  \\
\hline
\end{tabular}
\label{tab_targets}
\end{table*}

\subsubsection{Giants in the Bulge, globular clusters and local dwarf galaxies}

We aim to isolate samples of  RGB and RC stars, avoiding very luminous giants and AGB stars with a high intrinsic noise that do not permit exoplanet detection. Using local giants as a template, the bulk of exoplanets have been found around giants with $-1.5 < M_V <1.5$ and so we would aim to cover this region where possible.

\cite{sozzetti2006} have shown that a metallicity of $\rm [Fe/H]\approx-1$ only reduces $Q$ by a factor of $\sim2$, so the limiting magnitudes shown in Table~\ref{tab_bouchy} are still mostly applicable to these targets, although may be underestimates for some of the metal-rich bulge stars and overestimates for the most metal-poor starsin, for instance, the cluster Omega~Cen or the Sagittarius dwarf galaxy (Sgr~dSph). There are multiple rich target fields available in all these regions and {\it Gaia} data is easily able to pick out largely uncontaminated samples, such that an order of 100 targets are available over the HRMOS FoV. In both 47~Tuc and Omega~Cen, the range of target magnitudes easily encompasses the clump and the central part of the RGB. Single one-hour exposures in these fields should result in velocities that are limited by the instrumental precision, which for a precision of 10\,m\,s$^{-1}$ should identify exoplanets with masses of $0.7-2$\,M$_J$ in orbits of $100-1000$ days ($\sim0.5-2.5$\,au separation). Given a frequency for Jovian planets within 2.5\,au of field giants of $\sim25$\,\% \citep[e.g.][]{jones2021}, then even observing single fields of 100 objects in each region should yield robust statistical conclusions about the planetary frequency. Such a survey would take only order of 4 nights per region spread over 5 years of observation. 

A more ambitious programme is to observe multiple fields in the bulge (in Baade’s window). Several fields with $\sim100$ RGB and red clump targets have already been observed with FLAMES \citep[e.g.][]{hill2011}.  For these targets one hour exposures will just reach a 10\,m\,s$^{-1}$ precision for the brighter stars, but have SNR-limited precision of $\sim20$\,m\,s$^{-1}$ in the faintest, which is still sufficient to probe $2-5$\,M$_J$ planets in orbits of $100-1000$ days (which are observed in local field giants). Observing  4 fields should give a large enough sample to see whether the frequency of giant exoplanets was a function of metallicity within the bulge.

Finally,  we can consider observing giants in the nearest dwarf galaxies. Observations of clump stars will be almost impossible in Sgr~dSph and the Magellanic Clouds, but it should be possible to target the upper RGB ($-2< M_V <0$) using two hour exposures in dark conditions and reach a precision of $\sim20-30$\,m\,s$^{-1}$ that is limited by the SNR of the spectrum rather than the instrument precision. The frequency of $>3-8$\,M$_J$ exoplanets at $<2.5$\,au around field giants is a few percent. Observing even one carefully chosen field in each of these regions should determine whether the frequency is similar to within a factor of two.

\begin{table*}[t]%
\centering
\caption{Instrument requirement summary (Exoplanets)}%
\footnotesize
\begin{tabularx}{\textwidth}{lXX}
\noalign{\smallskip}
\hline
\hline
\textbf{Parameter} & \textbf{Value}   & \textbf{Justification}  \\
\hline
Resolving power ($R$)     & 80\,000     & Required SNR levels become much higher for lower resolutions. \\
%
Spectral range & As large as possible. Should also allow simultaneous monitoring of Ca\,{\sc ii} lines as well as a window at $\sim500$\,nm.  & Planet detection signal is proportional to how much wavelength range can be covered. Simultaneous monitoring of chromospheric diagnostics is important for distinguishing between planet signatures and magnetic-activity induced RV modulations.   \\
Multiplexing   & 100      & Median number of solar-type targets in the $\sim30$ nearby clusters is $\sim100$ within a 25 arcmin FoV. A higher multiplex could be used in  the bulge, and external galaxies, but 100 is a good compromise    \\
Stability      & 10\,m\,s$^{-1}$         & This level of stability is required to ensure detection of exoplanets down to below a Jupiter mass at a range or orbital periods and around stars at a range of masses. It is also important for distinguishing between RV jitter and genuine planetary signatures.    \\
Fibre spacing  & 30 arcsec         & The target density in even the most densely populated target OCs is about 4 per square arc minute.     \\
\hline
\end{tabularx}
\end{table*}

\frame{\vspace{-7mm}
\paragraph{Uniqueness}
Surveys by Kepler, TESS, CHEOPS and PLATO will discover populations of (mostly close-in,  $a<0.1$\,au) exoplanets using the transit technique, but they will only uncover a small fraction of the population because the orbit inclination must be close to 90 degrees. In addition, only a small fraction of the targets will be in clusters and the transit technique is limited to relatively nearby main sequence stars and is much less sensitive to planets around evolved stars. {\it Gaia} should detect more than $10^4$ exoplanets via astrometry. These planets will mostly have $M>1$\,M$_J$,  $0.5 < a < 5$\,au; they will have host stars closer than 600\,pc with $G<16$, the majority of which will be field main sequence stars of uncertain age, with perhaps only a few hundred detected exoplanets around giants \citep{perryman2014}. WFIRST (now the Nancy Grace Roman telescope) will use microlensing techniques to detect $\sim 10^3$ exoplanets around field stars towards the Galactic bulge \citep{penny2019}. A few hundred of these will have Jovian masses, but they will be mostly located at orbital separations of $1-10$\,au, overlapping with {\it Gaia}.  HRMOS is almost entirely complementary to these surveys – it will provide a much more complete survey of close-in giant exoplanets with $a<0.5$\,au (for solar-type stars) in targeted clusters of known age and composition. In addition it can survey for exoplanets around a large number of giant stars in clusters, at distances and brightness to which both {\it Gaia} and transit surveys will be insensitive to exoplanets.

Spectroscopic facilities around the world cannot duplicate the capabilities of HRMOS. There are many spectrographs with velocity stability that surpasses what is proposed, but these are either on smaller telescopes (e.g., HARPS) or can only observe one object at a time (e.g., ESPRESSO, Keck ANDES). On the other hand, multi-object spectrographs, such as WEAVE and 4MOST, lack the wavelength stability or the resolving power to do the work proposed here. The proposed Mauna Kea Explorer will have a wider FoV, more throughput and better multiplexing, but it has a resolving power of $40\,000$ and likely a velocity stability of only 
100\,m\,s$^{-1}$. Similar arguments apply to WST, which in its high resolution mode will not have a resolving power higher than $40\,000$.
}

\clearpage

%% file: clusters.tex
\section{Exploiting the full potential of star clusters at very high spectral resolution}\label{sect:clusters}

For decades, open and globular star clusters have been considered the best examples of simple stellar populations. However, over time, they have revealed their complexity: from the existence of multiple populations in globular clusters to the age spreads in star clusters belonging to large star-forming regions and the presence of halos and tidal streams in both open and globular clusters. Furthermore, star clusters are the best test beds for studying the effects of stellar evolution. Since open clusters are composed of coeval populations formed with the same initial chemical composition, it is possible to analyse in detail the effects of stellar evolution on the photospheric abundances of their member stars, studying, for example, the effect of mixing and transport mechanisms as a function of stellar mass. Understanding these mechanisms is crucial not only for stellar physics but also for understanding the applicability and limits of chemical tagging.

Here we list a number of questions that can be addressed with an instrument such as HRMOS: 

\frame{
\titlecol{Key Questions:}
\begin{itemize}
\item Do stars born together from the same molecular cloud share exactly the same chemical composition? To what extent can we use chemical abundances to trace stars back to their native environments? Can we use open clusters to design the best strategies for {\it chemical tagging}?  
\item How do the multiple stellar populations in globular clusters (GCs) form and evolve? What is the relationship between the properties of star clusters and their birth environments?
\item Do star clusters contain central black holes? Can we detect their effects? 
\item What are the physical mechanisms driving the extra mixing processes in the stellar interior? How do these mechanisms vary along the evolutionary tracks?  How does the efficiency of the transport processes depend on stellar metallicity, mass, and rotation?
\end{itemize}
}
\subsection{Unveiling the foundation of chemical tagging with open clusters}
\begin{figure}[t]
\centering
\resizebox{0.5\hsize}{!}{
\includegraphics{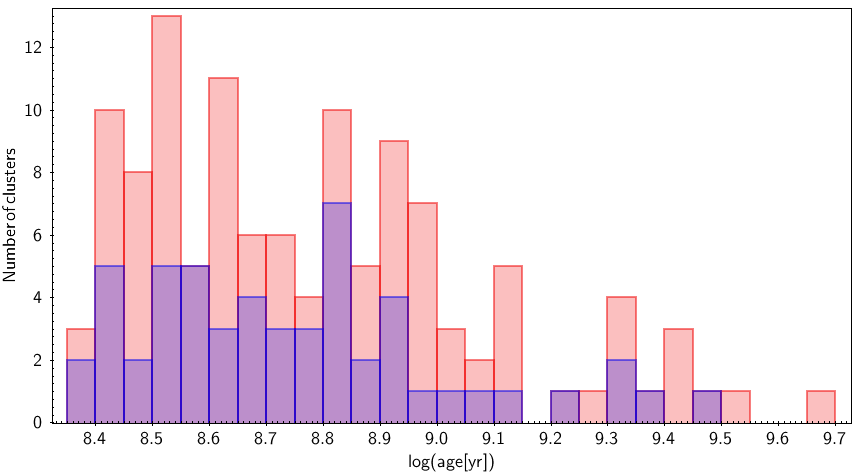}}
\resizebox{0.5\hsize}{!}{
\includegraphics{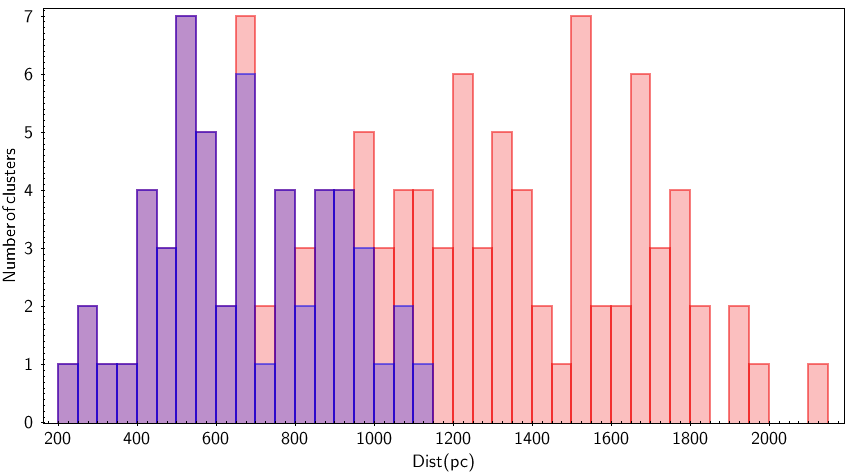}}
\caption{
Age distributions (upper panel) and distance distributions (lower panel) for the proposed samples of open clusters (in red the sample of 120 clusters, considering a limiting magnitude of $AB-mag=17$ of G-type dwarfs -  $\rm SNR\sim20-30$ (depending on the wavelength) in a one hour exposure time, and in blue the sample of 54 clusters, $AB-mag=15$ of G-type dwarfs - $\rm SNR\sim 70-100$ in one hour exposure time -$R=80\,000$).
\label{fig:age_dist}
}
\end{figure}

Open clusters are among the most versatile objects with a variety of applications in astronomy \citep{Friel2013, Krumholz2019}. On the one hand, with their well-measurable ages and distances \citep[e.g.][]{Bossini2019, cantatgaudin2020, Dias2021, hunt23, cavallo23}, they are the optimal test particles to constrain Galactic chemodynamical models \citep[e.g.][]{Palla2020, molero23}. Open star clusters, in particular, are recognised to be among the best tracers of the Galactic disc \citep[e.g.][]{Donor2020,magrini23}. Their study is based on the interplay between photometric data (generally ground-based but now also from {\sl Gaia}; and in the future from both Rubin/LSST\footnote{\url{https://www.lsst.org/}}, in the optical, and from the Nancy Grace Roman space telescope\footnote{\url{https://roman.gsfc.nasa.gov/}}, in the near-IR) to define the cluster colour-magnitude diagram, astrometric data and radial velocities for an accurate determination of cluster membership, and high-resolution spectroscopy to derive stellar parameters and detailed elemental abundances.

On the other hand, open clusters are also test beds for theories of star formation and evolution, being the best examples of simple stellar populations \citep[e.g.][and references therein]{Luck1994, Smiljanic2009, Drazdauskas2016, Krumholz2019, Krause2020, Magrini2021, Tsantaki2023}. However, there are several unresolved questions regarding open clusters. One of them is the chemical homogeneity of the members of the cluster, which has important implications in confirming the theoretical basis of the chemical tagging technique \citep[e.g.][and references therein]{blancocuaresma15, Bovy2016, Poovelil2020, cheng21, Spina22}.

An assumption often made in the studies of our Galaxy is, indeed, that the chemical makeup of a star provides fossil information on the environment where it was born. Under this premise, it should be possible to tag stars that formed from the same material through clustering in the chemical space. This methodology is known as ``chemical tagging'' \citep{freeman2002}. The chemical tagging technique makes the hypothesis that it is possible to use elemental abundances (with the exception of some light elements, where the photospheric composition is modified by stellar evolution) to reconstruct dispersed chemically homogeneous star clusters \citep[for masses up to 10$^5$ M$_{\odot}$, see e.g.][]{blandhawthorn2016}. The success of chemical tagging depends on the significance of two critical factors: {\it i)} the chemical diversity of the interstellar medium in space and time, and {\it ii)} the level of chemical homogeneity of stars formed from the same giant molecular cloud. Both of these factors need to be probed with the highest precision possible.

Therefore, probing the level of chemical homogeneity of star clusters is a key test to confirm the basis of chemical tagging. With current instrumentation, essentially all open clusters appear to be chemically homogeneous to better than 0.1\,dex. However, recent studies have shown that chemical anomalies can be present among members of the same association \citep[e.g.][]{liu2016, liu2019, oh17, bertellimotta17, ramirez19, nagar20, church20}. The frequency of these anomalies, their causes, and the levels of chemical variation for different elements are still under investigation \citep[e.g.][]{Spina21}. Furthermore, magnetic fields and star spots can alter the equivalent widths of absorption lines in the spectra of young stars \citep{yana19, baratella20, spina2020}. This can result in apparent chemical variations during stellar activity cycles, depending on the method of spectroscopic analysis and the line-list that is employed, with potential repercussions on our ability to use chemical tagging. Finally, in specific age ranges, atomic diffusion can generate differences in the photospheric composition of main-sequence stars compared to giants, as observed in the case of M67 \citep[see, e.g.][]{onehag2014,bertellimotta17, liu2019, Souto2019}. 

 It is possible to characterise these modulations of stellar chemical patterns by obtaining extremely precise differential abundances between members of star clusters. All this would provide a quantitative evaluation of the basic assumptions of chemical tagging. That is critical for probing to what extent chemical abundances can be used to tag stars to their native building blocks in our Galaxy. It will also improve current techniques of spectroscopic analysis by highlighting the limitations and inconsistencies caused by the simplistic assumption that stellar spectra are not affected by magnetic fields and stellar spots.

\frame{\vspace{-8mm}\paragraph{Uniqueness:} Very high-resolution spectroscopy, coupled with a wide spectral range -or to optimised spectral windows- will help to i) improve the precision of the elemental abundances; ii) put constraints on intra-cluster abundance variations; iii) define the inter-cluster differences; iv) design the best strategies for the chemical tagging analysis of field stars \citep[see][]{blancocuaresma15,Spina22}. All of these objectives have a fundamental importance that goes beyond chemical tagging testing. For example, we expect greater differences among the abundance patterns of different open clusters in the inner Galactic disc, since this region undergoes a faster chemical evolution, while we expect more homogeneous abundances in the outer Galactic disc, at least for clusters that were born {\it in situ} \citep[see][]{Poovelil2020}. A complete characterisation of the abundance patterns of distant open clusters (considering their bright red clump (RC) stars) will be possible with HRMOS, potentially providing new insight into the formation of our Galaxy.
}

\subsubsection{Requirements on open cluster science cases}

\begin{SCfigure}[0.7][t]
\centering
\resizebox{0.7\hsize}{!}{
\includegraphics{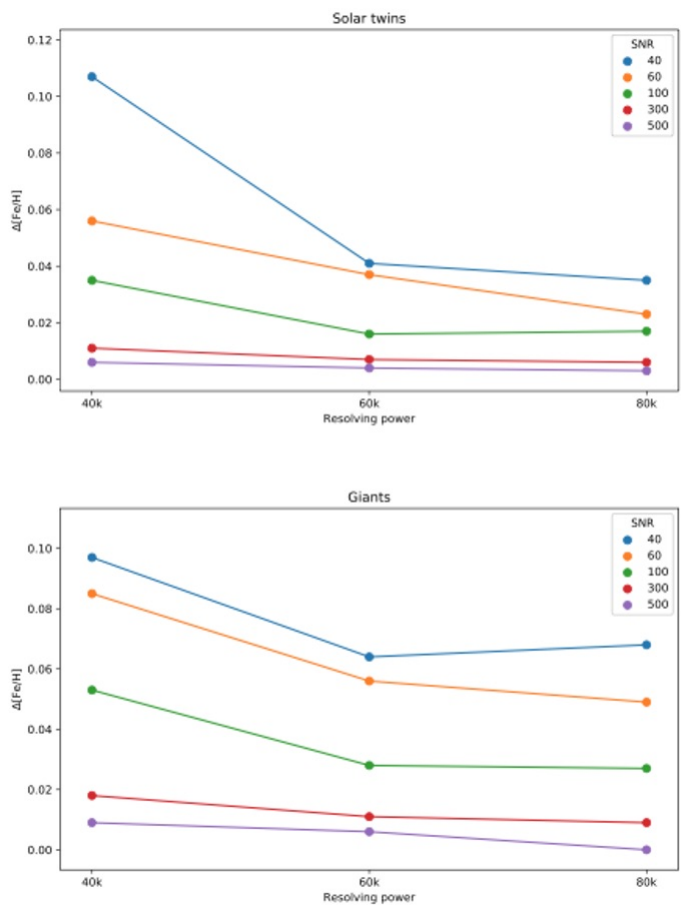}
}
\caption{
Fe abundance error as a function of resolving power and signal to noise based on analysis of simulated spectra of solar twins and giant stars in open clusters by Lorenzo Spina.
\label{fig:spina}
}
\end{SCfigure}

Differential line-by-line analysis of simulated spectra of solar twin stars and of RC stars, in which small variations in metallicity and in elemental abundances were added, have shown that spectra with $\rm SNR \sim 100$ can produce final uncertainties on [Fe/H] of the order of 0.02\,dex for solar twin stars and of 0.03\,dex for giant stars at R=80\,000 (see Fig.~\ref{fig:spina}). However, the effect of high resolution is even more important at intermediate SNR, where much higher precision can be achieved for R=80\,000 than for R=60\,000.

The recommended spectral windows should include a large number of Fe\,{\sc ii} lines, covering a wide range of excitation potential (EP) and EWs. The Fe\,{\sc ii} lines are located primarily within the two spectral windows $500-550$\,nm and $600-650$\,nm. These spectral windows also allow detection of Fe\,{\sc i} lines and lines of other elements of interest. The requirements for this case are summarised in Table~\ref{tab:oc}.

We estimated the number of open clusters, observable with the Very Large Telescope (VLT), in which we can target at the same time stars of FGK spectral types at the main sequence (MS) and the RC, to investigate the effects of stellar evolution. Taking into account the most favourable case, with a limiting magnitude of $AB-mag\sim17$ for G-type dwarf stars, we have a sample of 120 clusters ($\rm \log (age) > 8.4$). In this case, we will be able to achieve $SNR > 30$ in one hour in the redder part of the spectrum and $\rm SNR\sim20$ in the bluer part. 
On the other hand, with a $AB-mag\sim15$ we should get $\rm SNR \sim125$ in the red part of the spectrum and slightly lower in the bluer part ($\rm SNR\sim60$). This magnitude is reached in a typical G-type dwarf star in about 54 clusters, covering a wide range of ages, and located within 1.2\,kpc from the Sun.
Much higher SNRs will be achieved for giant stars belonging to the same clusters because they are brighter. 

In Fig.~\ref{fig:age_dist} we show the age and distance distribution of the selected samples of open clusters \citep[ages and distances from][]{cantatgaudin2020}.


\begin{table*}[t]%
\centering
\caption{Instrument requirement summary (star clusters)}%
\label{tab:oc}
\footnotesize{
\begin{tabularx}{\textwidth}{llX}
\noalign{\smallskip}
\hline
\hline
\textbf{Parameter} & \textbf{Value}   & \textbf{Justification}  \\
\hline
Resolving Power & $>$ 60,000 & The higher the R, the lower are the demands on the SNR. The performances at R=60,000 and R=80,000 are quite similar for SNR > 100.\\
Spectral range & $500-550$\,nm and $600-650$\,nm  &  Spectral windows which include several Fe I and Fe II lines, and lines of other important elements \\
Multiplexing   & $\geq$20 fibres   &   to observe at least 20 members of open clusters, across the evolutionary sequence    \\
Stability      & 0.05 km s$^{-1}$    &  Necessary to study the cluster internal kinematics and possibly to detect signatures of the central black hole  \\
Fibre spacing  & 30 arcsec    &  The target density in even the most densely populated target OCs is about 4 per square arc minute.         \\
SNR          &     $>100$    &  To reach uncertainties in [Fe/H] in dwarf and giant stars as low as 0.02 and 0.03 dex.       \\
\hline
\end{tabularx}
}
\end{table*}

\subsection{The origin of the multiple populations in globular clusters}

One of the hottest science topics related to globular clusters (GCs) is the existence and origin of their multiple populations. Multiple populations are characterised by an anomalous chemical composition with respect to field stars of similar metallicity, with light elements (anti-)correlations (see reviews by \citealt{gratton2004,gratton2012,gratton2019,bastian2018,milone2022}). The light elements involved are He, C, N, O, F, Na, Mg, and Al (and also Li, sometimes Si and heavier elements such as Ca and K). High-precision photometric diagrams of GCs show that around the 17\,\% of them (among the most massive) present variations in heavy elements \citep{milone2017b}, in the form of star-to-star differences in Fe, often correlated with variations in elements produced by the slow neutron capture process \citep{marino2009,marino2011,marino2015}. Recent observations have shown that variations in overall metallicity could be a much more widespread phenomenon that affect GCs \citep{milone2017b, legnardi2022}. The variations are expected to be as small as $\lesssim0.10$\,dex, and can be spectroscopically detected only by analysing large samples of stars in GCs with high-quality data; namely  sufficient resolution and high SNR \citep[e.g.][]{yong2016,marino2019a, marino2019b}.

\begin{figure*}
\centering
\resizebox{0.6\hsize}{!}{
\includegraphics[viewport=720 10 1300 480,clip]{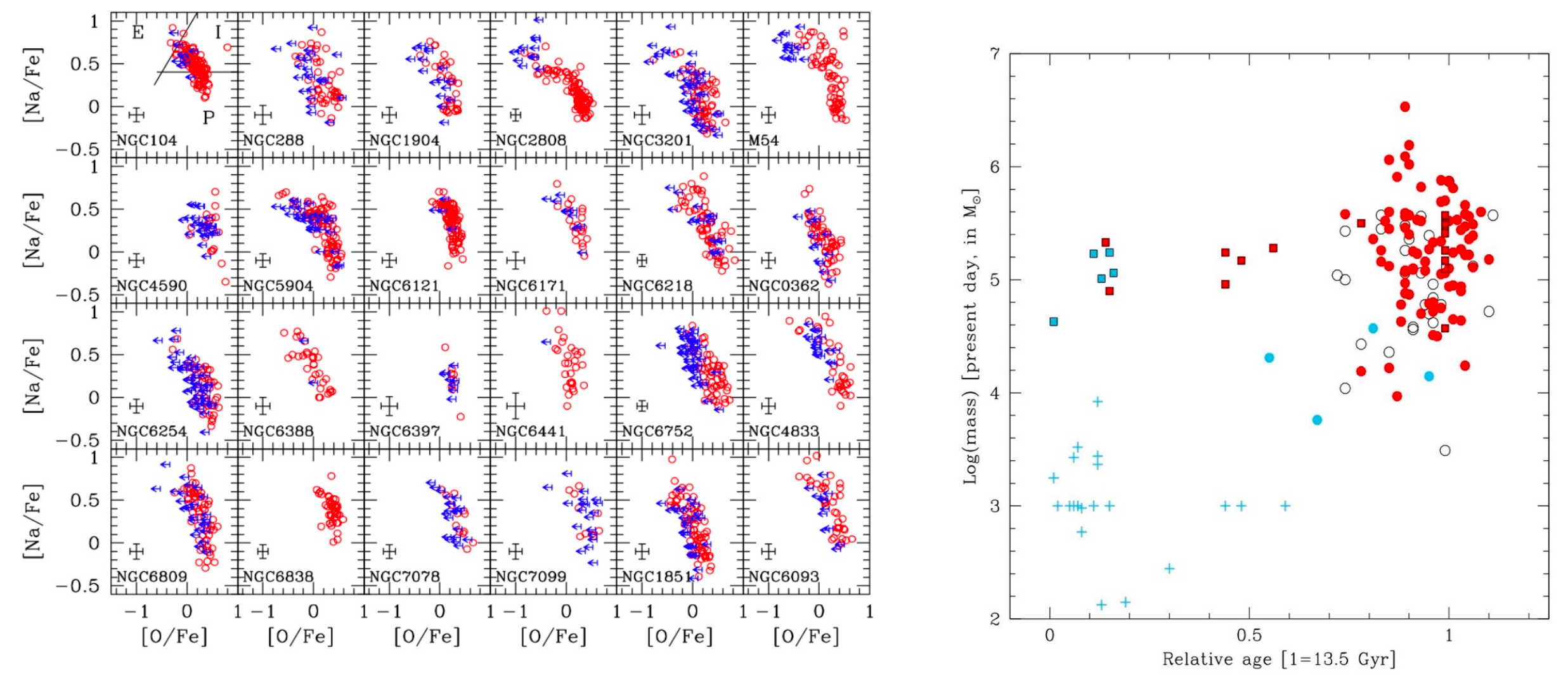}}
\caption{
Age-mass plot of star clusters in different environments.  Plus signs stand for open clusters in the MW, circles for GCs in the MW, and squares for GCs in the Magellanic Clouds and in the dwarf galaxy Fornax. Red symbols indicate the presence of multiple populations, light blue absence, and white with no indication yet \citep[figure taken from][]{bragaglia17}.
\label{fig:nao_anticorr2}
}
\end{figure*}

Almost all of the Milky Way GCs, and many old massive clusters in the Magellanic Clouds and in Fornax (but not open clusters), show indications of multiple populations (see Fig.~\ref{fig:nao_anticorr2}, taken from \citet{bragaglia17}). Their origin probably lies in multiple episodes of star formation (although the consensus is not unanimous), with subsequent generations forming out of material polluted by stars formed in previous generation(s). The nature of the polluters and the actual mechanism of formation of GCs is still debated.

\begin{figure*}
\centering
\resizebox{0.8\hsize}{!}{
\includegraphics[viewport=0 22 700 505,clip]{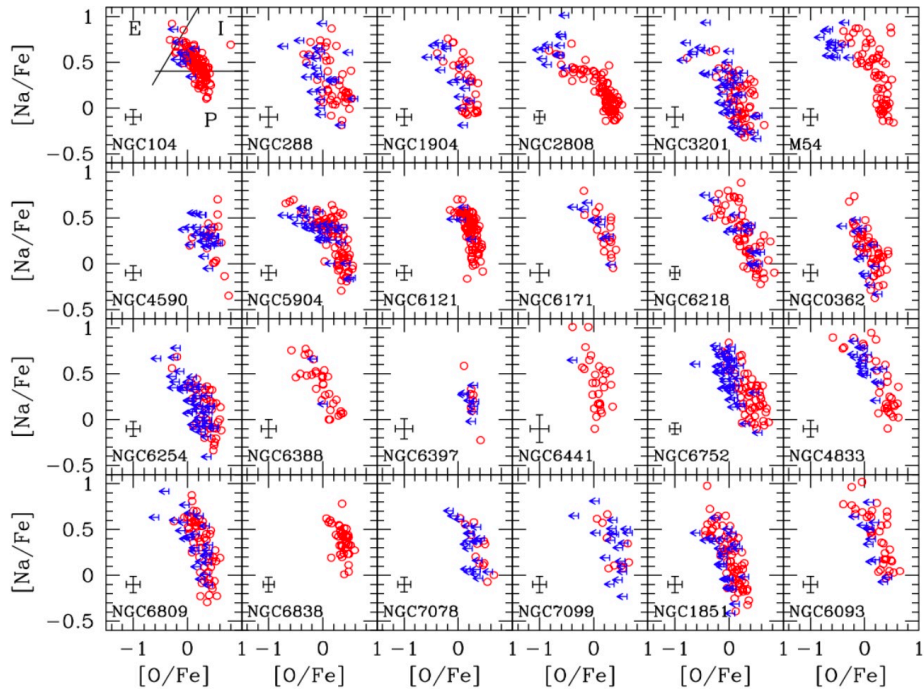}}
\caption{
Na-O anti-correlations in MW GCs (taken mostly from \citealt{carretta2009a,carretta2009b} and in general from the FLAMES GC survey by Carretta and collaborators). 
\label{fig:nao_anticorr1}
}
\end{figure*}

Observations have shown that every GC has its own history even if there are regularities (in particular relations with mass, see, e.g. \citealt{carretta2009a, carretta2009b,  milone2017a}). Figure~\ref{fig:nao_anticorr1} shows examples of Na-O anti-correlations detected in different GCs, in part, from the homogeneous studies, using FLAMES, by Carretta and collaborators \citep[e.g][]{carretta2009a,carretta2009b,bragaglia17}. More data are available from many different groups and large spectroscopic surveys (such as {\sl Gaia}-ESO, \citealt{pancino2017} and APOGEE, \citealt{meszaros2020}). It should be noted that the high multiplexing required to efficiently study star clusters needs to be weighed against resolution, wavelength coverage, and exposure time. As an example, more than 100 stars in the GC M4 has been studied by \citet{carretta2009a} with FLAMES-GIRAFFE in less than one hour and by \citet{marino2009} with FLAMES-UVES in about 30 hours. The latter study, due to the higher resolution and much larger wavelength coverage, reached a higher precision on abundance ratios, which is crucial when separating GC populations. 

 A new instrument that aims to contribute to the study of GCs must reach the best trade off between all factors mentioned above to be efficient and effective. A relatively large FoV (compared to the observed angular size of GCs), coupled with the ability of {\sl Gaia} astrometry to assign robust membership also in the external regions of GCs, will allow a complete radial coverage of the clusters.

High-precision measurements must be provided for Fe, to uncover and robustly measure possible small star-to-star variations in more GCs which are not uniformly found at the moment. For example, the case of the massive bulge cluster NGC~6388, for which \citet{carretta_bragaglia2022} do not find a significant Fe spread, but only the typical correlations and anti-correlations for light elements \citep{CarrettaBragaglia2023}. Furthermore, similar coverage should be extended to all other elements involved, in particular Li and those produced by neutron-capture processes. 

Lithium abundances offer crucial diagnostics as to the stellar source of internal pollution in GCs, that is, massive stars vs. intermediate-mass AGB  stars \citep{PasquiniMolaro1996, Bonifacio2002, pasquini2005, lind2009b, Mucciarelli2011}. Lithium production is theoretically expected (and observed) only in relatively massive AGB stars ($\rm \gtrsim 5 M_{\odot}$) \citep{VenturaDAntona2010}. The investigation of Li as a function of elements involved in p-capture reactions provides, hence, a very powerful constraint: observations should be performed by exploiting giants below the RGB bump luminosity (where all Li should have disappeared from the stellar surface), after the dilution occurring at the first dredge-up has been taken into account. This calls for an efficient spectrograph which is able to reach fainter magnitudes than what is currently feasible and which extends the study to GCs with a wider range of properties.

As far as neutron-capture elements are concerned, no systematic investigation of their behaviour has yet been carried out, with few exceptions including  NGC 362, NGC 6388, Omega Centauri, M22, NGC 1851 to name a few \citep[see][]{worley10, worley10a, worley13, marino2015}. The underlying reason for this lack of observations is that previous GC surveys (e.g., by Carretta and collaborators) have been analysing FLAMES-GIRAFFE (and UVES) spectra, which do not cover the blue wavelength domain ($< 500$\,nm) where most of the lines of heavy elements reside. This is of particular importance for the study of overlooked species, such as Pb. The determination of Pb abundances allows us to infer crucial information on the mass range of intra-cluster polluters, in particular where s-process element variations have been detected. The derivation of the relative ratios of the light to heavy s-process elements (first peak Y and Zr; second peak Ba and La; and Pb for the third peak) represents the most powerful tool to constrain the source of the polluters and their properties, in particular mass and metallicity.

The wavelength coverage to redder wavelengths up to $\sim800$\,nm is also important. This would allow us, for the first time, to carry out an extensive abundance study of rubidium (Rb), which has been predicted and observed in relatively massive AGB stars in the Magellanic Clouds, undergoing hot bottom burning \citep{GarciaHernandez2009}. This process has been deemed to likely be responsible for light-element variations and multiple populations in GCs.

In addition, high-resolution spectroscopy will allow us to analyse MgH spectral features for the coolest stars ($T_{\rm eff}< 4500$\,K), to be compared with abundances from Mg\,{\sc i} lines. This comparison can provide information on the He abundances of GC stars (see \citealt{reddy2020}, and references therein), assuming we fully understand the line formation of Mg\,{\sc i} and MgH lines, including effects caused by granulation, deviations from local thermodynamic equilibrium (LTE), and activity.

\begin{figure*}
\resizebox{\hsize}{!}{
\includegraphics{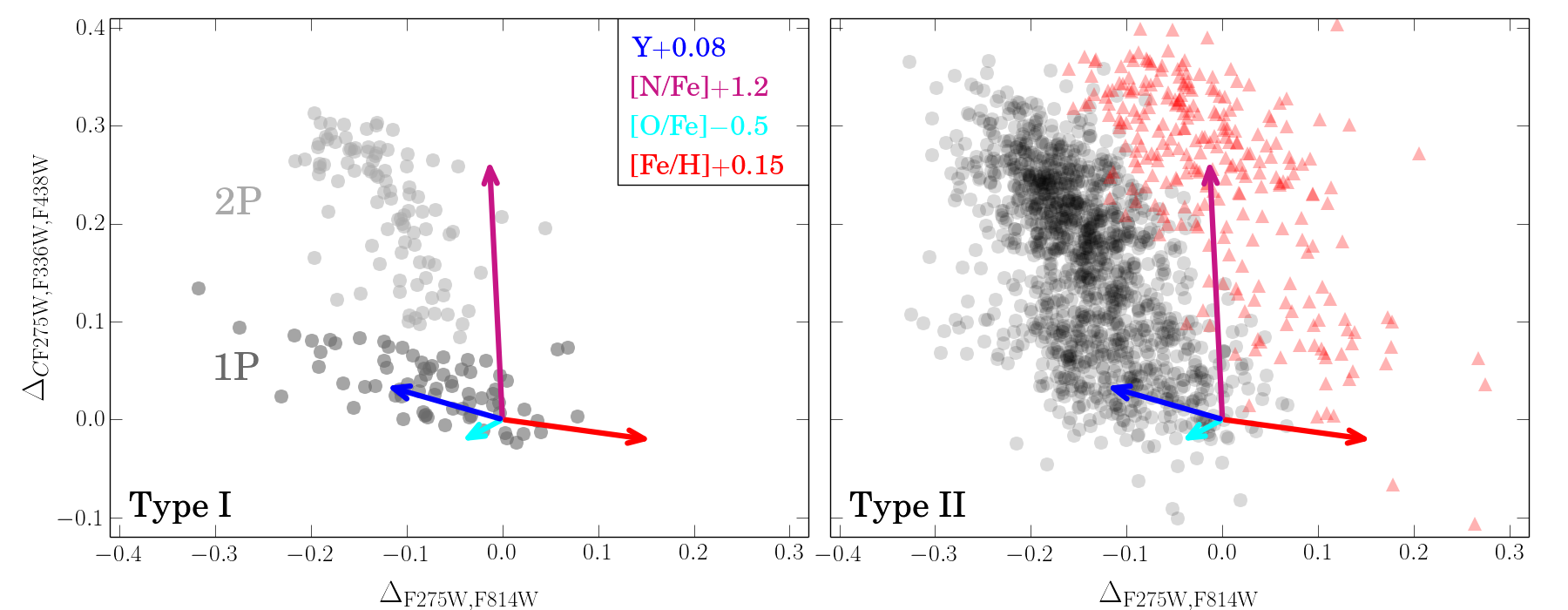}}
\caption{
A combination of HST bands (F275W, F336W, F438W, and F814W) allows the construction of multiple colour plots nicknamed ``Chromosome Maps'' \citep{milone2015}, where the position of stars is especially sensitive to the abundance of helium, nitrogen, and metallicity. Left and right panels show a typical Milky Way GC, NGC~3201, with first and second populations (1P and 2P) and a GC with a more complex population pattern, NGC~5286, associated with internal variations in heavy elements and C+N+O. 
\label{fig:hst_bands}
}
\end{figure*}

High-precision photometry from the Hubble Space Telescope (HST) Treasury programme \citep{piotto15} has recently revealed unexpected features of the multiple population pattern in the newly introduced ``Chromosome Map'' (ChM) diagrams \citep{milone2015, milone2017a}. So far, this is the most effective tool for isolating the different populations in a GC. Examples of ChMs are shown in Fig.~\ref{fig:hst_bands}, where it is clear that the appearance of stellar populations with different chemical abundance patterns shapes this diagnostic diagram. The ChMs clearly show the presence of the two main populations (1P and 2P in the Type~I GC NGC~3201) with more massive GCs displaying multiple maps, corresponding to internal variations in metallicity (e.g. the Type~II GC NGC~5286). Perhaps, the most unexpected discovery made using ChMs is the apparent chemical in-homogeneity of even the first stellar population (1P), that is, the one with elemental ratios similar to halo field stars. This feature is quite common among GCs. Recent observations on a few stars suggest that tiny spreads in overall metallicity, at a level of $\lesssim0.1$ dex, may in part explain the phenomenon \citep{marino2019a, marino2019b, lardo2023}.

The possibility of having internal metallicity variations even within the first population suggests that either the molecular cloud out of which these stars formed was not chemically homogeneous, or that part of the stars formed while supernovae (SNe) had already started to enrich the interstellar medium (ISM) of the nascent GC. Hence, determining high-precision chemical abundances of large samples of first population stars in all those GCs with hints of inhomogeneity is crucial to understand the nature of these stellar systems and reconstruct the early phases of their formation. To this aim, we require high-resolution, multiplexing capabilities, and optimisation for the observations in the central GC regions, where ChM are available for target identification.

 To fully understand what causes the multiple populations and how they are fundamentally tied with the cluster formation and early evolution mechanisms, we need to have large statistics in each GC, from the centre to the outskirts, for as many GCs as possible, for many species, with high precision, analysed in the most homogeneous way. We also need to couple information coming from spectroscopy and photometry (which, as illustrated in Fig.~\ref{fig:hst_bands}, is sensitive to variations in He, O, C, N that can act as very low-resolution spectroscopy over very large samples). Future asteroseismic missions would also bring fundamental information on the masses, He content, and rotation properties of first- and second-generation stars. 

\frame{\vspace{-7mm}
\paragraph{Uniqueness:} Future large spectroscopic surveys at intermediate resolution ($R \sim5000$ to $\sim20\,000$), such as WEAVE and 4MOST, will be able to obtain large samples of stars and cover the full extension of GCs, up to and beyond the tidal radius, possibly extending to tails. However, to obtain the full suite of important elemental abundances, those observations will have to be combined with higher-resolution, high-multiplexing spectroscopy of stars in the cluster core.
}

\subsubsection{Requirements on globular cluster science cases}

The main scientific aims to pursue with HRMOS in the field of GCs are:
\begin{itemize}
\item
To derive high-precision, homogeneous detailed elemental abundances in RGB, AGB, horizontal branch (HB), and main-sequence turnoff (MSTO) stars in selected GCs, focusing on light element abundances (Li, C, N, O, Na, Mg, and Al) and neutron-capture elements, including elements belonging to the third peak, such as Pb. The spectral coverage of HRMOS needs to allow for simultaneous measurements of all elements for many stars (for studies of the mechanisms of pollution and GC formation).
\item
To assess a (possible) internal spread/split in metallicity to understand if this is present only in Type~II GCs. This would demonstrate whether these higher mass GCs have experienced a more complex star formation history, possibly related to a different formation environment, or whether this is instead a normal feature of all GCs.
\item
To extend the study of the Magellanic Cloud clusters providing access to systems of younger ages, lower masses, and different formation environments \citep[see, e.g.][for previous attempts]{mucciarelli06, mucciarelli08, mucciarelli09, mucciarelli10}.
\end{itemize}

HRMOS is expected to be mounted at the UT2 of the VLT (where FLAMES is currently located) and to have a FoV with a diameter of about 30\,arcmin, which is optimal for observing the central part of GCs. Using the ETC, we derive that it will be possible to observe with  SNR$\sim$60, necessary for precise abundance analysis stars down to a magnitude of $V=18$ with about one night of observation. The requested separation between fibres needs to be at least the one of FLAMES.

At least 100 GCs are accessible and observable from the VLT:
\begin{itemize}
\item
146 Milky Way GCs with $Dec <+20$ (\citealt{harris1996} – plus the new discoveries by, e.g., {\sl Gaia})
\item
101 GCs with $Dec <+20$ \& $V_{\rm HB} < 18$ to sample at least the RGB, AGB and HB
\item 8 GCs with $Dec <+20$ \& $V_{\rm HB} < 18$ in which it is possible to reach the MSTO).
\end{itemize}

We have performed the analysis of simulated spectra of a giant star at the typical metallicity of GCs ($\rm [M/H]=-1.5$). Our study has shown that an internal precision in [Fe/H]\footnote{$\Delta$Fe, considered as the r.m.s. of the differences in the line-by-line abundances with respect to an ideally perfect spectrum} of about $0.02$\,dex can be reached with $R\sim60\,000$ and $\rm SNR\sim150$, or with $R\sim80\,000$ and $\rm SNR\sim100$ (see Fig.~\ref{fig:marino}), in case the wavelength coverage is large enough to allow the analysis of a sufficient number of Fe lines. This nominal precision can be achieved by performing a line-by-line differential abundance analysis, as already done, for example, in \cite{yong2013}.

\begin{figure}
\centering
\resizebox{0.65\hsize}{!}{
\includegraphics{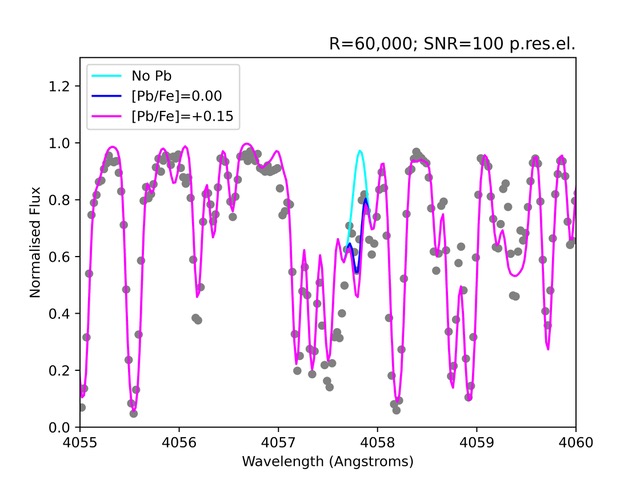}}
\caption{
Synthetic spectra of a typical giant star in a GC (4200 K, 1.5, -1.5) with different Pb abundances (No Pb, $\rm [Pb/Fe]=0$, and $\rm [Pb/Fe]=0.15$) at $R=60\,000$ and $\rm SNR=100$.
\label{fig:pb_abundances}
}
\end{figure}

Simulated spectra in the blue wavelength regions have been analysed to find the best conditions to measure the Pb lines in GC stars. At the resolving power of $R=40\,000$, the blending is severe with the Mn\,{\sc i} line at 405.79\,nm. We estimate that the combination of SNR and resolving power is not sufficient to detect differences at a level equal to, or better than, 0.1\,dex. Therefore, to reveal differences as small as 0.1\,dex in [Pb/Fe], a resolution of $R=60\,000$ is needed. This is shown in Fig.~\ref{fig:pb_abundances} for the case of $\rm SNR=100$\footnote{$\rm [C/Fe]=-0.3$ for the simulated spectra, which is a typical value for a GC giant star}.

In conclusion, a minimum resolution of $R=60\,000$ is requested for both scientific aspects tested here, in the general framework of the origin of multiple populations of GC. This high spectral resolution needs to be combined with the ability to obtain a high SNR$\sim100-150$ to meet the requirement in the precision of elemental abundances in the science cases, summarised in Table~\ref{tab:gc}.

\begin{center}
\begin{table*}[t]%
\centering
\caption{Instrument requirement summary (Globular clusters).}%
\label{tab:gc}
\footnotesize
\begin{tabularx}{\textwidth}{llX}
\noalign{\smallskip}
\hline
\hline
\textbf{Parameter} & \textbf{Value}  &  \textbf{Justification} \\
\hline
Resolving power ($R$)     & $>60\,000$  &  $R=80\,000$ is mandatory  for weak lines of heavy elements in the blue      \\
Spectral range & Down to $\sim380$\,nm &  A blue window to include Pb, Th and U lines (see also Sect.\ref{nucleo}). Several windows selected to maximise the number of Fe lines and including elements of interest. A red window to allow for both C isotopic and Rb abundances. \\
Multiplexing   & $\geq20$ fibres  & to observe at least 20 member stars of each GC        \\
Stability      & 0.05\,km\,s$^{-1}$    & necessary to study the cluster internal kinematics and possibly to detect signatures of the central black hole          \\
Fibre spacing  & 30\,arcsec    &   for the typical crowding of GCs       \\
SNR          &     $> 100-150$  & 100-150 to detect possible dishomogeneity in [Fe/H], and up to 200 for weak lines of heavy elements in the blue           \\
\hline
\end{tabularx}
\end{table*}
\end{center}
\subsection{Star clusters as probes of  mixing processes in stellar interiors}

High-resolution spectroscopic studies of the photospheric chemical composition (in terms of both elemental and isotopic abundances) of stars, whether they belong to clusters or are isolated, provide key clues to understanding their evolution, as well as that of their host galaxies. In fact, they allow us to probe a large diversity of stellar properties (i.e. different metallicities, ages, and masses for stars belonging to different Galactic stellar populations). In particular, the combination of data from astrometry, asteroseismology, and interferometry with spectroscopic observations provides essential information to follow and understand the magneto-hydrodynamic transport processes occurring in stellar interiors and to constrain their efficiency as a function of the stellar properties. In turn, stellar evolutionary models that include a more refined treatment of transport processes produce more reliable stellar yields. Once implemented in chemical evolution models and chemo-dynamical simulations, these yields lead to a better understanding of chemical enrichment processes on galactic scales.
The combination of high-spectral resolution with high multiplexity in an 8-m-class telescope makes HRMOS the best (and unique) instrument to study stellar evolution. 

Deciphering the evolution of stars requires understanding the transport processes that occur in their interiors. Surface chemical properties allow us to decrypt the signatures of these mixing processes. Therefore, it is very important to have precise constraints on these mechanisms at all stages of the evolution of stars. On the one hand, stellar clusters are very interesting to study transport processes because they provide independent constraints on the masses and radii of member stars as well as on the distance to the clusters and their ages. On the other hand, field stars allow us to collect larger statistics needed to discuss the efficiency of each mechanism as a function of mass and metallicity in different regions of the Milky Way.

The theory of stellar structure and evolution remains very uncertain when it comes to predicting the extent of the different regions inside stars (core, convective envelope, and burning shells). Magneto-hydrodynamical processes occurring in the radiative interior of stars, related or not to the stellar rotation, have an impact on the surface abundances (e.g., atomic diffusion, rotation-induced mixing, mass loss, thermohaline instability) and on the stellar structure (e.g., rotation-induced mixing, internal gravity waves and mass loss). Individually, these processes modify the lifetimes as well as the chemical and asteroseismic properties of stars at different phases of their evolution, also affecting the determination of stellar ages. The modelling and study of these physical processes have become more and more efficient with the development of hydrodynamical simulations (in one, two, or three dimensions). Over the last decade, the following issues have been addressed with (magneto-)hydrodynamical simulations: the transport of internal angular momentum \citep[e.g.][]{Denissenkov10, mathis18, Petitdemange23}; the fossil and dynamo magnetic fields to understand their role in convective envelope and radiative zones \citep[e.g.][]{duez10, duezmathis10}; the transport of angular momentum by internal gravity waves \citep[e.g.][]{mathis09} and their interaction with magnetic fields in stellar interior \citep[e.g.][]{rogers10}; and finally thermohaline instability \citep[e.g.][]{Denissenkov10, Brown13, Sengupta18}.

When combined, the chemical abundances derived from spectroscopic surveys and the stellar structure properties revealed by asteroseismic surveys for a large number of stars help us to understand the impact of hydrodynamical processes as a function of stellar masses or metallicity at different evolutionary stages \citep[e.g.][]{lagarde2019}. This leads to a deeper understanding of stellar structure and evolution. 

In the classical theory of stellar evolution, the only expected mixing episode between the main sequence and the tip of the RGB is the first dredge-up. In low- and intermediate-mass stars, it leads to an increase in the photospheric abundances of $^{13}$C and $^{14}$N and a decrease in that of Li and $^{12}$C in proportions that vary as a function of the initial stellar mass and metallicity. However, observational data show a different reality. The first observation of the carbon isotopic ratio for two K-giants by \citet{day73} revealed much lower values than expected for $\rm ^{12}C/^{13}C$ in Arcturus and $\alpha$ Serpentis, at $7.2\pm1.5$ and $12\pm2$, respectively. This discrepancy was confirmed by \citet{lambert81} with observations of C, N, and O abundances in a sample of 32 subgiant and giant stars of G- and K-type. In addition to the first dredge-up, low- and intermediate-mass stars show a further increase in N and decrease in C and $\rm ^{12}C/^{13}C$ just after the luminosity bump \citep{charbonnel98}. Various studies of relatively small samples of red giants in clusters and in the field \citep[e.g.][]{gratton2000, Smiljanic2009, tautvaisiene2013, morel14, takeda19, Aguilera23, McCormick2023} have confirmed these post-dredge-up variations, making the carbon isotopic ratio one of the best chemical indicators to constrain the mixing efficiency in giant stars. Although current large spectroscopic surveys provide chemical properties of heavier elements, the C isotopic ratio has been determined only for a small sample of evolved field stars and for a small sample of open and globular clusters, with unclear evolutionary states and much less accurate stellar masses than may be determined with asteroseismology.

 Addressing this goal requires systematic determination of C isotopic ratio for large samples of evolved stars with accurate evolutionary status (complementary with \textsl{Gaia} and \textit{Kepler}/TESS/PLATO) covering large mass and metallicity ranges. Observations should be made for a large number of stars in the field as well as in open and globular clusters.

Moreover, the description of physical mechanisms that occur in evolutionary stages from the main sequence, turn-off, sub-giant, and up to the giant phases, such as atomic diffusion \citep[e.g.][]{michaud86, michaud10}, rotation-induced mixing \citep[e.g.][]{talon98, Maeder00, palacios03,palacios06, ekstrom12, amard18}, mixing by internal gravity waves \citep[e.g.][]{garcialopez91}, or a combination of some of these processes \citep[e.g.][]{talon2005, Charbonnel08}, remains uncertain and is not able to reproduce surface abundances of all chemical elements (e.g. Li) and stellar structure properties \citep{smiljanic2010, Cummings2017, Charbonnel2020, Magrini2021, Eggenberger2022, Dumont2023}. A thorough statistical comparison between models including theoretically constrained (from numerical simulations) and observationally constrained (from asteroseismic and spectroscopic data) descriptions of transport processes opens up a new promising path for our understanding of the transport of matter and angular momentum in stellar interiors. Indeed, statistical studies of large samples of stars could highlight the efficiency and possibly the nature of magneto-hydrodynamical transport processes according to metallicity, stellar mass, age, and evolutionary state.

 Addressing this goal requires also the determination of abundances of light elements (e.g., Li, C, N, O) and iron-peak elements sensitive to radiative acceleration in field stars and in members of open and globular clusters.

The combination of an 8-m class telescope, achieving high SNR, very high spectral resolution, and multiplexing capability makes HRMOS the most promising instrument for measuring light element abundances and isotopic ratios. These quantities will allow us to build a consistent view of  transport processes occurring in the stellar interior for a large number of field stars and members of open and globular clusters. This investigation will put together all the pieces of the stellar chemical puzzle and enable a better understanding of hydrodynamical mechanisms along the evolutionary path of stars.

In principle, it would be extremely interesting to measure isotope ratios of light elements  in non-evolved stars as well, so that they could be used as tracers of Galactic chemical evolution. 
The CNO isotopic ratios measured in unevolved stars and/or in molecular gas across the Milky Way disc and in external galaxies can be used to constrain the significance and extent of rotation in stellar models, the impact of nova nucleosynthesis on the chemical evolution of galaxies, and the shape of the prevailing stellar initial mass function in galaxies \citep[][and references therein]{romano2019,romano2022}.
We mention that the SNR required to measure e.g. $^{12}$C/$^{13}$C ratio of unevolved stars is extremely high, and can only be achieved by targeted observations on selected samples (see next section for a detailed estimation of the achievable SNR).

\subsubsection{Requirements on the measurement of carbon isotopic abundance ratios} 
\label{sec:reqCiso}

\begin{figure*}
\centering
\resizebox{0.95\hsize}{!}{
\includegraphics{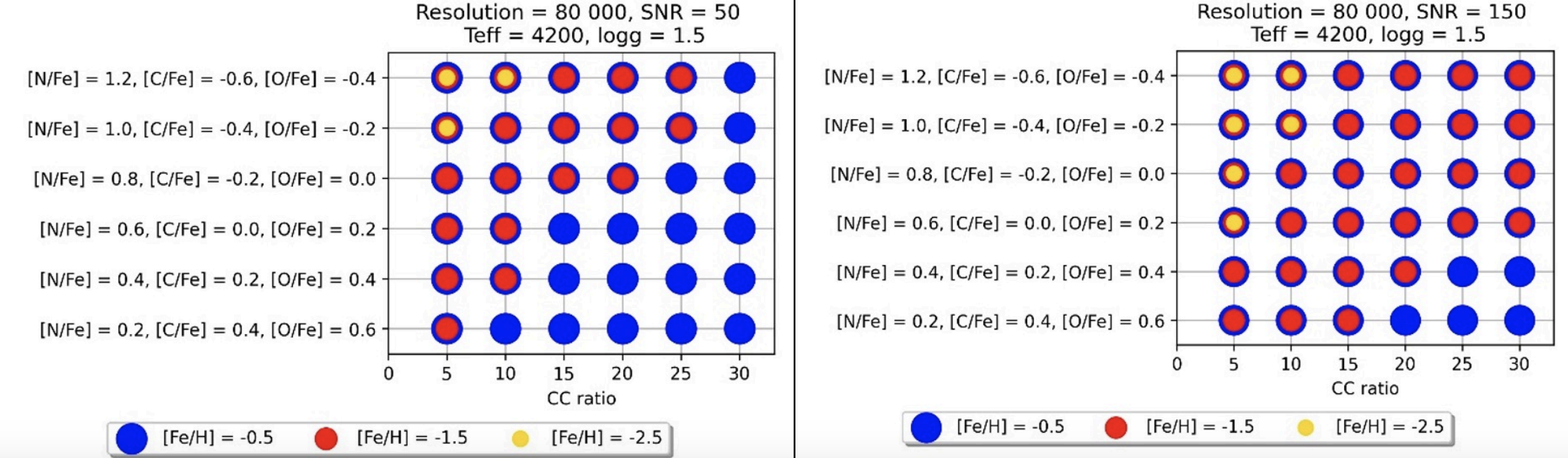}}
\resizebox{0.95\hsize}{!}{
\includegraphics{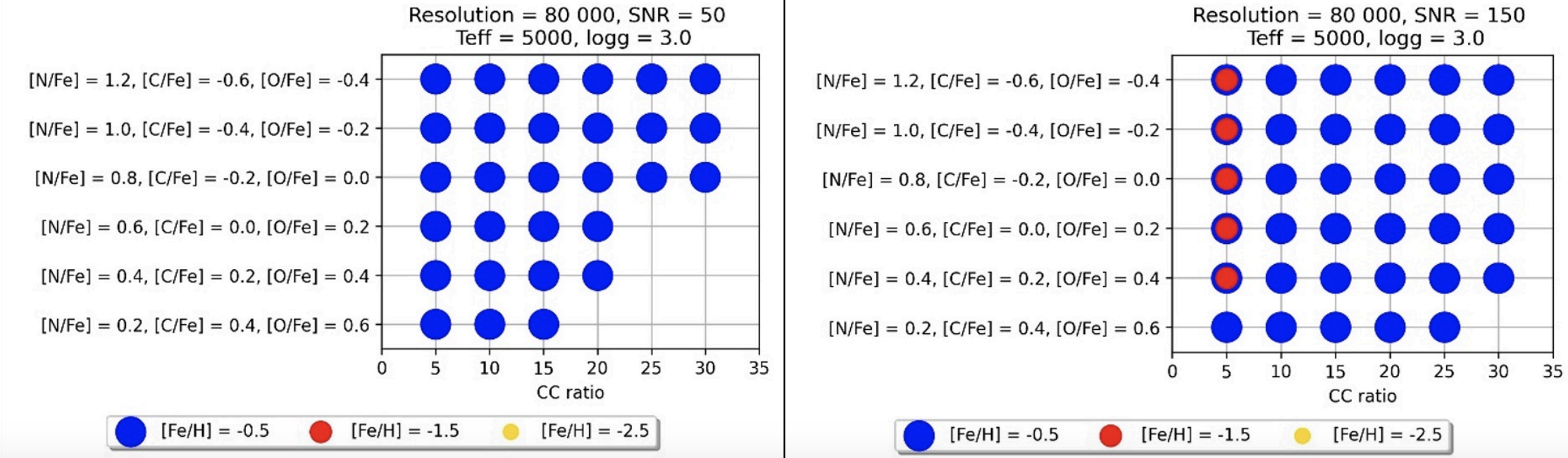}}
\caption{
Results of the analysis of synthetic spectra of a giant star with $T_{\rm eff} = 4200$\,K and $\log g = 1.5$ (shown in upper panel) and of a subgiant star with  $T_{\rm eff} = 5000$\,K and $\log g = 3.0$ (shown in lower panel). The spectra are similated at $R = 80\,000$ for two values of SNR, 50 (left panel) and 150 (right panel). 
\label{fig:synthetic_cratio}
}
\end{figure*}

In this section  we tested the requirements for measuring the  $^{12}$C/$^{13}$C isotopic ratio from the molecular lines of the CN located at $\sim 800$\,nm.  We computed synthetic spectra at $R\sim80\,000$ and for two SNRs (50 and 150). We used the stellar parameters ($T_{\rm eff}-\log g$) typical of subgiant and a red clump stars, assuming six sets of CNO abundances (see Table~\ref{tab:stars_c_iso}) together with carbon isotopic ratios from 5 to 30 changing with a step of 5. 
We also tested the possibility to use slightly lower spectral resolution, confirming that it is still possible to perform isotopic measurements with $R=50,000-60,000$, at least at higher metallicity. 
The results for $R=80,000$ are displayed in Fig.~\ref{fig:synthetic_cratio}. 

We found that it is possible to measure the carbon isotopic ratio in different stellar parameter regimes. The  spectra with higher SNR might allow us to extend the measurement towards lower metallicity.  The intrinsic nature of the molecular bands from which isotopic ratios are measured makes them less visible in the hotter stars. Stars with higher temperature and $\log g$ significantly have a reduced depth of the $^{13}$C features. As seen in Fig.~\ref{fig:synthetic_cratio}, an increase in surface temperature by 800\,K (from 4200 to 5000\,K) and in $\log g$ by 1.5 (from 1.5 to 3.0) significantly reduces our ability to determine $^{12}$C/$^{13}$C ratios to  stars having $\rm [Fe/H] < -0.5$. In main sequence stars, with higher temperature and log g, it will be more difficult to measure the C isotopic ratio. 
The requirements for this case are summarised in Table~\ref{tab:isotopes}.

\begin{center}
\begin{table}[t]%
\centering
\caption{Stellar atmospheric parameters and CNO abundances used for spectra simulations in the investigation of the carbon isotopic ratios science case.}%
\footnotesize
\begin{tabular}{cccccccc}
\noalign{\smallskip}
\hline
\hline
$T_{\rm eff}$ & $\log g$ & [Fe/H] & $V_{\rm t}$      & $V_{\rm rot}$  & [C/Fe] & [N/Fe] & [O/Fe] \\
  $\rm [K]$     & [cgs]    &        & [km\,s$^{-1}$] & [km\,s$^{-1}$] &  \multicolumn{3}{c}{start / stop / step}         \\
\hline
$4200$     &  $1.5$      &  $-2.5/-1.5/-0.5$     & $1.6$    &   $2$  & $-0.6/0.4/0.2$ & $1.2/0.2/0.2$ & $-0.4/0.6/0.2$ \\
$5000$     &  $3.0$      &  $-2.5/-1.5/-0.5$     & $1.6$    &   $2$  & $-0.6/0.4/0.2$ & $1.2/0.2/0.2$ & $-0.4/0.6/0.2$ \\
\hline
\end{tabular}
\label{tab:stars_c_iso}
\end{table}
\end{center}

\frame{\vspace{-7mm}
\paragraph{Uniqueness}
HRMOS will be the only facility that combines high multiplexing with very high spectral resolution, allowing spectral analysis in relatively crowded fields such as open and globular clusters. The MOS capability and its field of view make this instrument particularly suitable for the study of nearby clusters, within a few kpc. Therefore HRMOS is uniquely placed in the current/planned instrumentation landscape for the study of member of star clusters along the whole evolutionary sequence and thus to investigate stellar evolution effects that cause variations of isotopic ratios. 
}

\begin{table*}[t]
\centering
\caption{Instrument requirement summary (isotopic abundance ratios).}%
\label{tab:isotopes}
\footnotesize
\begin{tabularx}{\textwidth}{llX}
\noalign{\smallskip}
\hline
\hline
\textbf{Parameter} & \textbf{Value}  &  \textbf{Justification}  \\
\hline
Resolving power ($R$)     & $>50\,000$   & for isotopic ratios of C      \\
Spectral range & several windows  & Most important features are located: C$_2$ at 513.55\,nm and 563.52\,nm for the carbon abundance determinations; 630.0\,nm for [O\,{\sc i}];  CN bands for N abundances at $647.0-649.0$\,nm, the red $^{12}$CN and $^{13}$CN bands ($797.0-802.0$\,nm which are essential for isotopic ratios in open clusters (especially $^{13}$CN at 800.5\,nm). For the other elements, as Fe, Ca, Si, there are less stringent constraints on the spectral windows since they have more lines. \\
Multiplexing   & $\geq20$ fibres  &     to allow both cluster and field observations    \\
Stability      & n/a      &       \\
Fibre spacing  & 30 arcsec    & sufficient to observe in environments with the typical crowding of OCs and GCs      \\
SNR          &     $> 50 - 150$     &   higher SNR allow to measure C isotopic ratios also in  more  metal poor giants and in subgiants     \\
\hline
\end{tabularx}
\end{table*}

\clearpage

\subsection{Dynamics and kinematics of star clusters}

Recent years have been transformational for studies of the dynamics of GCs \citep{Varri2018}. Indeed, {\sl Gaia} astrometric data \citep{GDR2_helmi2018}, together with proper motions measured with HST (e.g., \citealt{Libralato2022} and other papers from the HSTPROMO team), and dedicated spectroscopic surveys \citep{Ferraro2018,Kamann2018} have unveiled a new kinematic complexity in these systems \citep{Bianchini2018,Sollima2019,Sollima2020}, which were previously mainly described with dynamical models having stringent simplifying assumptions. On the one hand, more sophisticated models are needed to reproduce the observed kinematical richness of GCs. On the other hand, more accurate and precise data could help disentangle some of the degeneracies and clarify the role of important dynamical ingredients in the formation and evolution of these stellar systems.

In this framework, the importance of kinematic measurements is multifaceted. First, measurements of line-of-sight velocities (and of proper motions) of stars are a powerful way to separate cluster members from field stars. They also allow one to assess the amount of rotation within the system and to quantify the variation of the velocity dispersion as a function of the distance from the centre. The presence of binaries can affect these measurements and, therefore, it is particularly important to characterise the binary population of GCs \citep{Rastello2020,Aros2021}. Kinematic measurements are therefore crucial to determine the dynamics of GCs, to measure their mass, and to detect the possible presence of a population of stellar-mass black holes, which do not contribute to the light but affect the motions of visible stars.

\begin{figure*}[t]
\centering
\resizebox{0.95\hsize}{!}{
\includegraphics{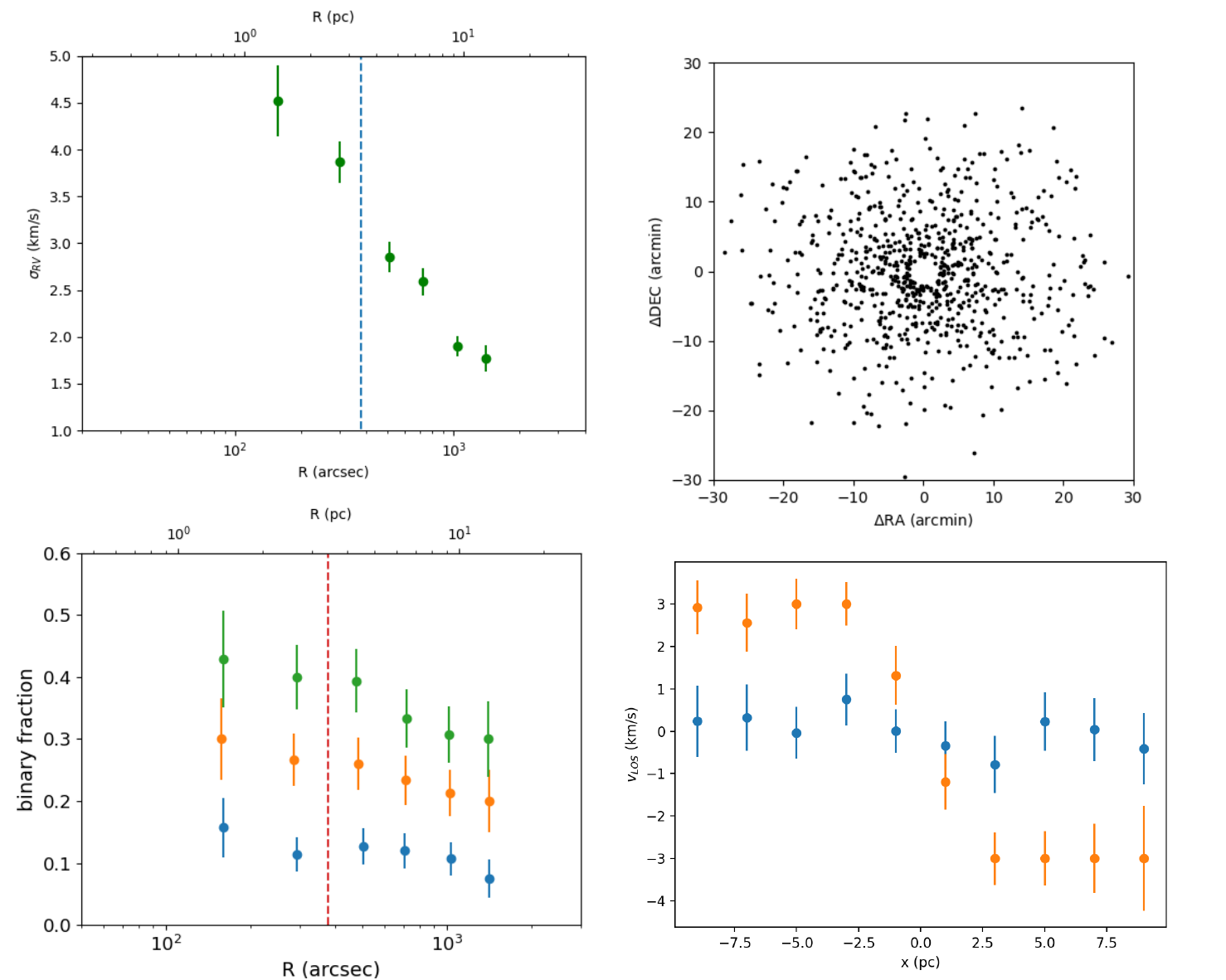}}
\caption{A simulated observation of the GC M4 with HRMOS.  {\sl Top-right:} the distribution of the observed stars. {\sl Top-left:} the variation of the velocity dispersion $\sigma_{RV}$ with the distance from the GC centre. {\sl Bottom-left:} the observed binary fraction as a function of the distance of the GC, assuming a binary fraction $f_{\rm bin} = f_{\rm bin}(0)(R/R_{HM})^{0.25}$, with $f_{\rm bin}(0)$= 0.1 (blue dots), 0.2 (orange dots), 0.3 (green dots). {\sl Bottom-right:} the mean velocity along the line of sight at different distances from the GC centre for a slow ($v_{\rm rot}/\sigma_{RV}=0.04$, blue dots) and a fast rotating GC ($v_{\rm rot}/\sigma_{\rm RV}=0.2$, orange dots), where $v_{\rm rot}$ is the rotational velocity of the GC. 
 \label{fig:kinematicGC}
}
\end{figure*}
\subsubsection{Requirements on kinematic of star cluster science case}

To define the requirements of the spectrograph needed to address this science case, which are reported in Tables~\ref{tab:oc} and \ref{tab:gc}, in particular stability,  we simulated the observations of a globular cluster. We used a snapshot of the simulation carried out by \cite{Heggie14} and assumed the distance and the extinction of M4.
In Fig.~\ref{fig:kinematicGC}, we show the results of a simulated observation of 750 stars in the magnitude range $15<V<19$ in two different epochs, assuming a precision in RV of 50\,m\,s$^{-1}$. In particular, we investigate the capabilities of this instrument to a) derive the velocity dispersion as function of the distance from the cluster centre; b) measure the rotational velocity of the cluster; c) estimate the binary fraction and the effects on the measurements of the velocity dispersion.

%% file: gal_archaeology.tex
\section{New Paradigms for Galactic Archaeology}

HRMOS will be operational when major medium-resolution spectroscopic surveys, such as the ones carried out with the WEAVE, 4MOST, and MOONS spectrographs, are well underway or even completed. The mission of HRMOS will then be to expand the parameter space for studies of Galactic Archaeology. 

The path that an instrument such as HRMOS can open up is far-reaching. It promises formidable results in our understanding of the origin of the elements, allowing in some cases the measurement of isotopic abundance ratios and thus the separation of nucleosynthesis channels. Additionally, it will provide insight into the formation and evolution of the Galaxy with, for instance, the measurement of `absolute' ages from nucleocosmochronology, which is possible only with high-resolution and high-SNR spectra reaching the blue part of the visible range.  

Another aspect that should not be overlooked is the potential to increase both the quality and amount of medium-resolution survey data products, as HRMOS can provide a very high-resolution training set for a data-driven analysis of the spectra of these other surveys. On the one hand, HRMOS will be able to provide data for volume-limited regions of our Galaxy, with extremely precise abundances of many elements, including many of the heavy elements that cannot be measured at medium resolution. These data will be of great value in themselves. On the other hand, these samples will be the basis for training sets, acting similarly to the use of the results of the APOGEE and GALAH survey to improve the quality and quantity of abundances derived from the LAMOST data \citep[see, e.g.][]{ho17}.

\frame{
\titlecol{Key Questions:}
\begin{itemize}
\item What is the age of the oldest stars? What are their implication on cosmology and on the age of the Universe? 
\item  What is the accretion history of the Galactic halo? What are the properties of the first stars? What are the origins of the heaviest elements?
\item How can we enhance the value of the products of the medium resolution surveys?  
\end{itemize}
}

\subsection{Cosmological ages: new constraints from nucleocosmochronology}
\label{nucleo}

Stellar ages lie at the heart of several research topics in astrophysics, from cosmology to stellar evolution. Ages are typically inferred through empirical (e.g., gyrochronology, stellar activity, lithium depletion, chemical clocks) or model-dependent methods (e.g., isochrone fitting, asteroseismology), which rely on several assumptions and are affected by observational biases. The only {\it semi}-fundamental method to measure the age of stars whose physics is almost completely understood is nucleocosmochronology. Nucleocosmochronology uses the decay of long-lived actinides, uranium ($^{238}$U; $\tau_{1/2}$ = 4.47 Gyr) and thorium ($^{232}$Th; $\tau_{1/2}$ = 14.05 Gyr) to estimate stellar ages \citep{cowan1991, Shah23}. The only assumption that must be made is about the starting abundance of these elements, which is generally based on scaling other $r$-process abundances (for example, Eu).

The U\,{\sc ii} lines are very weak, and thus its abundance can be well determined only in metal-poor r-process enhanced stars. However, uranium is important in the assembly of one of the most robust chronometers known. The absolute ages of the stars are determined by using the observed present-day abundance ratios and theoretical production ratios of U and/or Th with respect to co-produced r-process elements, for example, U/Th, U/X, and Th/X, where X refers to a lighter stable r-process element, such as Eu, Os, or Ir. However, because of the different production mechanisms between heavy and lighter neutron-capture elements, sometimes the U/Eu or Th/Eu ratio fails. The most reliable ratio is the Th/U ratio, for which \citet{Shah23} have presented a new set of absorption lines, all possibly observable with HRMOS. For Th, the lines are located at $\lambda$$\lambda$ 401.913~nm, 408.652~nm, and 409.475~nm, while for U they are located at $\lambda$$\lambda$ 385.957~nm, 405.004~nm and 409.013~nm. By appropriately selecting targets among r-process enhanced stars, HRMOS could allow the simultaneous measurement of their Th and U abundances. Clearly, these are rare targets and their observations would be combined with a more systematic survey of halo stars. 

Another important goal would be to apply nucleocosmochronology to members of GCs \citep[see, e.g.][]{pagel93, valcin20}. First, GCs are among the oldest objects in the Milky Way, and thus their ages would give valuable lower limits to the age of the Universe. Second, nucleocosmochronology in GCs offers the opportunity to calibrate other methods of age determination, such as isochrone fitting and asteroseismology. Third, given that the analysis can be carried out for multiple members of a given GC, the resulting age is expected to be much more precise than those determined for field stars.

Between U and Th, Th is easier to measure in a star that is not extremely enhanced in r-process elements, as was done for giant stars in the GC M15 \citep[][]{sneden00}. It is worth mentioning that this technique has been applied not only to older stellar populations, but also to populations in the Galactic disc \citep{delpeloso05a, delpeloso05b, delpeloso05c}. 

We finally mention that the comparison of ages from isochrones with those from nucleocosmochronology would have important implications also for cosmology and for measuring the Hubble constant \citep[see, e.g.][]{cimatti23}.

\begin{figure}
\centering
\resizebox{0.7\hsize}{!}{
\includegraphics{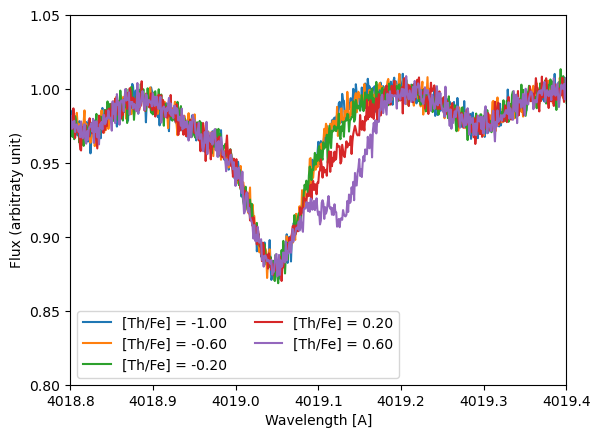}}
\caption{Synthetic spectra of a giant star ($T_{\rm eff}=4500$\,K, $\log g =2.5$) at a metallicity $\rm [Fe/H] = -2.5$ and with various [Th/Fe] ratios. All spectra have $\rm  SNR =200$ and $R=80\,000$.
 \label{fig:Th_80000}
}
\end{figure}
\begin{figure}
\centering
\resizebox{1.0\hsize}{!}{
\includegraphics{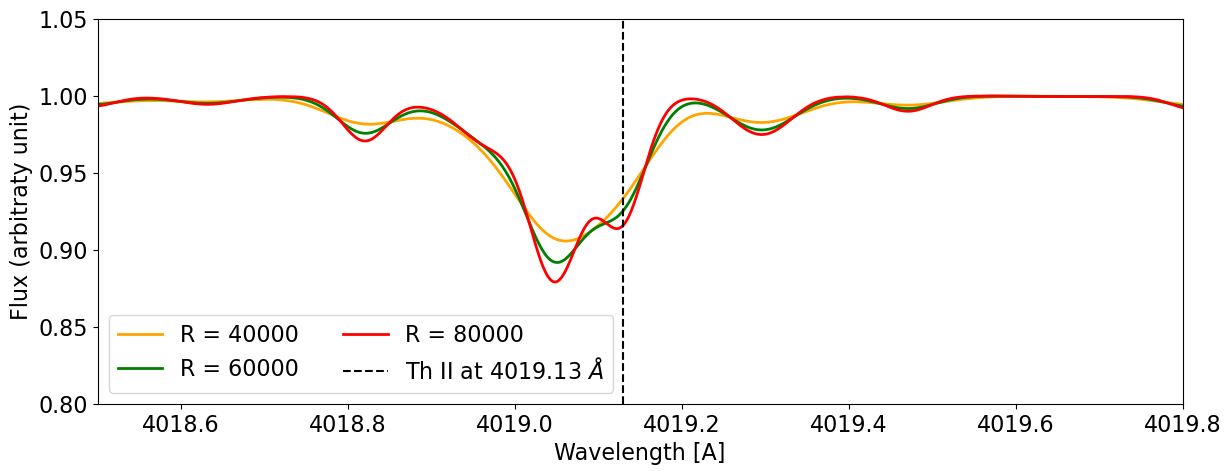}}
\caption{Synthetic spectra of a giant star ($T_{\rm eff}=4500$\,K, $\log g =2.5$) at a metallicity $\rm [Fe/H] = -2.5$ and with  [Th/Fe]=0.6 ratios at three resolutions. The dashed line indicates the wavelength of the Th~{\sc ii} line. 
 \label{fig:Th_all}
}
\end{figure}

\subsubsection{Requirements on nucleocosmochronology}

To define the requirements of the spectrograph needed to address this science case,  we generated synthetic spectra around the Th\,{\sc ii} line at 401.913~nm  for a giant star ($T_{\rm eff} = 4500$\,K, $\log g = 2.5$) at a metallicity $\rm [Fe/H] = -2.5$ and with various [Th/Fe] ratios (see Fig.~\ref{fig:Th_80000}). These spectra have been generated with $\rm SNR = 200$ and $R=80\,000$. Under these conditions, the measurement of Th abundances would be feasible. For a star of $mag (AB)=14$ \citep[the brightest giants in M15 have magnitudes between 12 and 14,][]{Fahlman85}, a $\rm SNR=200$ in the blue (around 390\,nm) can be achieved with exposure times of 4-5 hours. This is easily feasible in less than one night of observations. On the other hand, Fig.~\ref{fig:Th_all} shows the same spectrum as in Fig.~\ref{fig:Th_80000} with [Th/Fe]=0.6 and three different resolutions. Only with $R=80\,000$ can we clearly separate the Th~{\sc II} line from the nearby Co~{\sc I} line. 
The main requirements for this scientific case are described in Table~\ref{tab:gc}.

\subsection{Exploring the Galactic halo at high-resolution}

\begin{figure}[t]
\centering
\resizebox{0.999\hsize}{!}{
\includegraphics{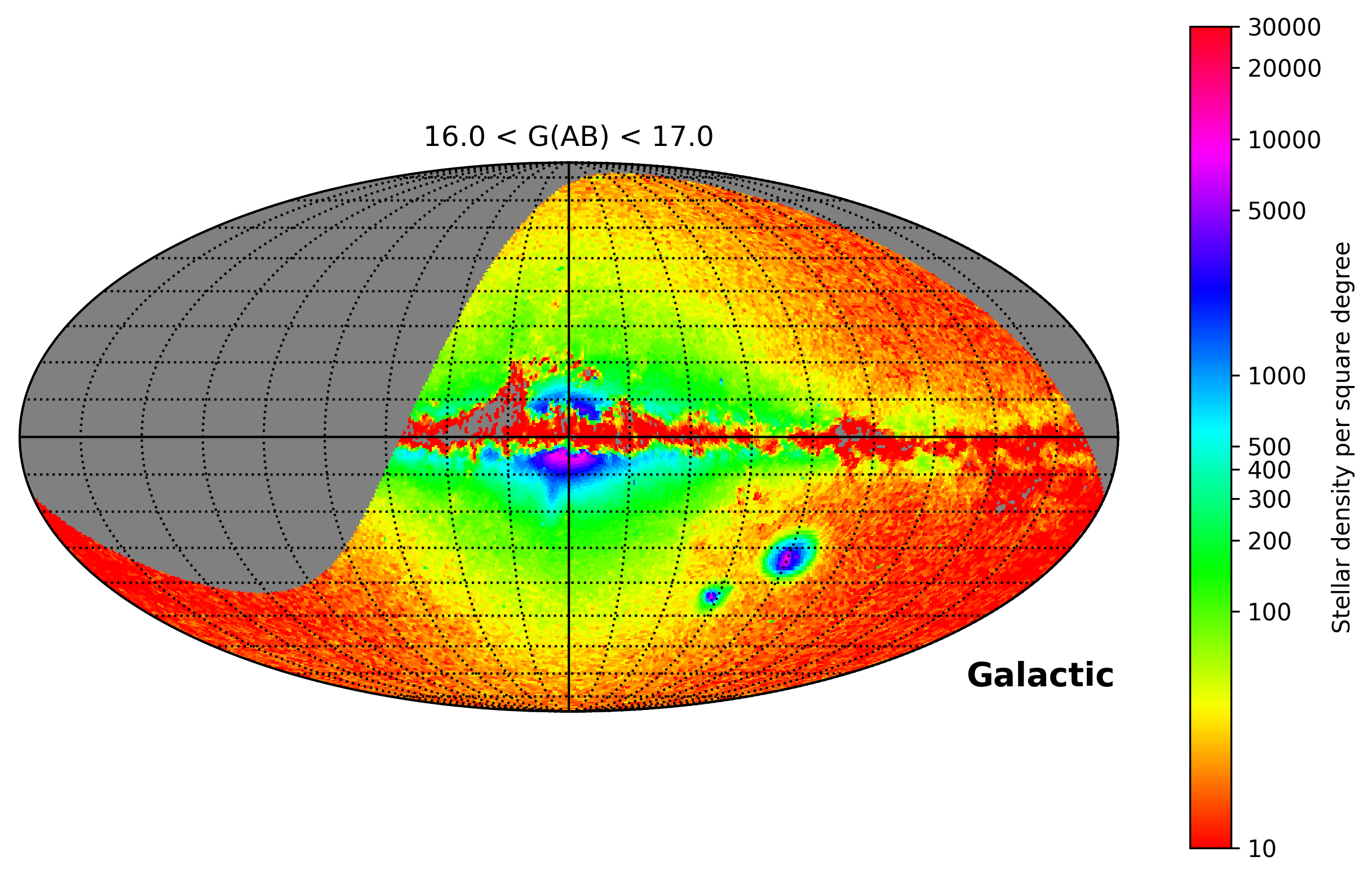}}
\caption{
The density distribution of halo candidate stars from Gaia DR3 in the magnitude bin $G$ = 16-17. 
\label{fig:mollweide}
}
\end{figure}

The study of the halo is of paramount importance for understanding the formation mechanisms of the Milky Way and for studying nucleosynthesis in stars of very low metallicity. Figure~\ref{fig:mollweide} shows a selection of halo stars based on Gaia DR3, including only stars with transverse velocity $>$ 180 km s$^{-1}$. The density of stars in the magnitude bin $G$ = 16-17 is shown, which corresponds to targets that can be reached with one hour of observation and SNR between 25 and 50. Except for the outermost regions and the regions high on the Galactic plane, in this magnitude range we expect to have between 30 and 100 halo targets per square degree. This results in a density of about 4-15 targets per HRMOS field, to which a few brighter targets can still be added. Although we are not dealing with large numbers, being able to obtain high-resolution spectra of 10-20 targets at a time would revamp our view of the halo, allowing us to address various issues including the characterisation of the first stars and the determination of abundances and isotopic ratios of heavy elements in metal-poor stars.

\subsubsection{Disentangling the properties of the first stars}

Characterising the abundance patterns of the most metal-poor stars of the Milky Way provides a wealth of information about the early Universe. The extremely metal-poor (EMP) stars in our Galaxy (i.e., those with $\rm [Fe/H] \lesssim -3.0$) are probably the most ancient fossil records of the chemical composition of the interstellar medium in the early stages of the Milky Way formation and thus indirectly also of pre-Galactic states \citep[e.g.,][]{Agarst2000, Caffau2020, Bonifacio2021, Smiljanic2021}. They allow us to study the nature of the first stars and supernovae, the relevant nucleosynthesis processes responsible for the formation and evolution of the elements, star and galaxy formation processes, in particular the assembly process of the stellar halo.

The first stars are those formed from primordial material produced by Big Bang nucleosynthesis \citep[see, e.g.,][]{Bromm2013}. The first stars were likely much more massive and have long since died \citep{hirano2014}, although there have been suggestions of elusive low-mass first stars. The chemical imprints of the first stars should be detectable in the second generation of stars, including long-lived low-mass stars. Although these are extremely rare, one telltale sign is the extremely low metal content. However, so far only eight stars with $\rm [Fe/H] < -5$, the so-called hyper metal-poor stars\footnote{Following the nomenclature introduced by \citet{beers2005araa}.} with less than 1/100,000$^{th}$ the solar iron-to-hydrogen ratio, are known \citep{frebel2015}; these are some of the most chemically ancient objects known \citep[see, e.g.][and references therein]{Vanni2023}. These stars exhibit a large scatter in their relative chemical abundance ratios, which demands a wide variety in the properties of the first supernovae such as, for example, mass, explosion energy, rotation, mass cut, and explosion mechanism \citep{nomoto2013}. More data at the lowest metallicities are needed to better constrain the diversity of their properties, such as enhancements in carbon, which will result in a better understanding of the parent first stars.

In extremely metal-poor stars, we can also find signatures of different types of r-process(es) that polluted the interstellar medium at different timescales \citep{eichler2019}, provided that we can measure neutron capture elements such as Hf and Eu. Examples of possible sources of r-process material include neutron star mergers and magnetorotational SNe \citep{cescutti2015,cote2018,simonetti2019}. This is a case intrinsically linked to the science case on the origin of the elements.

There are several past, present, and future surveys that aim to improve the number statistics on ultra-metal-poor stars. Candidates are identified through photometric and/or lower-resolution spectroscopic observations, most of which give only rough indications of metallicity rather than precise measurements. Obtaining detailed chemical characteristics requires measuring the full range of elements, from light elements such as lithium to heavy elements such as europium, for these metal-poor stars. Follow-up observations to determine the detailed chemical abundance patterns of these candidates is essential to gain full insights into the range of nucleosynthesis processes that were present in the early Universe, as well as to probe the physical properties and mass distribution of the very first stars.

\subsubsection{Origin of the heavy elements: constraints from elemental and isotopic abundances in metal poor stars}

Nuclei heavier than the Fe peak with $Z>30$ are mainly synthesised by successive captures of neutrons. The main neutron-capture processes are the slow (s-) and rapid (r-) processes, depending on whether the timescale for neutron capture is faster or slower than that for radioactive beta decay \citep{burbidge1957}. There is possibly also an intermediate (i-) process operating at different metallicities \citep[see e.g.][and references therein]{Karinkuzhi2023}. These modes of nucleosynthesis are complex, their contribution may vary in time, and there are several open questions regarding their details\citep[see related discussions in e.g.][and references therein]{Diehl2022, ArconesThielemann2023}. The production sites of these heaviest elements remain unclear and have been driving large theoretical efforts \citep[][and references therein]{molero23}.

The first peculiar chemical signature observed in halo stars was the large spread in the abundances of neutron-capture elements \citep{mcwilliam1998}. The first studies made with 8-meter-class telescopes with high-resolution spectra \citep[providing detailed chemistry for about 30 elements,][]{francois2007,honda2004} and subsequent investigations of stars with extremely low metallicities \citep[e.g.][]{roederer2014,yong2013fepoor} confirmed that such a spread in heavy-element abundances is real. At the same time, lighter elements, such as the $\alpha$- and iron-peak ones, do not show such large scatter.

\begin{figure}[t]
\centering
\resizebox{0.65\hsize}{!}{
\includegraphics{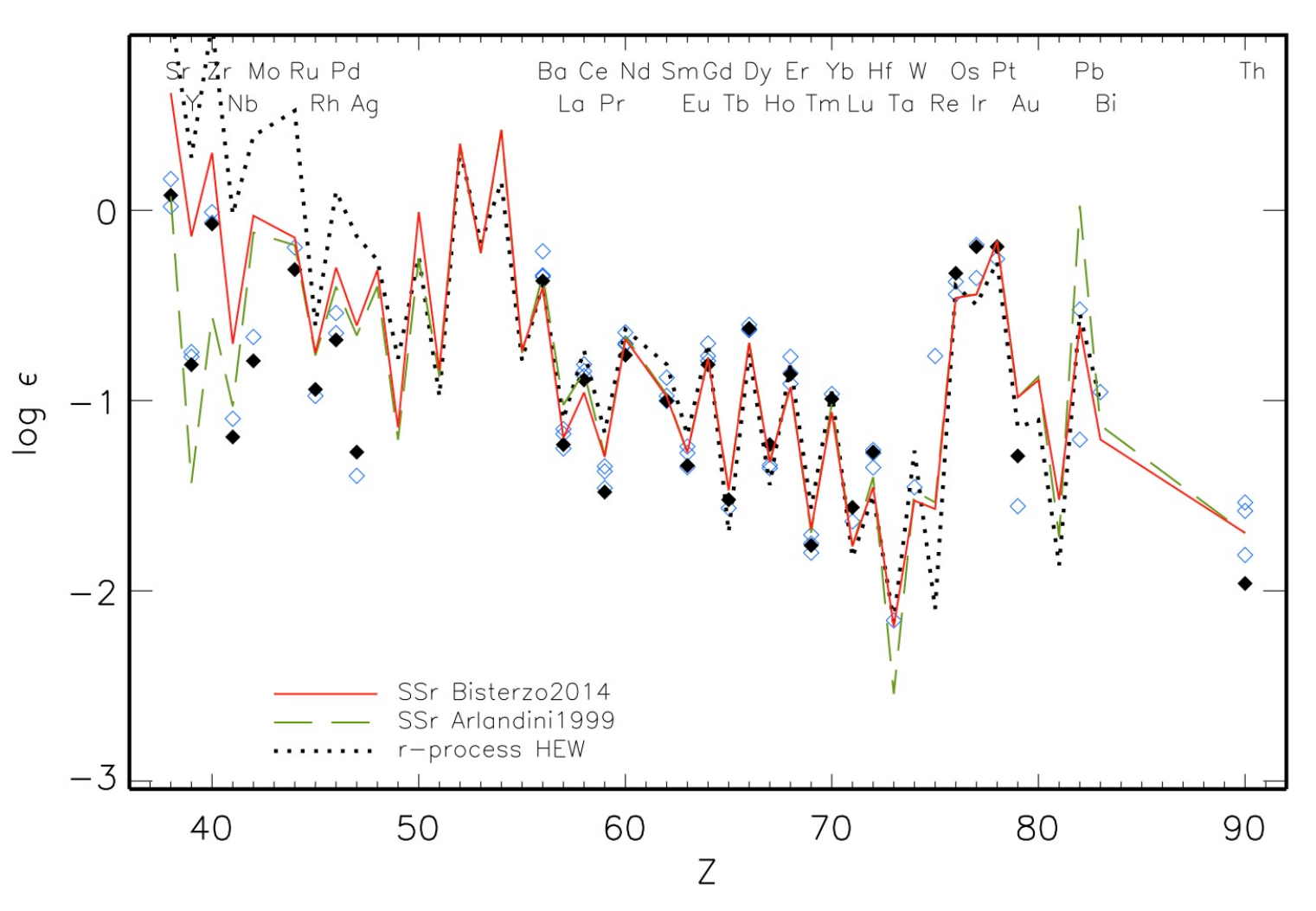}}
\caption{
The heavy element abundance variations in metal-poor stars \citep{mashonkina2014}.
\label{fig:mashonkina}
}
\end{figure}

The spread is likely produced by stochastic processes driven by the rarity of r-process events \citep{argast2004,cescutti2008, cavallo23}. The way this dispersion decreases toward higher metallicity is an important constraint to the rate of r-process events. As the Galaxy becomes more metal-rich over time, s-process signatures in stars begin to increase relative to the r-process \citep{francois2007,mashonkina2003}, as seen in the halo and thick disc. There are variations in star-to-star compositions that suggest different r- and s-process regimes for a given metallicity. Moreover, the exact moment when the s-process starts to dominate over the r-process is still debated (see, e.g., Fig.\ref{fig:mashonkina}).

A way of detecting r- and s-process signatures in a star is to measure the isotopic fractions in heavy elements. The pure s- and r-process isotope ratios are different in most heavy elements. These differences are detectable through small changes in absorption line asymmetry. For example, the r- and s-process are responsible for five of the seven stable isotopes of Ba (the lightest two, $^{130,132}$Ba, arise in the so-called p-process). Whereas the s-process can synthesise all five Ba n-capture isotopes, the r-process cannot synthesise the even isotopes, $^{134,136}$Ba. Therefore, the ratio of odd/even isotopic Ba is an indication of the s-/r-process fraction.

At low metallicities ($\rm [Fe/H] \leq -2.5$) there is a large scatter in abundances of neutron-capture elements that is not seen, for example, for $\alpha$-elements. The light-to-heavy abundance ratio (e.g. [Sr/Ba]) can not be reproduced assuming only r-process enrichment at low metallicity. It could instead be a contribution of s-process from fast rotating massive stars or a weak r-process component. Measurement of the isotopic ratios of Ba could help \citep[see][]{cescutti2021} as the ratio is different depending on whether the s- or r-process dominates. 

Further, there is much debate and large gaps in our understanding of the astrophysical sites and the production mechanism of neutron-capture elements (and their isotopes). The high neutron flux densities required to synthesise the r-process elements suggest that the formation sites are related to the birth or death of neutron stars. Proposed sites include rare types of core-collapse supernovae \citep[e.g.][]{winteler2012,siegel2018} and neutron star binary or neutron star-black hole mergers \citep[e.g.][]{lattimer1977,metzger2010}. Additionally, sites with lower neutron fluxes, such as normal core-collapse supernovae, may contribute to elements in the light r-process range, up to Ba \citep[e.g.][]{arcones2011,wanajo2013}.

\begin{figure}[t]
\centering
\resizebox{0.6\hsize}{!}{
\includegraphics{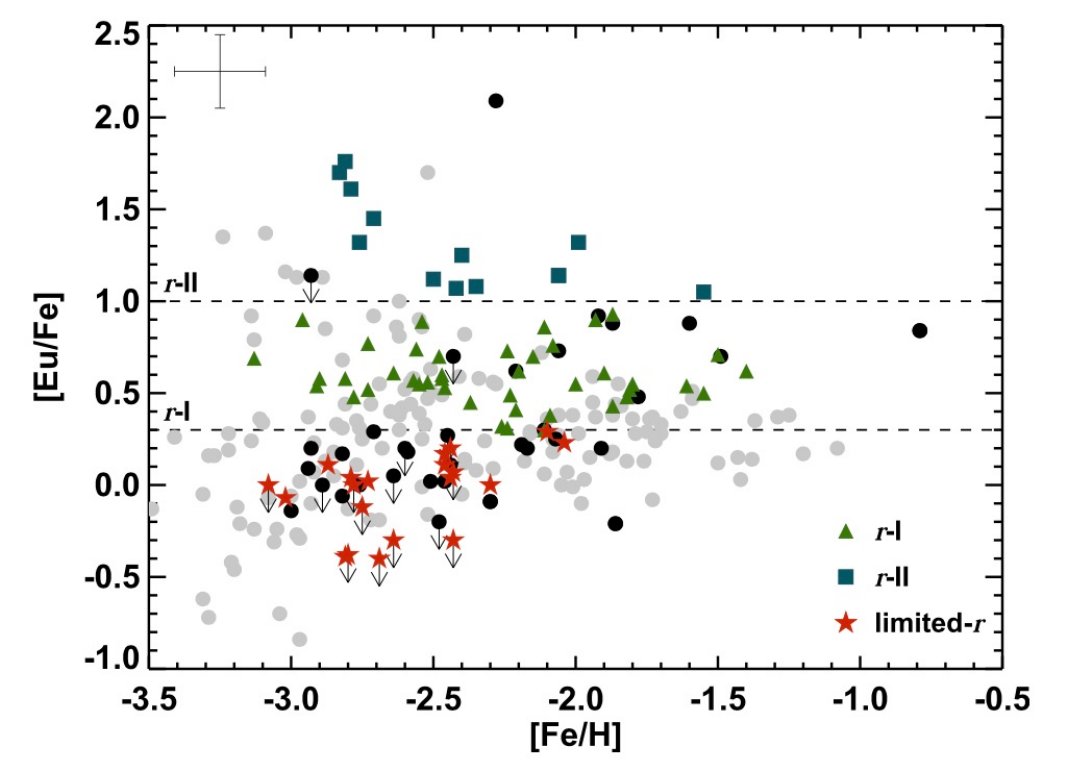}}
\caption{
The r-process element Eu as a function of metallicity \citep{hansen2018}.
\label{fig:hansen}
}
\end{figure}

The number of stars with homogeneous and accurate abundance measurements of the r-process is much too low. The currently available small but detailed samples have provided several key insights, but have also added to the mysteries. The r-process occurs relatively early, enriching metal-poor stars with $\rm [Fe/H] \approx -3 $ \citep{sneden1994,hill2002,frebel2007}. The r-process nucleosynthesis appears to be ubiquitous in many environments \citep[e.g.][]{roederer2010,roederer2013,roederer2017}, and the r-process element production must be rare compared to core-collapse supernovae \citep[e.g.][]{ji2016,macias2018}. Different types of r-process element enhancement have been detected in stars (see Fig.~\ref{fig:hansen}). One group (r-II stars, see \citealt{beers2005araa}) exhibits enhancement in the heavy r-process elements (like Eu) while another (limited-r stars) exhibits larger abundances for light neutron-capture elements like Sr, compared to the heavier species \citep{barklem2005,hansen2018}.

Full characterisation and modelling of the heavy neutron capture elements still awaits homogeneous analysis of larger samples, including the high-quality data needed to measure the isotopic ratios. Understanding the origin of these elements provides key constraints on topics such as stellar age dating, chemical evolution, and galaxy formation across different environments. The presence of r-process-rich halo stars at extremely low metallicities shows that they are clearly important clues to the earliest enrichment processes in the universe.

\subsubsection{Requirements - deriving the properties of metal-poor stars in the halo} 

For metal-poor stars, coverage of relatively blue wavelengths is essential. Because of the low metal content, these spectra are relatively featureless. Therefore, observations must be carried out to detect the strongest Fe lines. These are located at 344.06\,nm, 358.12\,nm, 371.99\,nm, 373.71\,nm, 382.04\,nm and 385.99\,nm (the first two may be beyond HRMOS reach). Further, other elements including Li, C, N, O, alpha and Fe-peak elements, as well as neutron-capture elements, are key to characterising metal-poor stars. In potential r-process enhanced candidates, the U line at 385.9\,nm would be a key detection.

Therefore, very high spectral resolution and high signal-to-noise ratio are critical. Figure~\ref{fig:marino} shows the error in Fe for a simulated low-metallicity star as a function of resolution and signal-to-noise ratio. A $SNR>100$ at $R = 80\,000$ or $SNR>200$ at $R = 60\,000$ are the minimum requirements, consistent with the science case for the abundances of heavy elements.

Heavy elements usually have very weak lines, which require spectra with high $R$ and high SNR. Furthermore, if one wants to extract isotopic abundances to differentiate between the s- and r-process components, one must look at the small asymmetry of the lines, which is usually challenging to measure even for the strongest features. This necessarily requires a high resolution. 
To address the case on the origin of the heavy elements, we require observations of stars for which there are neutron-capture element measurements for several heavy element species, covering the first three peaks of the s-process elements, r-process elements, and elements that may be from the i-process. Specifically, these elements are mandatory, in order of atomic number: Sr, Y, Zr, Pd, Ba, La, Hf, Pb, Nd, and Eu. The line profiles of these elements may be weak and/or blended.
Therefore, high-resolution and high signal-to-noise spectroscopy becomes critical. The strongest lines of these elements are located at bluer wavelengths. Therefore, the wavelength coverage around  400\,nm  is a critical requirement.

In order to constrain theoretical models, uncertainties in the elemental abundance smaller than 0.1\,dex are requested. Table~\ref{tab:orazi} shows the change in equivalent widths and the error in Pb abundances for different simulated spectra.

\begin{table}[t]%
\centering
\caption{Pb errors at different values of the resolving power and SNR (per resolution element) based on analysis of simulated spectra by Valentina D’Orazi. \label{tab:orazi}}%
\tabcolsep=0pt%
\footnotesize
\begin{tabular*}{20pc}{@{\extracolsep\fill}cccc@{\extracolsep\fill}}
\noalign{\smallskip}
\hline
\hline
$R$ & SNR  & $\sigma$(EW)  & $\Delta$[Pb/H] \\
             & [res.el$^{-1}$]  &      [pm]          &             \\
\hline
40\,000 & 100  &  0.80  & 0.20    \\
40\,000 & 300  &  0.44  & 0.10   \\
60\,000 & 100  &  0.56  & 0.13    \\
60\,000 & 300  &  0.32  & 0.07    \\
\hline
\end{tabular*}
\end{table}

Furthermore, HRMOS should provide MOS capability to enable a significant sample to be observed in a timely manner. The higher the multiplex, the more efficiently large samples can be observed. Figure~\ref{fig:mollweide} presents the density per square degree of candidate halo stars from Gaia DR3: limiting to magnitude $G=17$, we can exploit an average of $4-15$ fibres per field with a diameter of 25\,arcmin (see also Fig.~\ref{fig:mollweide}).

\long\def\comment#1{}

\comment{Detecting the r- and s-process signatures directly in a star requires measuring the isotopic fractions. This requires measuring the asymmetry of some of the strongest heavy-element features. Figure~\ref{fig:wenyuan} shows the Ba line at 455.4\,nm with isotopic ratios \citep{wenyuan2018}.

Further, at the bluer wavelengths, the stellar spectrum can be crowded with many other line features. Therefore, a high resolution is essential to simply resolve the lines of interest, and more so for isotopic abundance ratios. An ideal resolution is closer to $R\approx 100\,000$. To make precision abundance measurements to push the limit of our knowledge on the origin of heavy elements, a spectral resolution of over $R = 60\,000$ is needed. Additionally, because the lines are weak and often also affected by issues such as sky contamination, a high signal-to-noise ratio is important. A minimum $SNR = 100$ per pixel is needed for the strongest features, whereas for the weakest features, a $SNR > 300$ is needed. We note that the data in Fig.~\ref{fig:wenyuan} was of $R \approx 70\,000$ at $\rm SNR \approx 100$ to obtain a precision better than 0.1\,dex.
\begin{figure*}[t]
\centering
\resizebox{0.8\hsize}{!}{
\includegraphics{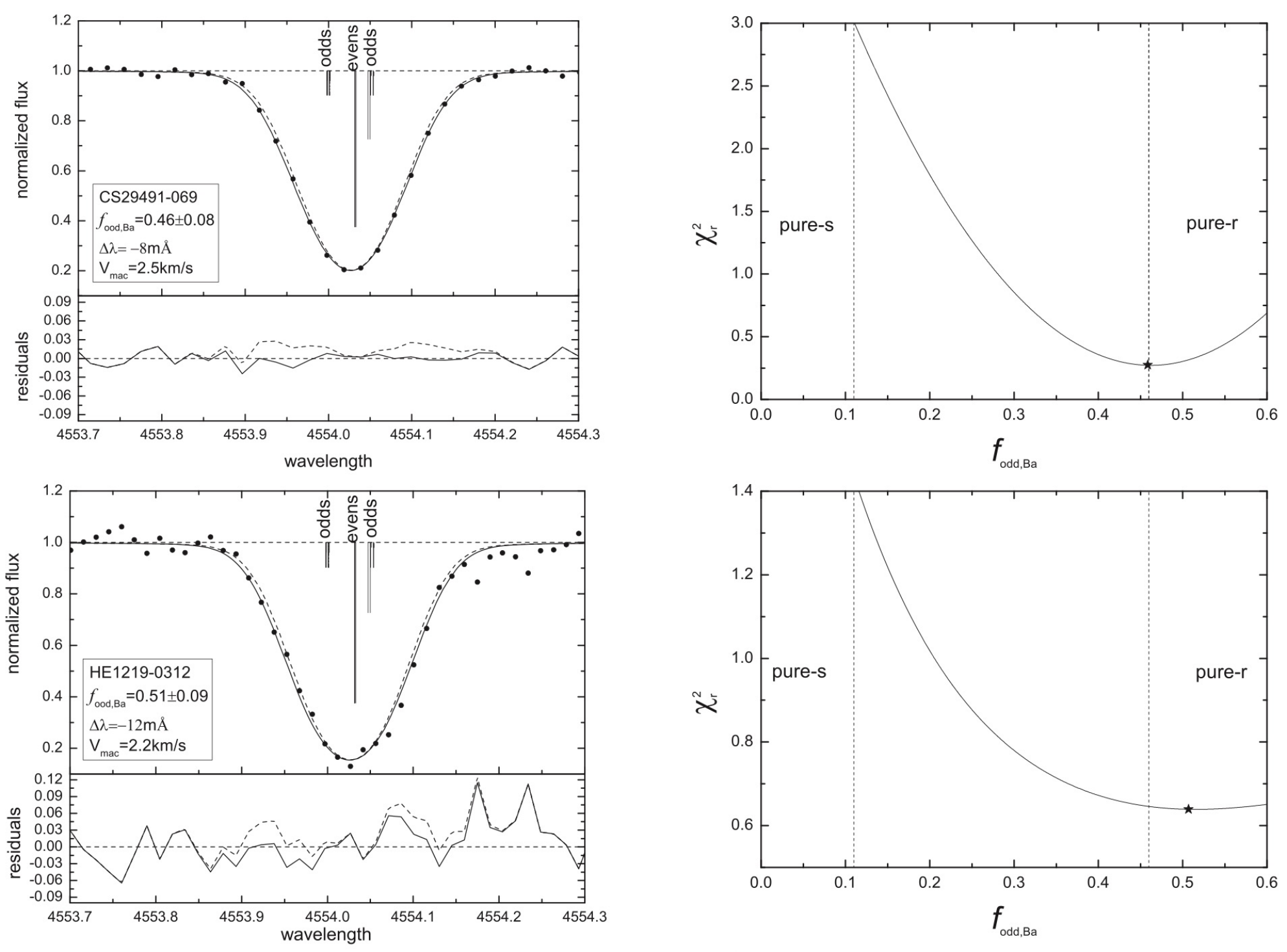}}
\caption{
Synthesis of the Ba line at 455.4\,nm accounting for the isotopic ratios  \citep{wenyuan2018}
\label{fig:wenyuan}
}
\end{figure*}
}

\begin{figure*}[th!]
\centering
\resizebox{0.8\hsize}{!}{
\includegraphics[trim={1cm 2cm 1cm 1cm},clip, scale=1]{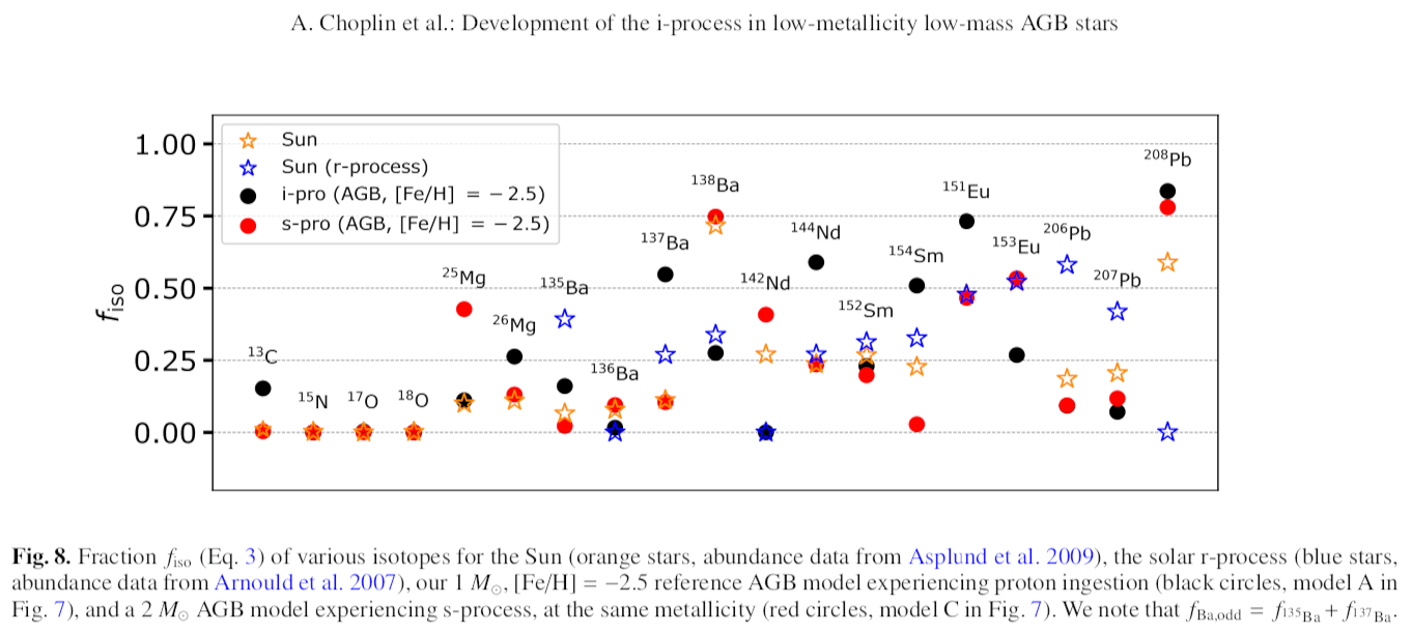}}
\caption{Isotopic fraction of various isotopes for the Sun (orange stars, abundances from \citealt{asplund2009}), the solar $r$-process (blue stars, abundances from \citealt{arnould2007}) and AGB models experiencing $s$-process (red circles) and $i$-process (black circles) from \cite{choplin2021}.
\label{fig:isotopes}
}
\end{figure*}

\begin{figure*}[th!]
\centering
\resizebox{0.6\hsize}{!}{
\includegraphics[trim={0cm 0cm 0cm 0.39cm},clip, scale=0.3]{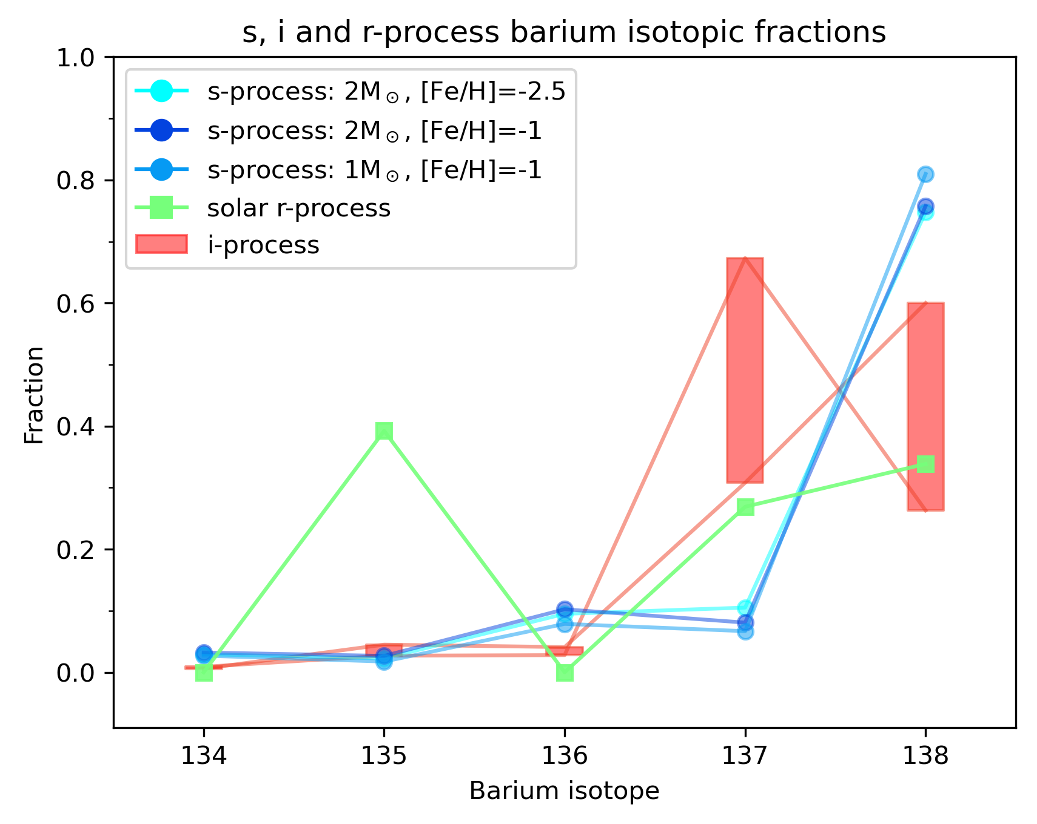}}
\caption{Predicted Ba isotopic fractions. Large (but correlated) uncertainties (represented by the red rectangles) are present for the $i$-process predictions of $^{137}$Ba and $^{138}$Ba. Adapted from \cite{martinet2023}.
\label{Fig:Ba-isotopes}
}
\end{figure*}

Finally, a unique contribution of high spectral resolution is the ability to contribute to the measurement of isotope ratios, which are key in giving strong observational constraints on nucleosynthesis processes. Different isotopic mixtures are predicted for the $s$-, $r$- and $i$-processes. Schematically, as can be seen in Fig.~\ref{fig:isotopes}, even isotopes are produced mainly by the $s$-process, while the contributions of the $r$-process to the odd isotopes are significant. For Ba, on average, the isotopic fractions are as illustrated in Fig.~\ref{Fig:Ba-isotopes}.

These predicted differences in the isotopic mixtures produced by the different nucleosynthetic processes are challenging to test. Isotopic abundances are notoriously difficult to measure from atomic lines, because the isotopic shifts of atomic lines are very small ($\sim 0.2$\,pm) compared to the line width. However, in contrast to the spectral lines of even isotopes, the lines of odd isotopes are affected by hyperfine splitting (HFS), which significantly broadens the lines ($\sim 5$\,pm).

For example, in the case of Ba, it is possible to measure the even-to-odd isotope abundance ratio
$f_{\rm Ba,odd}
= (N({\rm ^{135}Ba}) + N({\rm ^{137}Ba}))/N(\rm Ba)$ 
from the two strong resonance lines at 455,4 and 493,4\,nm, and to relate it to a given nucleosynthetic process (either $s$-process, or $i/r$-processes). This has been applied to the Ba isotopes
 in \cite{Cowley-Frey-89, Magain-Zhao-93, Lambert-AllendePrieto-02, Mashonkina-Zhao-06, Mashonkina-08, Collet-09, Gallagher-10, Gallagher-15, Jablonka-15, Meng-16, Wenyuan-18, Mashonkina-19} with contradictory conclusions. 
\cite{Short-06} have confirmed that the Ba\,{\sc ii} resonance line at 455.4\,nm (used to measure $f_{\rm Ba, odd}$)  is formed under non-LTE conditions in metal-poor stars. Actually, only a handful of measurements are available from high-resolution, high-SNR spectra analysed in non-LTE conditions in CEMP stars, and even less with 3D model atmospheres. However, atmospheric granulation induces asymmetry and enhancement in strong lines that are not reproduced by 1D models \citep[e.g.][]{ludwig2009,freytag2012,pereira2013,amarsi2018}, the effects of which are mostly larger than those of the isotopic mixture, making the use of 3D non-LTE models mandatory for this purpose.

\begin{figure*}[th!]
\begin{center}
\includegraphics[width=.45\textwidth]{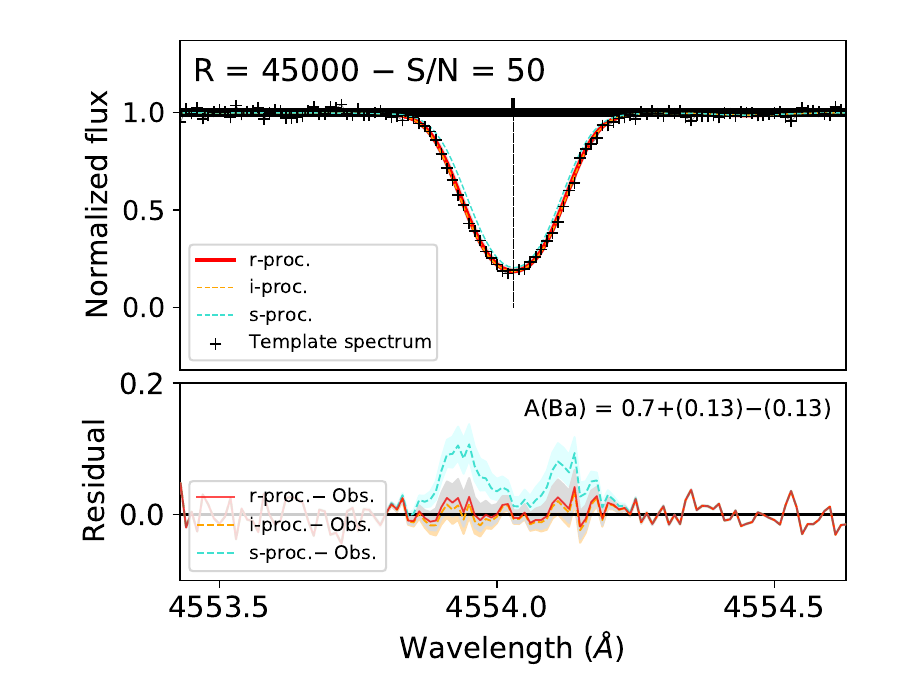}
\includegraphics[width=.45\textwidth]{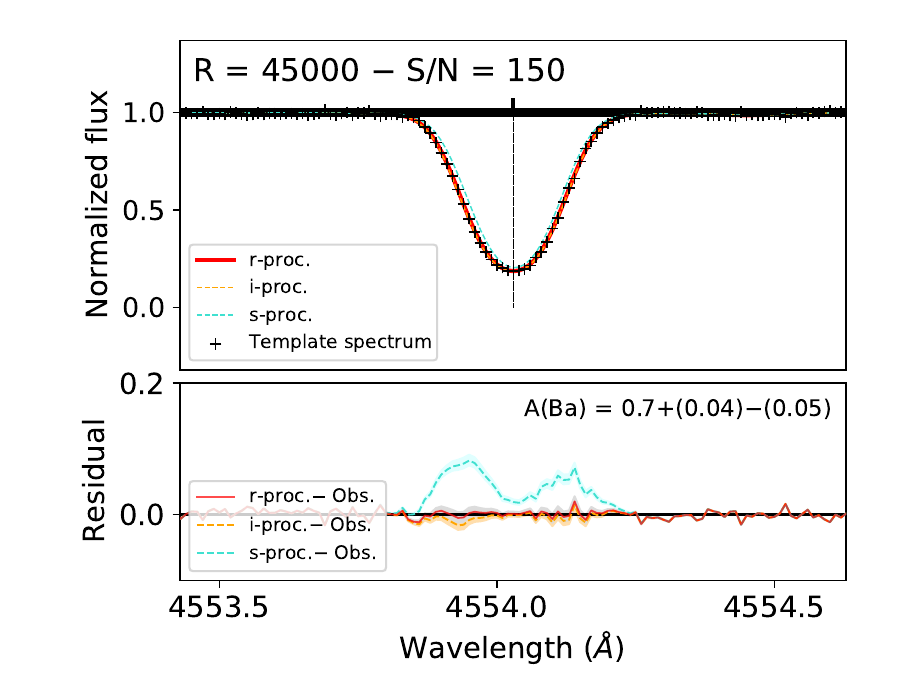}
\includegraphics[width=.45\textwidth]{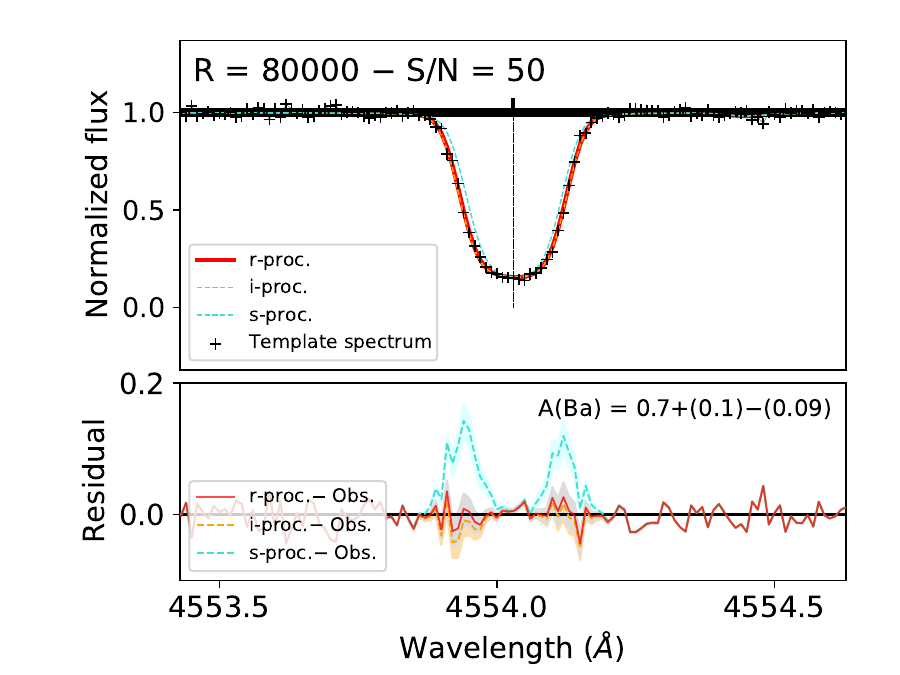}
\includegraphics[width=.45\textwidth]{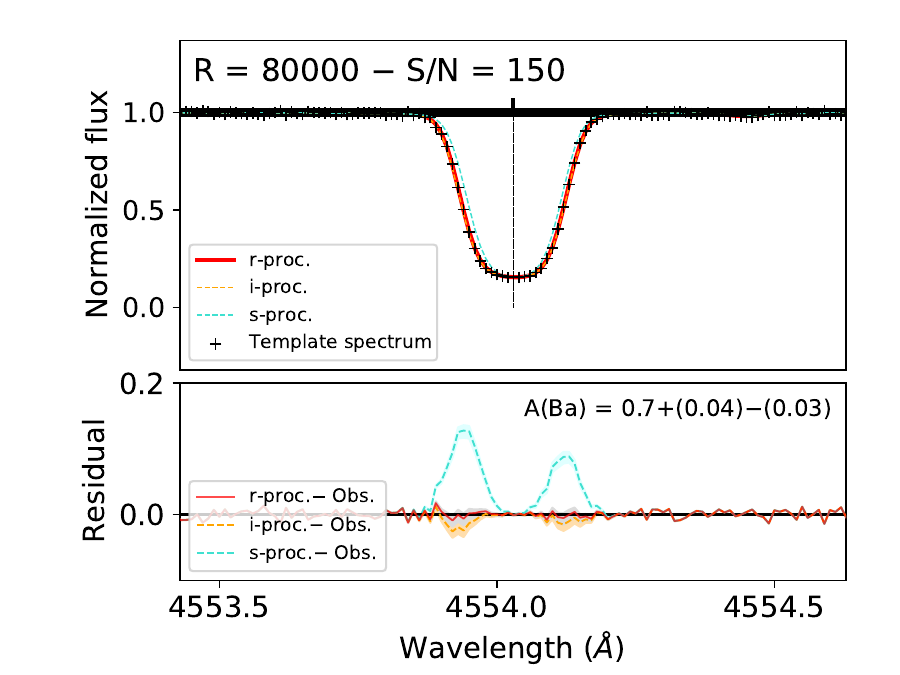}
\end{center}
\caption{{\it Upper panels in each of the 4 subplots:} The Ba resonance line at 455.4\,nm, at different SNR and spectral resolutions, as indicated in each panel. The black crosses correspond to a synthetic spectrum (``template spectrum") corresponding to a typical metal-poor $r$-process enriched star, degraded to the desired resolution and SNR. The Ba abundance ($A(\mathrm{Ba})=0.7$) has been determined from the subordinate lines at 585.4, 614.2 and 649.6\,nm. The red, orange, and blue curves correspond to synthetic spectra computed with the very same Ba abundance, but for isotopic mixtures corresponding to the $r-$, $i-$ and $s-$processes, respectively. {\it Lower panels: } The residuals computed from the difference between the template spectra and the $r$-, $i$- and $s$-synthesis. The coloured areas correspond to the uncertainties on the residuals, due to the uncertainties in the abundance of Ba derived from the subordinate lines; they obviously depend on the SNR of the template spectrum as well as on the resolution.}
\label{fig:Ba_SNR_R}
\end{figure*}

Figure~\ref{fig:Ba_SNR_R} illustrates, in the case of the Ba resonance line at 455.4\,nm, the sensitivity to SNR and $R$ of the ability to distinguish between the specific isotopic signatures caused by the three main nucleosynthetic processes. A spectrum with $\rm SNR \sim 50$ is the minimum required at a resolution of $R=80\,000$, to be able to distinguish between the different nucleosynthetic process signatures shaping this Ba resonance line. We note that this line is the most suitable Ba line for the determination of the isotopic composition. In the illustrated example, the template spectrum is best reproduced by an $i$- or $r$-process, resulting in the smallest residuals throughout the line. Actually, Ba allows us to distinguish between the $s$- and the $r$-process, but not between the $r$-process and the $i$-process. Using the same technique, europium lines can potentially help to further disentangle the $i-$ from the $r$-processes. In the general case (weaker, potentially blended Ba or Eu lines) a resolution of $R=80\,000$ and a $SNR>100$ are most certainly required. 

\begin{table*}[ht]%
\centering
\caption{Instrument requirement summary (Galactic Archaeology)}%
\footnotesize
\begin{tabularx}{\textwidth}{llX}
\hline
\hline
\noalign{\smallskip}
\textbf{Parameter} & \textbf{Value}   & \textbf{Justification}  \\
\hline
\hline
Resolving power ($R$)     & 80\,000     & Optimal resolution for isotopic ratio measurements and trade-off with SNR    \\
Spectral range & $380-800$\,nm  & Specific to locations of key spectral lines, including windows in the bluer part of the visible spectral range, such as, e.g., windows around the FeI lines at 371.99, 373.71, 382.04 and 385.99 nm, the Th II lines and 401.913, the U II lines at 385.957, 405.004 and 409.013 nm.   \\
Multiplexing   & $20-100$      &  A minimum of 20 fibers, considering the typical density of the halo fields.    \\
Stability      & n/a         &  Typically expect 100\,m\,s$^{-1}$ for Galactic kinematics and orbit computation.     \\
Fibre spacing  & n/a         & Typically expect FLAMES level of fibre spacing     \\
\hline
\end{tabularx}
\end{table*}

\newpage
\subsection{Enhancing the value of medium-resolution surveys}

The goal of Galactic archaeology is to unravel the lost substructures of the proto-galaxy by identifying its stellar “building blocks” and substructures. Chemical and dynamical information in stellar substructures are signposts to an array of evolutionary events, from in-falling satellites and dissolving stellar aggregates to the effects of spiral arm resonances. Disentangling the relative contribution of all these phenomena is fundamental for determining the physical sequence of events that built the Milky Way and are likely also important in other large spiral galaxies \citep[see, e.g.,][and references therein]{Bonaca2020, Buder2022, daSilva2023, Horta2023, Giribaldi2023}.

Already with {\sl Gaia} data, complemented by ground-based spectroscopic surveys, many new discoveries have been made - from the detection of a perturbed thin disc \citep{antoja2018,blandhawthorn2019}; of new stellar streams, clusters, and a complex structure in the local velocity distributions \citep{cantatgaudin2018,quillen2018, Shih22, ibata21}; to flaring in the outer disc \citep{thomas2019, ramos21} and the discovery of the Enceladus merger and other remnants in the inner halo and thick disc \citep{belokurov2018, helmi2018, donlon20, donlon22, ciuca23}. A multitude of additional new findings are expected, such as stellar streams, over-densities of stars, variations with Galactocentric radius and azimuth of the global properties of the disc structure that have hitherto not been possible to detect. High-resolution spectroscopy of stars in known substructures, and in potential new discoveries yet to come, is an essential complement to provide critical detailed abundance information that can separate different populations based on chemistry.

In the era of large spectroscopic surveys, many millions of stars will be observed in the major structural components of the Milky Way. However, the focus for the currently funded future facilities has shifted toward low- ($R < 10\,000$) to medium ($R \approx 20\,000$) resolution spectroscopy. Although these facilities will open up new discovery space, there remains a lack of matching multi-object facilities at high resolution ($R \approx 80\,000$), from which the parameter space can be expanded, especially from a chemical point of view, with decreased uncertainties, improving precision. 
The contribution of HRMOS will be decisive and essential. It will enable us to precisely calibrate medium spectral resolution surveys, providing the best possible training set for data-driven models for spectral analysis. It will also enable its own high-resolution volume limited survey and follow-up of rare and interesting objects identified by large surveys.  
 
The large, medium-resolution spectroscopic surveys will base their analysis on samples analysed in a physics-driven manner, which will function as training sets for the overall analysis that will be conducted in a data-driven manner.
Developments over the last decade mean that there are now several well-tested codes that are capable of extracting stellar parameters and individual elemental abundances within seconds to minutes. The most applied systems are The Cannon \citep{ness2015},  The Payne \citep{ting2019}, and more recent codes that make use of neutral networks \citep[e.g.][]{nepal23}, 
with customised versions that cater to specific aspects of abundance determinations \citep[e.g.][]{casey2016, leung2019a, xiang19}.  
A critical aspect for the success of machine-driven techniques is the availability of high-quality labels for smaller samples of reference objects \citep[e.g.][]{Heiter2015, Hawkins2016, Giribaldi2021, Giribaldi2023b}. We can envisage that having a sample of stars observed by HRMOS at higher resolution that will raise the level of precision of the lower-resolution surveys, due to the higher precision and the larger quantity of measurable elements, thus providing high-quality labels for their training sets. 

In addition, machine learning methods can be used to infer chemical signatures that are otherwise not easily detectable in low- or medium-resolution spectra. The power of these techniques has been demonstrated in the determination of detailed chemical abundances from low-resolution spectra. For example, \citet{xiang19} and \citet{wheeler2020}, derived a range of individual chemical elements from low-resolution ($R = 1800$) LAMOST spectra using overlapping samples from the GALAH and APOGEE surveys ($R=20\,000 - 30\,000$) for training. Nevertheless, one needs to be aware that blending and saturation may affect the quality of abundances determined in this way from such data \citep{Karinkuzhi2021}. 
Actually, by having a high-resolution calibration sample, low-resolution samples can be rejuvenated with the use of data-driven machine learning techniques.

For instance, the science possible with the 4MOST survey data \citep[4MIDABLE-HR/LR - The 4MOST Bulge and disc High/Low-Resolution Surveys,][]{bensby2019,chiappini2019} calibrated by observations at higher resolution is vast. Performing the spectral analysis with data-driven techniques, whose training set is obtained at higher resolution, could help shed light on the origin of the bulge and bar, the most complex and least understood components of the Milky Way. Adding more abundances might help to identify the possibly primordial bulge component and distinguish it from, or link it to, the structures in the solar neighbourhood.

\subsection{Targeted follow-up of rare objects}

Large surveys will produce important samples of “unusual” or “weird” objects. The understanding of such samples will then require a thorough examination with high-resolution spectroscopy, carried out on an 8-meter class telescope and covering a large range in wavelength. Therefore, the requirement would be to follow-up all these rare objects. It is important to note that direct follow-up observations within the large surveys themselves, to further characterise these findings, are not possible as that would alter and influence the efficiency and the overall selection function of the surveys. Therefore, there is direct need for an instrument with multiplex ability, with high resolving power, connected to a telescope that has a higher light collecting power than the typical 4-meter telescopes that most large spectroscopic surveys will use (e.g. WEAVE and 4MOST). This will provide both higher-resolution and higher-SNR observations, which will allow a better characterisation of the properties of these rare objects.

Among these rare objects that will benefit from an instrument such as HRMOS will certainly be post-AGB stars. The surface of a post-AGB star retains the signatures of the rich nucleosynthesis during the AGB phase. This makes post-AGB stars formidable probes for examining the elements produced by the star and subsequently gaining insight into how these elements are created. Although these scenarios for the creation of elements have general validity, there are several crucial missing pieces and unsolved questions that prevent our advancement in the basic understanding of the production of elements and isotopes in low- and intermediate-mass stars. Critical uncertainties exist in the theoretical modelling of processes, especially convection, convective-driven mixing processes, and mass loss, that govern the chemically rich AGB phase for these stars and affect the predicted stellar yields of elements and isotopes.
The high-resolution observations of post-AGB stars can provide a comprehensive data set of crucial chemical element abundances (e.g., C N, O, Fe-peak elements, and s-process elements including Pb), which will allow a better understanding of AGB and post-AGB stars. This will also provide a gateway to better understand the physical processes that can affect stellar yields.

\begin{figure}[ht]
\centering
\resizebox{0.6\hsize}{!}{
\includegraphics{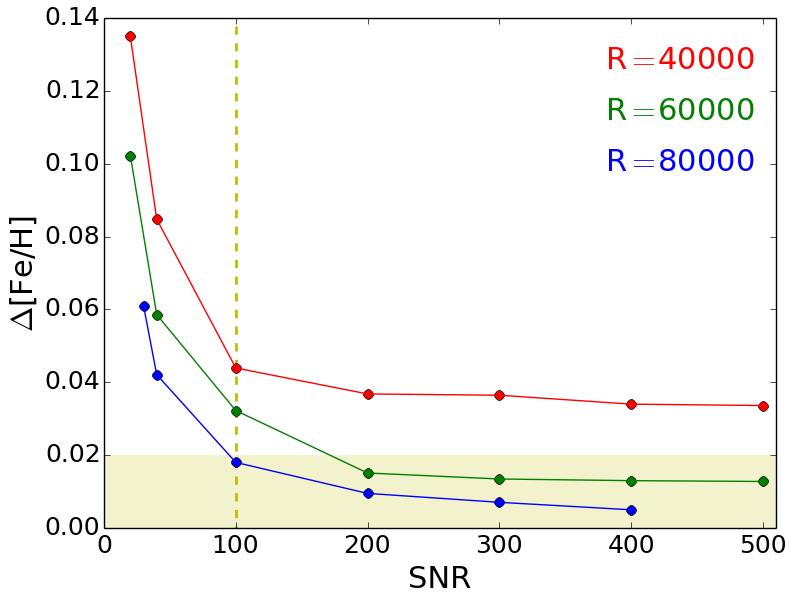}}
\caption{
Precision in the Fe abundances as a function of the SNR for three different resolutions. The shaded region indicates when the required precision ($<0.02$\,dex) has been achieved. The dashed line indicates $\rm SNR=100$. Based on analysis of simulated spectra of globular clusters by Anna F. Marino.
\label{fig:marino}
}
\end{figure}

\subsubsection{Requirements - Exploiting large surveys}

To derive stellar labels, especially for individual element abundances and isotopic ratios, in order to generate a model for data-driven techniques, the data quality requirement is crucial.
The higher the precision of these labels, the more robust the model that the machine will use to derive new labels for the remaining larger sample of stars from other surveys. Therefore, the critical aspect for the success of the machine-driven techniques is the quality of the labels for a small sub-sample of the survey stars.

According to Fig.~\ref{fig:marino}, a $\rm SNR \gtrsim 100$ at $R = 80\,000$ or, a $\rm SNR \gtrsim 200$ at $R = 60\,000$ are typical minimum requirements for the metal-poor stars. 
The sample size to calibrate a large survey to higher resolution precision depends on the number of parameters to be determined. To obtain stellar parameters as well as individual elements spanning light elements to heavy n-capture elements, of about 2\% of the sample observed at medium resolution with 4MOST \citep[see, e.g.][for transferring lables from APOGEE to LAMOST giant stars]{ho17}. Considering the number of targets that the high-resolution 4MOST survey for the bulge and disk will observe \citep[i.e. about 4 million stars][]{bensby2019}, a training set of about 80\,000 stars will allow efficient transfer of labels from the very high resolution of HRMOS to the medium resolution sample.

\frame{\vspace{-7mm}
\paragraph{Uniqueness}
The current planned MOS facilities for the 2030 timescales, with high multiplex capabilities, have low to moderate spectral resolution. Therefore, HRMOS is unique at the highest resolution. ANDES at the ELT is, conversely, a high-resolution spectrograph providing the necessary resolution, wavelength coverage, and signal-to-noise ratio (given the larger telescope aperture); however, it is a single-object instrument. HRMOS offers the unique opportunity to provide a great legacy value by combining higher resolution with MOS capabilities. The MOS capability compensates for the lower magnitude limit even for the ELT instruments. This makes HRMOS superior to current and planned facilities:

\begin{itemize}
\item The currently planned large surveys such as 4MOST and WEAVE do not provide sufficiently high spectral resolution to address the key science outlined here.
\item ANDES/ELT does not provide the MOS facility to gather sufficiently large samples at high resolution to calibrate the lower resolution surveys. It is better placed to observe fainter distant targets for niche individual objects, which is in synergy with HRMOS.
\item HRMOS data combined with data-driven analysis methods will enable re-calibration of high multiplex surveys, while also providing the capability for niche-targeted observations.
\end{itemize}

Therefore, HRMOS is uniquely placed in the current/planned instrumentation landscape.
}

%% file: dwarf_gal.tex
\clearpage
\section{The Milky Way satellite galaxies}

 The Milky Way satellite galaxies are ideal laboratories to study galaxy formation and evolution in a large variety of environments. These systems span over four orders of magnitude in stellar mass \citep[e.g.][]{mcconnachie2012,battaglianipoti2022}, and have a wide range of star formation and chemical enrichment histories \citep[e.g.][]{tolstoy2009,simon2019}, offering the unique opportunity to resolve some of the degeneracies present when trying to understand chemical evolution (e.g. nucleosynthetic yields, mass loss due to winds, star formation histories, time delay functions of different sources, and merger histories). Furthermore, these ancient galaxies are analogues to absorption systems seen at high redshifts \citep[e.g.][]{cooke2015,skuladottir2018}. These nearby systems can thus be used as independent probes of stellar physics, galaxy evolution, and cosmology. In order to fully exploit these galaxies, it is necessary to get high quality  spectra (high SNR and high spectral resolution) for large statistical samples ($\gtrsim100$) in these faint, challenging systems.

\frame{
\titlecol{Key Questions:}
\begin{itemize}
\item What does primarily drive the chemodynamical evolution of different galaxies?
\item How did the earliest chemical enrichment depend on environment and galaxy properties? 
\item Are there different dominant types of SN\,Ia in different galaxies?
\item What are the nucleosynthetic sites of the neutron-capture elements?
\item Are the stellar binarity fraction and orbital characteristics dependent on galaxy characteristics?
\item Do the Milky Way satellite galaxies contain substructures, or evidence of past mergers?
\end{itemize}
}



\begin{table}[ht]%
\centering
\caption{Milky Way satellite galaxies visible from Paranal, Chile, their distances, stellar masses and dynamical masses \citep[from][]{mcconnachie2012}.
\label{tab:mwsatellites}
}%
\footnotesize
\begin{tabular}{ccccc}
\noalign{\smallskip}
\hline
\hline
Satellite & Distance  & M$_{star}$           & M$_{\rm dyn}$        \\
          & [kpc]     & [10$^6$ M$_{\odot}$] & [10$^6$ M$_{\odot}$] \\
\hline
LMC & 51 & 1500 & $>10^4$\\
SMC & 64 & 460 & $>10^3$\\
Sagittarius & 23 & 21 & 190 \\
Fornax & 147 & 20 & 56 \\
Sculptor & 86 & 2.3 & 14 \\
Sextans & 86 & 0.44 & 25 \\
Carina & 105 & 0.38 & 6.3 \\
UFD ($>40$) & $>20$ & $<0.1$ & $<1$ \\
\hline
\end{tabular}
\end{table}

\subsection{Potential targets} 

\begin{figure*}
\centering
\includegraphics[width=0.42\linewidth]{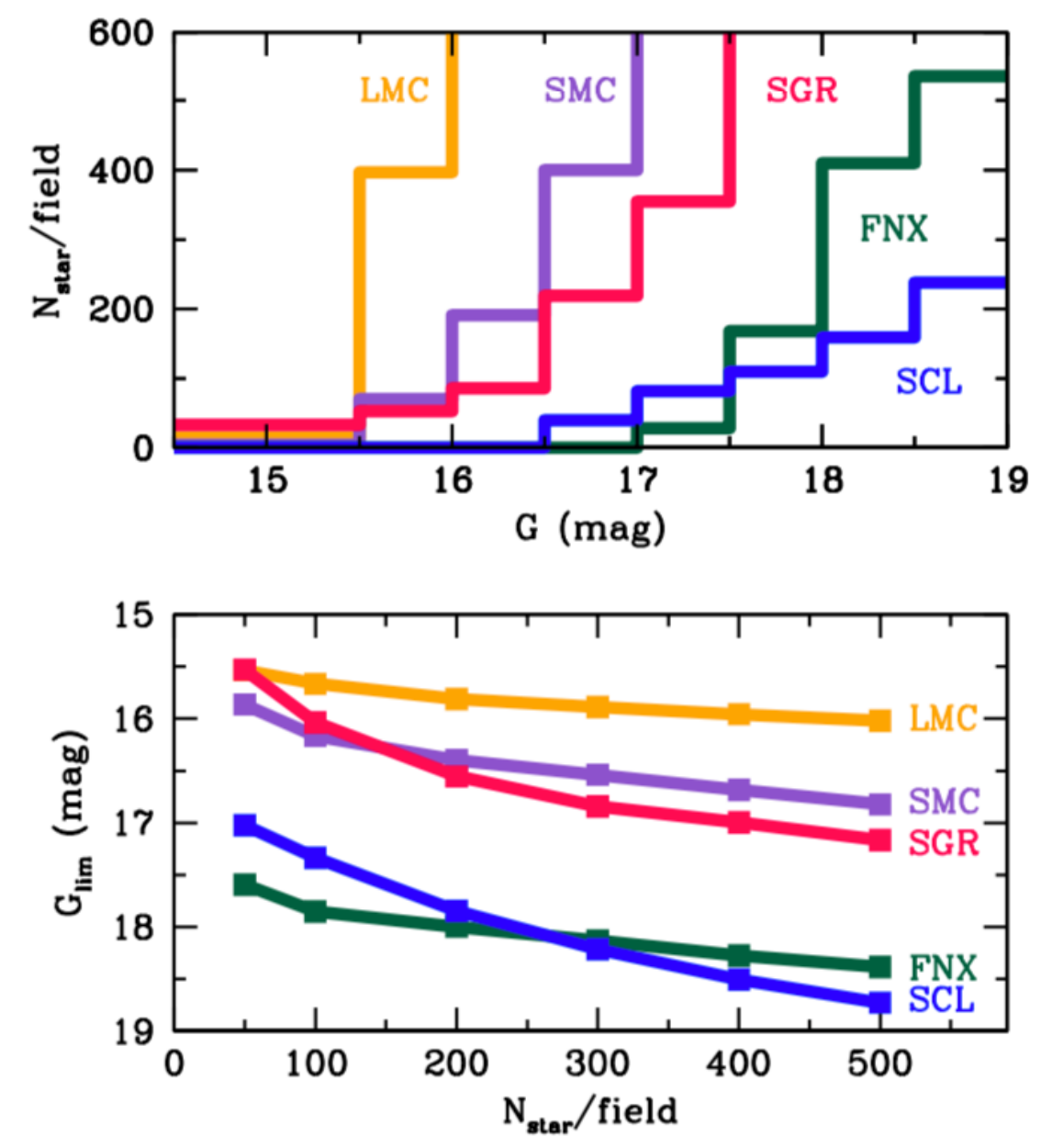}
\hspace{0.5cm}
\includegraphics[width=0.32\linewidth]{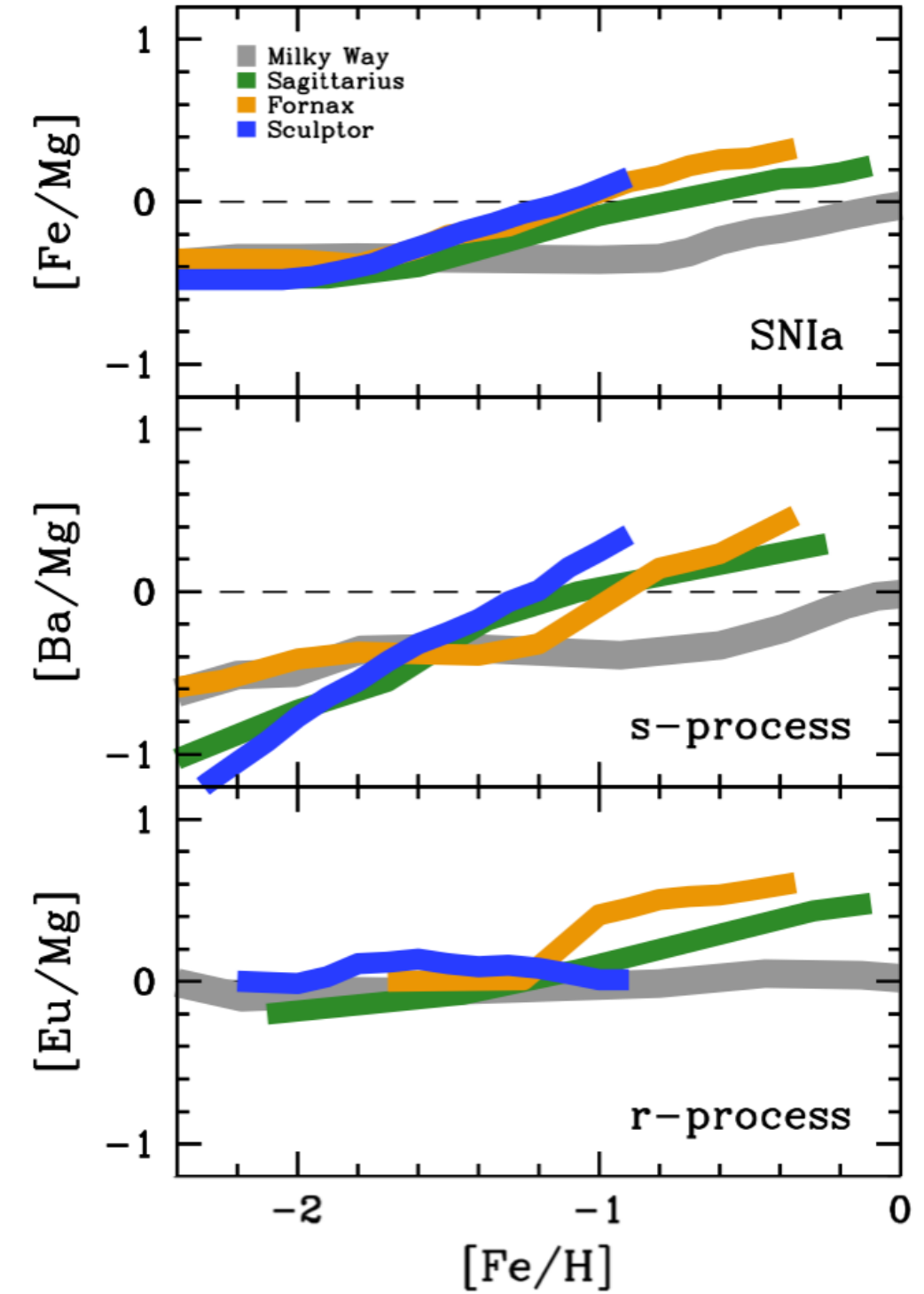}
\caption{
\textit{Left:} Target density of RGB stars for the largest satellite galaxies. \textit{Top panel:} Number of stars in one central field of 25\,arcmin diameter, per G magnitude bin. \textit{Bottom panel:} The limiting magnitude, $G_\textsl{lim}$, required to have a given number of stars per FoV. All estimates are based on Gaia DR2 (ra, dec, pm, G and BPRP). Distances ($d$) are adopted from \citet{mcconnachie2012}. \textit{Right:} Schematic figure of average abundance trends in the Milky Way (grey) and three dSph galaxies, Sagittarius (green), Fornax (yellow) and Sculptor (blue). Adapted from \citet{skuladottirsalvadori2020}.
\label{fig:mwsat_1}
}
\end{figure*}

The Southern hemisphere is rich of dwarf galaxies, within $d<150$\,kpc, see list of potential targets in Table~\ref{tab:mwsatellites} along with their main properties. In particular, the Milky Way's five largest nearby satellites are located in the Southern hemisphere (Fig.~\ref{fig:mwsat_1}): the LMC ($d=51$~kpc), the SMC ($d=64$\,kpc), the Sagittarius dSph ($d=26$\,kpc), Fornax (FNX; $d=147$\,kpc), and Sculptor (SCL, $d=86$\,kpc), where distances, $d$, are adopted from \citet{mcconnachie2012}. In addition to these, over 40 smaller dwarf spheroidal (dSph) and ultra-faint dwarf galaxies (UFDs) are visible from Paranal \citep[e.g.][]{mcconnachie2012,battaglia2022}. These UFDs typically have on the order of $\lesssim100$ stars, and most of which would typically be visible within one field-of-view of HRMOS. The location of the ESO VLT provides the opportunity to investigate chemical abundances, radial velocities, and binary fractions, contrasting over galaxies with a large range of properties \citep[e.g.][]{tolstoy2009,mcconnachie2012}. Furthermore, the Magellanic Clouds and Sagittarius cover a large region of the sky, and are thus excellent candidates for studies of spatial variations of chemodynamical properties and of binary fractions. The Magellanic Clouds offer the unique opportunity to simultaneously study very different mass regimes at low metallicities, from young main sequence stars, down to old RGB stars \citep[e.g.][]{cioni2019}.

\subsection{Earliest chemical enrichment}

Dwarf galaxies are intrinsically metal-poor, and thus ideal to study the physics of first stars, and the chemical enrichment of pristine environments. It is currently unknown how strongly the earliest chemical enrichment depends on environmental factors, or galaxy properties such as mass, but already some differences have emerged.

\begin{figure*}
\centering
\includegraphics[width=0.43\linewidth]{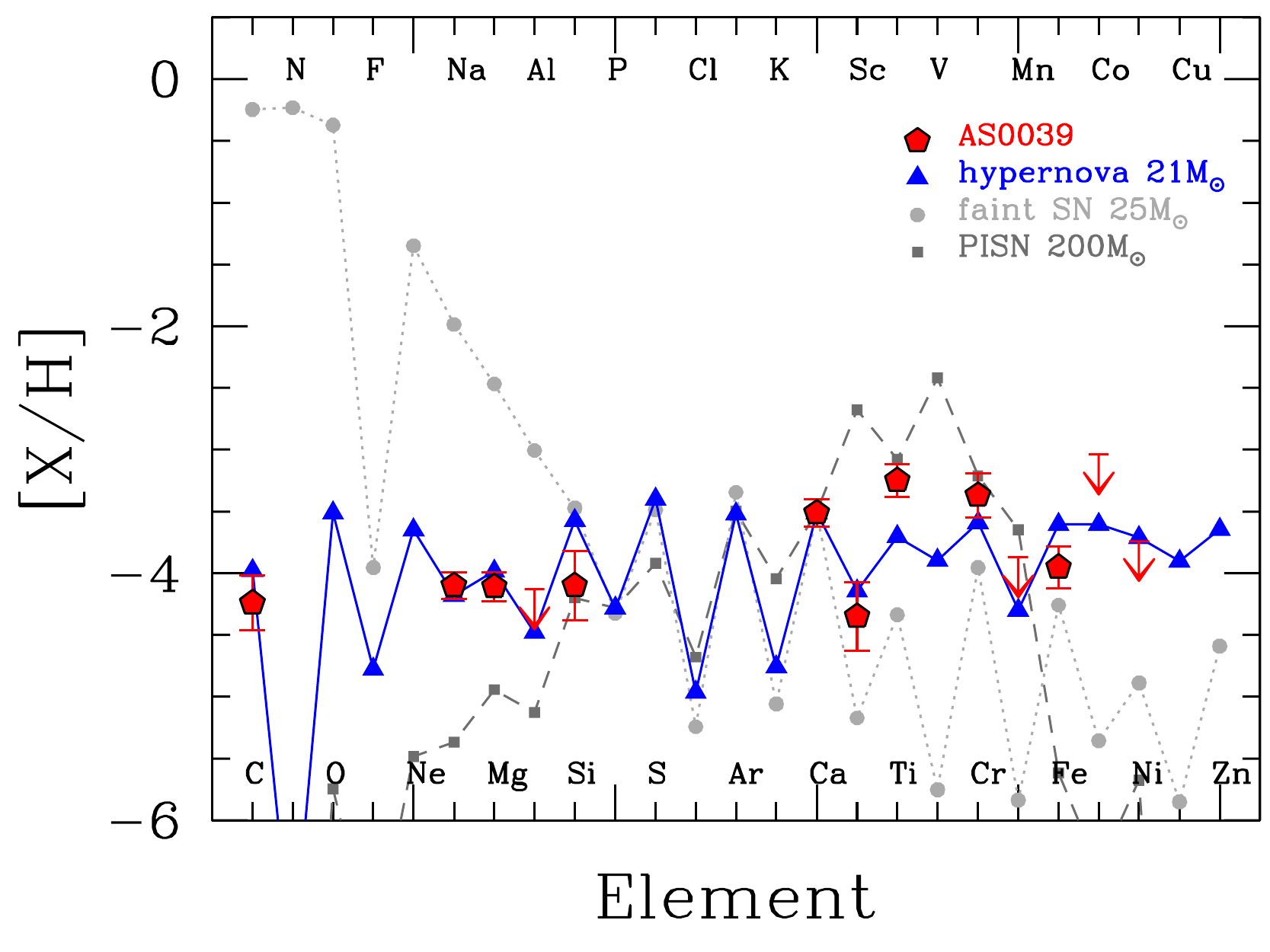}
\hspace{0.5cm}
\includegraphics[width=0.47\linewidth]{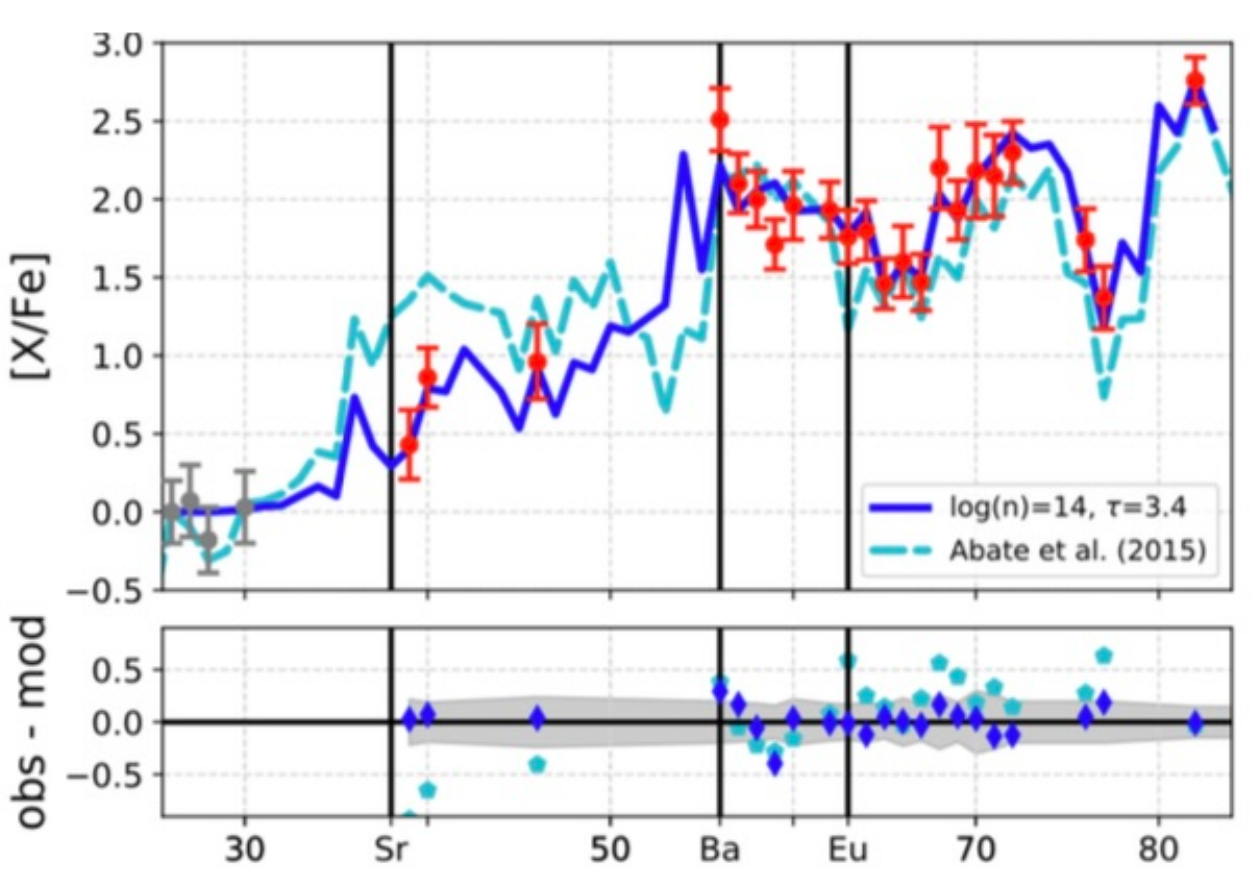}
\caption{
\textit{Left:} The abundance pattern of the Sculptor star AS0039 compared to theoretical yields. The best-fitting model is a zero-metallicity, 21 M$_{\odot}$ hypernovae, of energy $10 \times 10^{51}$\,erg. From \citet{skuladottir2021}. \textit{Right:} Red points show the measured abundances of the CEMP-i star CES31062-050. Best fit CEMP-i models are shown in Blue (solid line, diamonds), while the best fitted s-process is shown in cyan (dotted line, pentagons). From \citet{hampel2019}.
\label{fig:mwsat_2}
}
\end{figure*}

\begin{enumerate}
\item The fraction of carbon-enhanced metal-poor stars without Ba enhancement (CEMP-no; e.g. \citealt{beers2005araa}) shows indications of being significantly lower in dSph galaxies, compared to the smallest UFDs and the Milky Way halo \citep[e.g.][]{salvadori2015,ji2020,lucchesi2020,skuladottir2023}.
Furthermore, CEMP-no stars in dwarf galaxies might have differences in their detailed chemical evolution pattern, that is, high Sr, Y, Zr, over Ba ratios \citep{skuladottir2015,susmitha2017,spite2018} while the Milky Way halo CEMP-no stars follow the C-normal population in their [Sr,Y,Zr/Ba] abundances. As CEMP-no stars are thought to directly sample the material ejected by zero-metallicity stars, with low explosion energy \citep{iwamoto2005,hegerwoosley2010,rossi2023} this is a direct probe of the earliest conditions in these galaxies.

\item The abundance pattern of [Sr/Ba], at [Ba/H]$<-2.5$, is markedly different in dSphs compared to UFDs, where the UFDs systematically have lower values of [Sr/Ba] \citep{mashonkina2017}. The cause of this is still poorly understood, but gives insight into the earliest formation of these elements (i.e. the r-process, and the weak r/s-processes).

\item The energy distribution of the first stars is still very poorly constrained \citep{koutsouridou2023}, and the dwarf galaxies could play a fundamental role to improve this \citep{rossi2023}. In particular, the imprints of first stars exploding as high energy SN might be guarded in dSph galaxies, see Fig.~\ref{fig:mwsat_2} \citep{skuladottir2021,skuladottir2023}.

\end{enumerate}

Mapping the differences and similarities between the chemical abundances of metal-poor stars, in galaxies with a range of properties, is therefore necessary to fully understand the earliest chemical enrichment. This will also give insights into the formation of the metal-poor halo of the Milky Way, $\rm[Fe/H]<-2$, and constrain the properties of the smaller galaxies it was built from.

For this purpose, very high-resolution spectra at wavelengths shorter than 500\,nm are essential, to accurately measure the key chemical diagnostics available for this metal-poor population. 4MOST will provide excellent catalogues of metallicities in dwarf galaxies (4DWARFS; \citealt{skuladottir2023dwarfs}) and the Magellanic Clouds \citep[1001MC; ][]{cioni2019}, providing opportunities for very efficient follow up.

In addition, massive zero-metallicity stars in the range $140\leq M/M_\odot \leq260$ are predicted to explode as pair-instability supernovae (PISN), enriching the environment with a very unique signature, for example a prominent odd-even effect \citep{hegerwoosley2002,kozyreva2014,takahashi2018}. The descendants of such stars are very rare, typically less than 0.5\,\% of the total stellar population \citep[e.g.][]{debennassuti2017}, but large surveys such as 4MOST will provide candidates for such survivors, which then will need very high-resolution  follow-up to confirm the absence of key elements such as Cu, and Zn, and get accurate, high precision abundances to contrast against models \citep{salvadori2019,aguado2023,xing2023}. A very high-resolution multi-object spectrograph, such as HRMOS is therefore fundamental to identify {\it bonafide} descendants of massive primordial PISNe, and understand the properties of the massive first stars.

\subsection{Delayed nucleosynthetic sources}

Two of the key open questions in nuclear physics have been identified as \citep{schatz2022}: 1)~\textit{What do nucleosynthesis signatures tell us about the nature of the progenitor
systems and explosion mechanism of thermonuclear supernovae?;} and 2)~\textit{What are the astrophysical sources of the heavy elements, what are their relative contributions, and how have these evolved over the history of the Milky Way and the Universe?} To answer these pressing and challenging questions, high-quality spectra of stars in diverse systems are crucial, and here the Milky Way satellite galaxies will play a fundamental role. 

The star formation in small galaxies is inefficient, which amplifies the impact of time-delayed nucleosynthetic sources \citep[][]{hill2019,kirby2019,reichert2020,skuladottirsalvadori2020}, as is shown in Fig.~\ref{fig:mwsat_2}. The delayed time distributions of SNIa, AGB stars, and neutron star mergers (NSM) via their kilonovae, cause the influence of their chemical signatures to increase with time (and [Fe/H]). This makes dwarf galaxies ideal to study the nucleosynthetic yields of such events, and to constrain their time-delay distributions, and their level of impact depending on environment, that is, star formation histories and mass loss due to stellar winds and SN explosions.

\subsubsection{Supernovae Type Ia}
The possible progenitors of SNIa are highly debated, and are likely to be many and varied \citep{maoz2014}. Their progenitors can be probed via the predicted yields \citep[e.g.][]{seitenzahl2013,kirby2019} and their time-delay distributions \citep[e.g.][]{ruiter2009}. Studies of individual dwarf galaxies, indicate that the dominant class of SNIa is different for galaxy which have extended vs truncated star formation histories \citep{kirby2019}. To fully investigate this and provide meaningful constraint against theoretical models, accurate and high-precision complete abundance patterns of $\alpha$-elements and the iron-peak elements are needed. This can best be done with an HR multi-object spectrograph which enables measurements of key high-precision elemental abundances such as V, Mn, Fe, and Ni, see Fig.~\ref{fig:mwsat_3}.

\subsubsection{The s- and i-processes}
The Magellanic Clouds and dwarf galaxies provide unique test-beds to study the products of metal-poor AGB stars, both via the slow (s) and the intermediate (i) neutron-capture processes \citep[e.g.][]{hasselquist2021}. Having high precision abundances (Fig.~\ref{fig:mwsat_3}) in galaxies of a range of characteristic, metallicities and star formation histories is crucial to constrain stellar models at low metallicity, and probe the role of various suggested i-process sites. A study of heavy elements in the Sculptor dSph, which only includes a limited number of elements (Y, Ba, La, Nd, Eu). shows that a significant contribution of the i-process is needed to explain the build-up of heavy elements in this galaxy \citep{skuladottir2020}, but whether that is true for the other satellite galaxies is currently unknown. However, one of the few CEMP stars reported in the Sagittarius dSph is a CEMP-r/s star \citep{sbordone2020}, possibly showing imprints of the i-process, and similar results are obtained in the Carina dSph \citep{hansen2023}. Thus indicating that the i-process might be particularly important in these moderately sized galaxies. Figure~\ref{fig:mwsat_2} shows the abundance pattern of a Milky Way CEMP-i star, along with models for the i- and s-process. To distinguish between these processes, elements at $Z>30$ that are lighter than Ba, and heavier than Eu are of special importance (Fig.~\ref{fig:mwsat_3}).

\subsubsection{The r-process}
\begin{figure*}
\centering
\resizebox{0.54\hsize}{!}{
\includegraphics{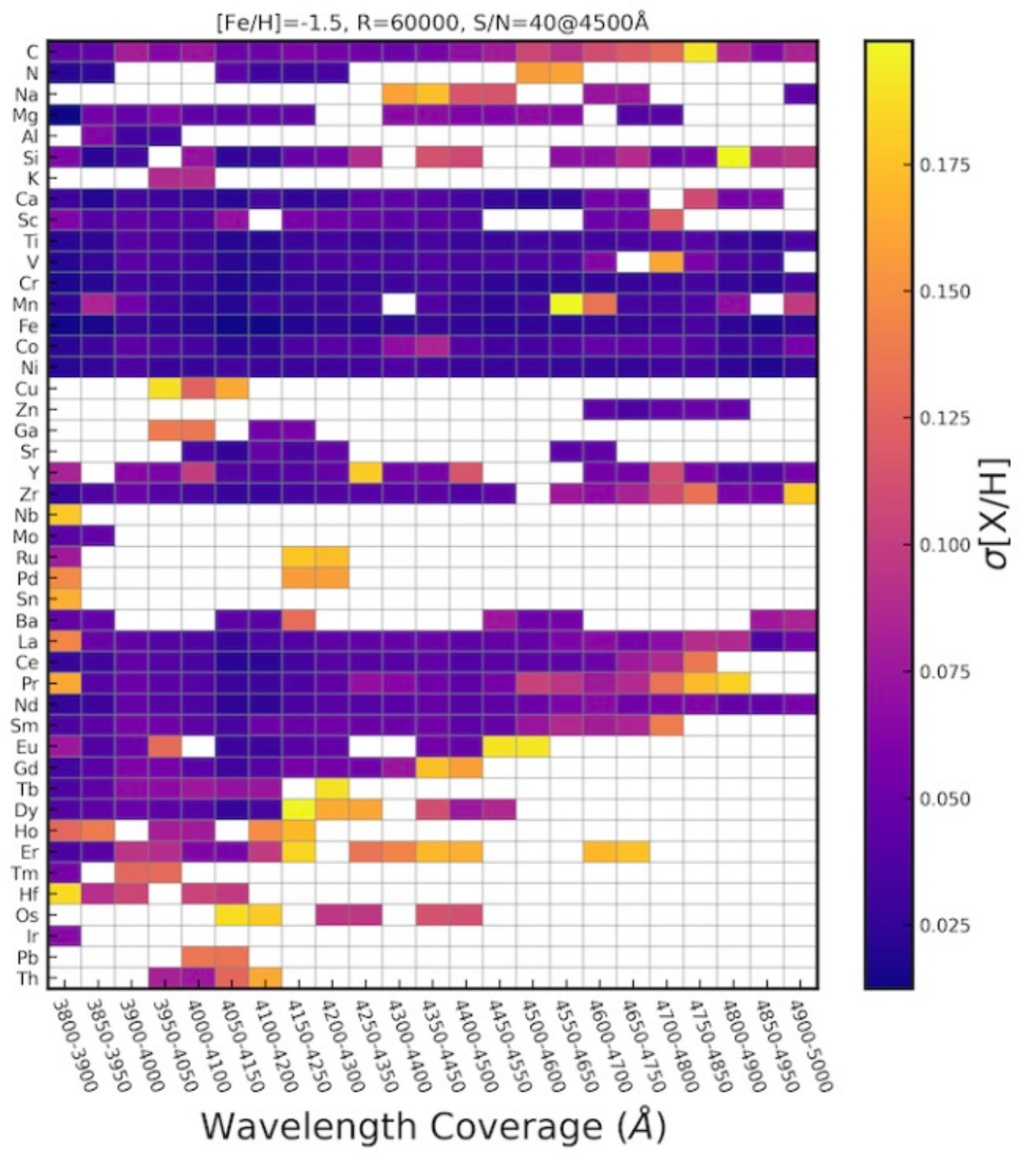}}
\caption{
Abundance information as a function of wavelength windows, for a $\rm[Fe/H]=-1.5$ giant, from spectra with $R=60\,000$, and $\rm SNR=40$ at 450\,nm. Colours shows the highest achievable theoretical precision. Based on the work of \citet{sandford2020}.
\label{fig:mwsat_3}
}
\end{figure*}

\begin{figure*}
\centering
\resizebox{0.6\hsize}{!}{
\includegraphics[]{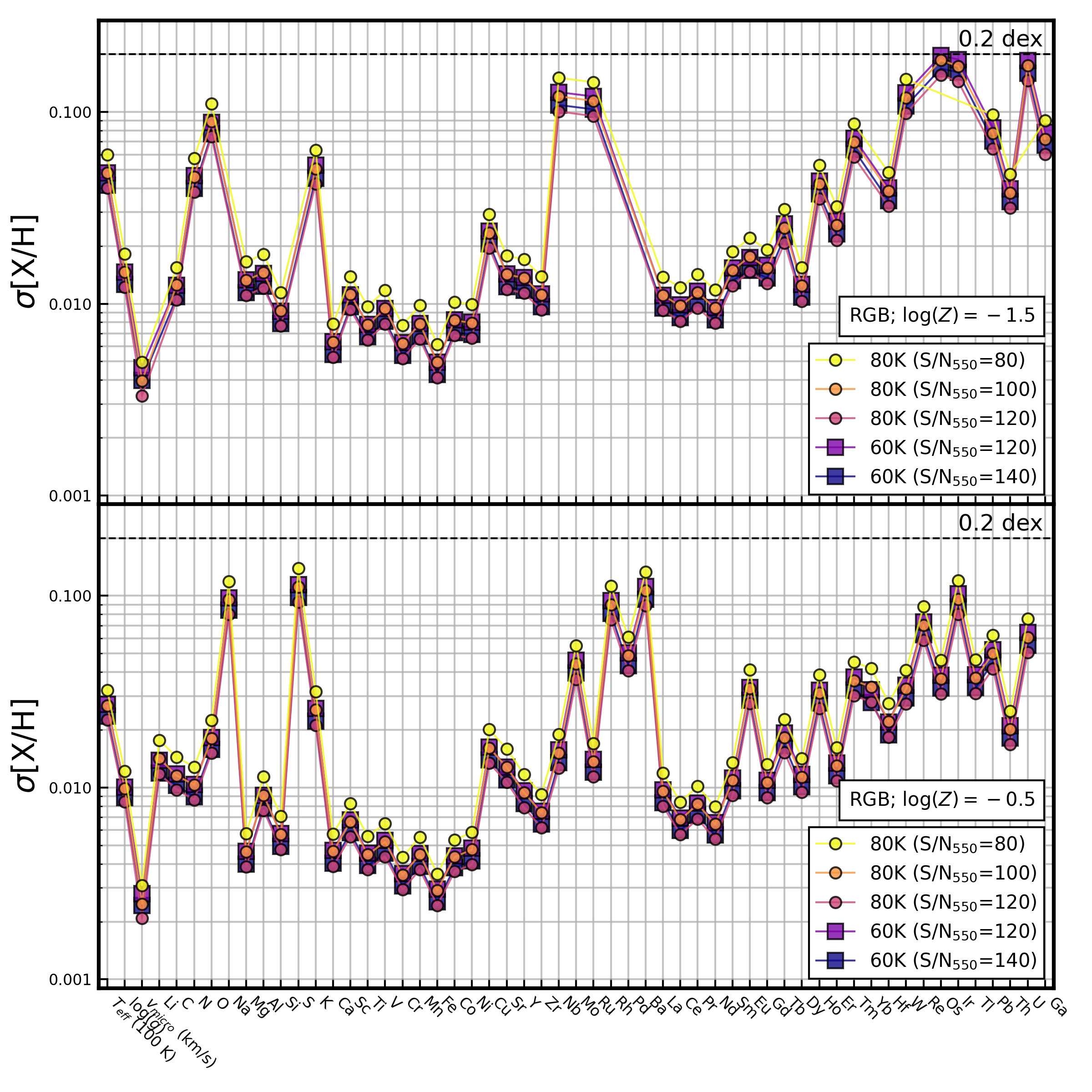}}
\resizebox{0.6\hsize}{!}{
\includegraphics[]{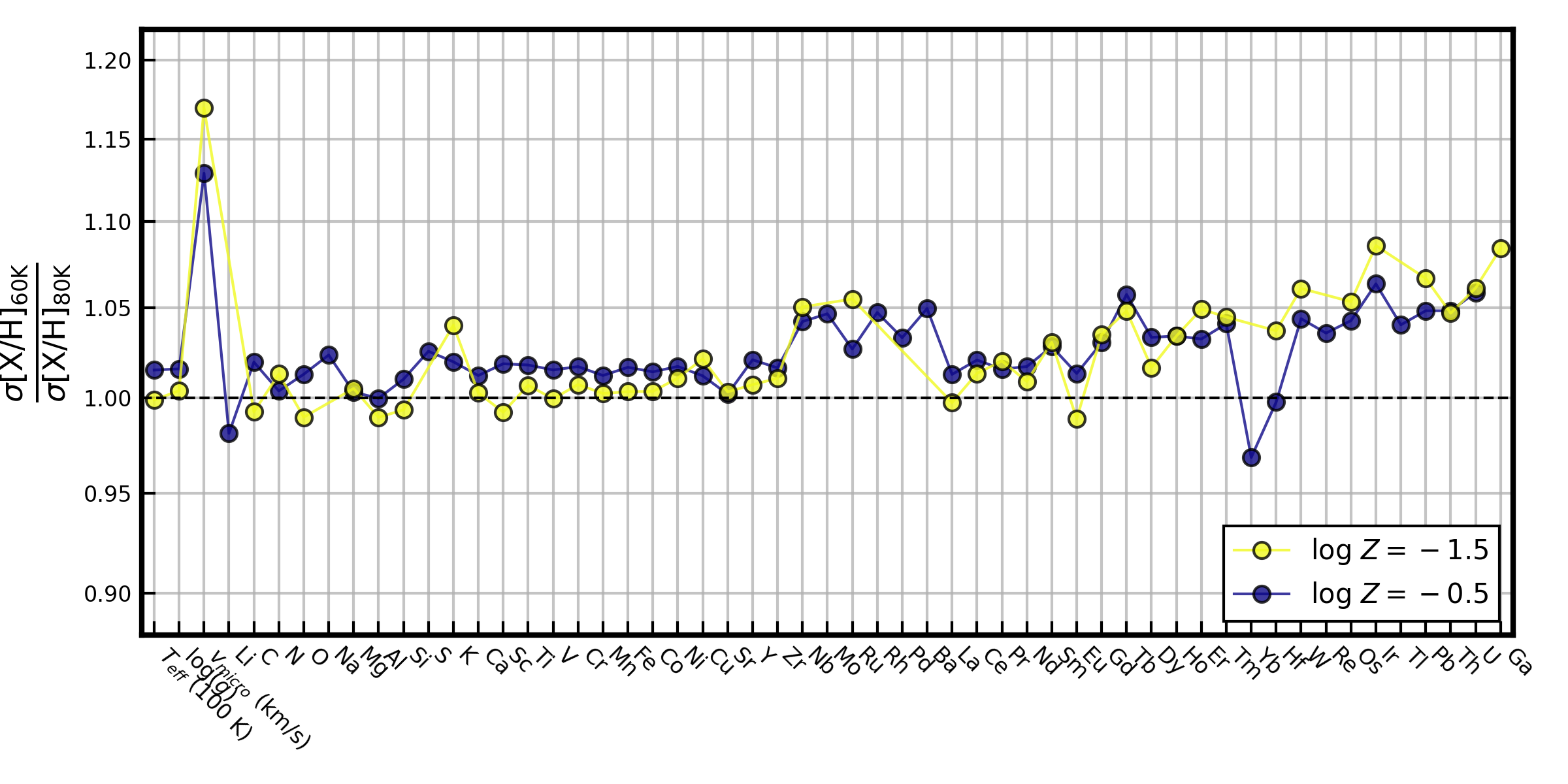}}
\resizebox{0.6\hsize}{!}{
\includegraphics[]{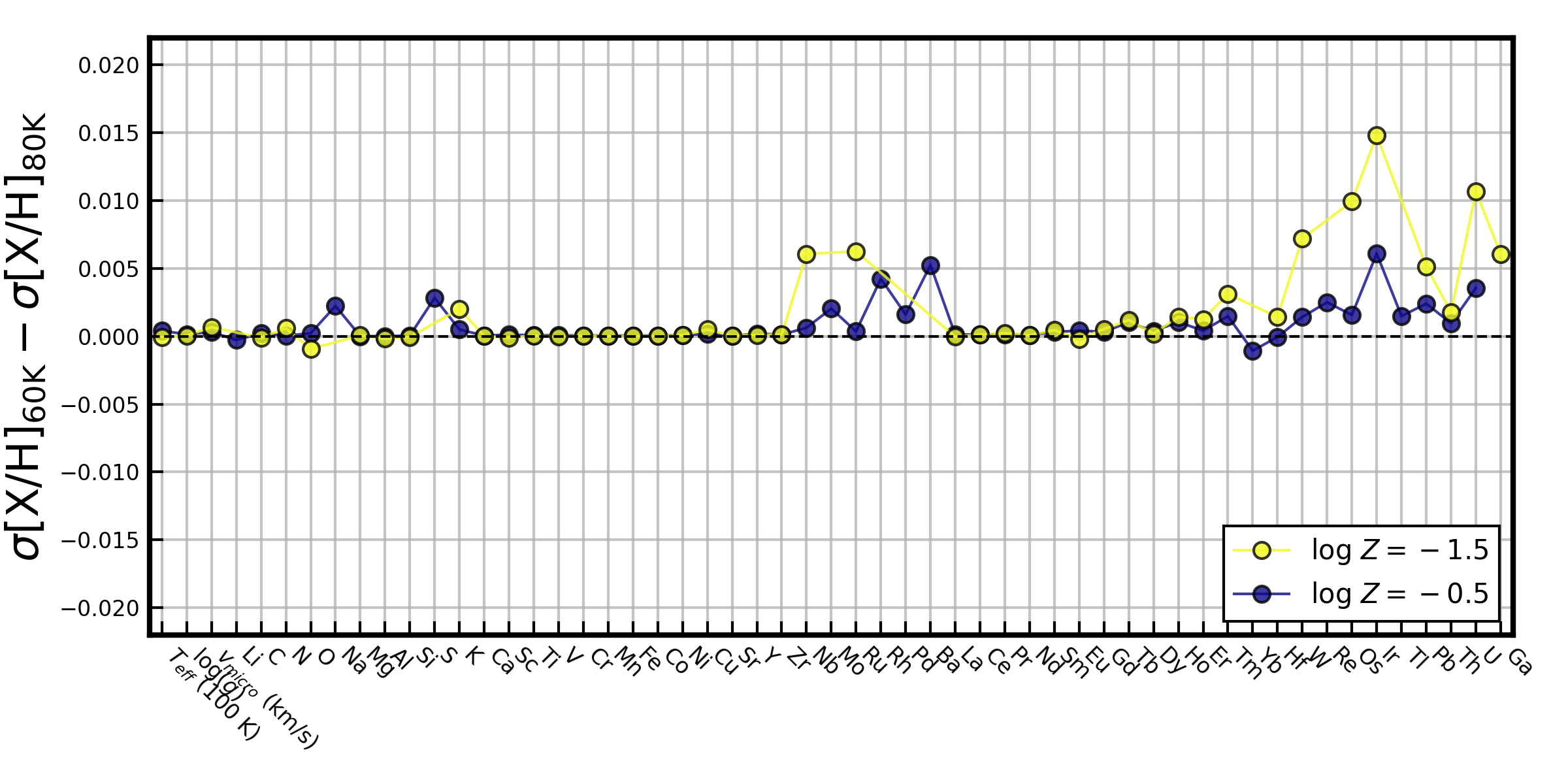}}
\caption{
Abundance precision for two giant stars at $\rm [Fe/H]=-0.5$ and $-1.5$, assuming four simultaneous wavelength windows [nm]: $373.5-406.5$ [$\lambda_\textit{cen}=390$], $498.5-541.5$ [$\lambda_\textit{cen}=520$], $632.5-687.5$ [$\lambda_\textit{cen}=660$], and $824-896$ [$\lambda_\textit{cen}=860$]. SNR are given per resolution element at 550\,nm. Based on the work of \citet{sandford2020}.
\textit{Upper panels:} precision shown for different resolution $R$ and SNR (see labels). 
\textit{Lower panels:} Difference in the precision achieved in observations of two $G=16.2$ giants at $\rm [Fe/H]=-0.5$ and $-1.5$, with $R=60\,000$ (SNR$\sim$120 per resolution element) and $R=80\,000$ (SNR$\sim$100).}
\label{fig:mwsat_4}
\end{figure*}

Recently, NSM have been shown as very prominent sites of the rapid (r) neutron-capture process \citep[e.g.][]{watson2019}. However, chemical evolution models of the Milky Way are unable to reproduce the observed chemical abundances of r-process elements if realistic time delay functions of such events are taken into account \citep[e.g.][]{cote2018}. The third panel of Fig.~\ref{fig:mwsat_1} (right) show these complications, as the Milky Way and Sculptor have a near flat trend of $\rm [Eu/Mg]\approx 0$ with [Fe/H], indicating that a prompt r-process is dominating, while Sagittarius and Fornax show a clear increase of [Eu/Mg] with [Fe/H], indicative of a significant contribution from a delayed r-process. This can either be explained by two separate r-process sites (quick and a delayed site), or a bimodal distribution of NSM events (see detailed discussion in \citealt{skuladottirsalvadori2020}; also \citealt{beniaminipiran2019}). Whether the abundance pattern of the r-process is dependent on the delay-time is currently completely unknown. To answer that, it is essential to have a HR multi-object spectrograph with sufficient efficiency in the blue (see Figs.~\ref{fig:mwsat_3} and \ref{fig:mwsat_4}) to be able to observe stars in the largest satellite galaxies (see Fig.~\ref{fig:mwsat_1}). Furthermore, studies of systems with different star formation histories will be needed to better constrain the time scales of NSM and other possible r-process sites.


\subsection{Binary stars in Local Group galaxies}

\begin{figure}
\centering
\resizebox{0.65\hsize}{!}{
\includegraphics{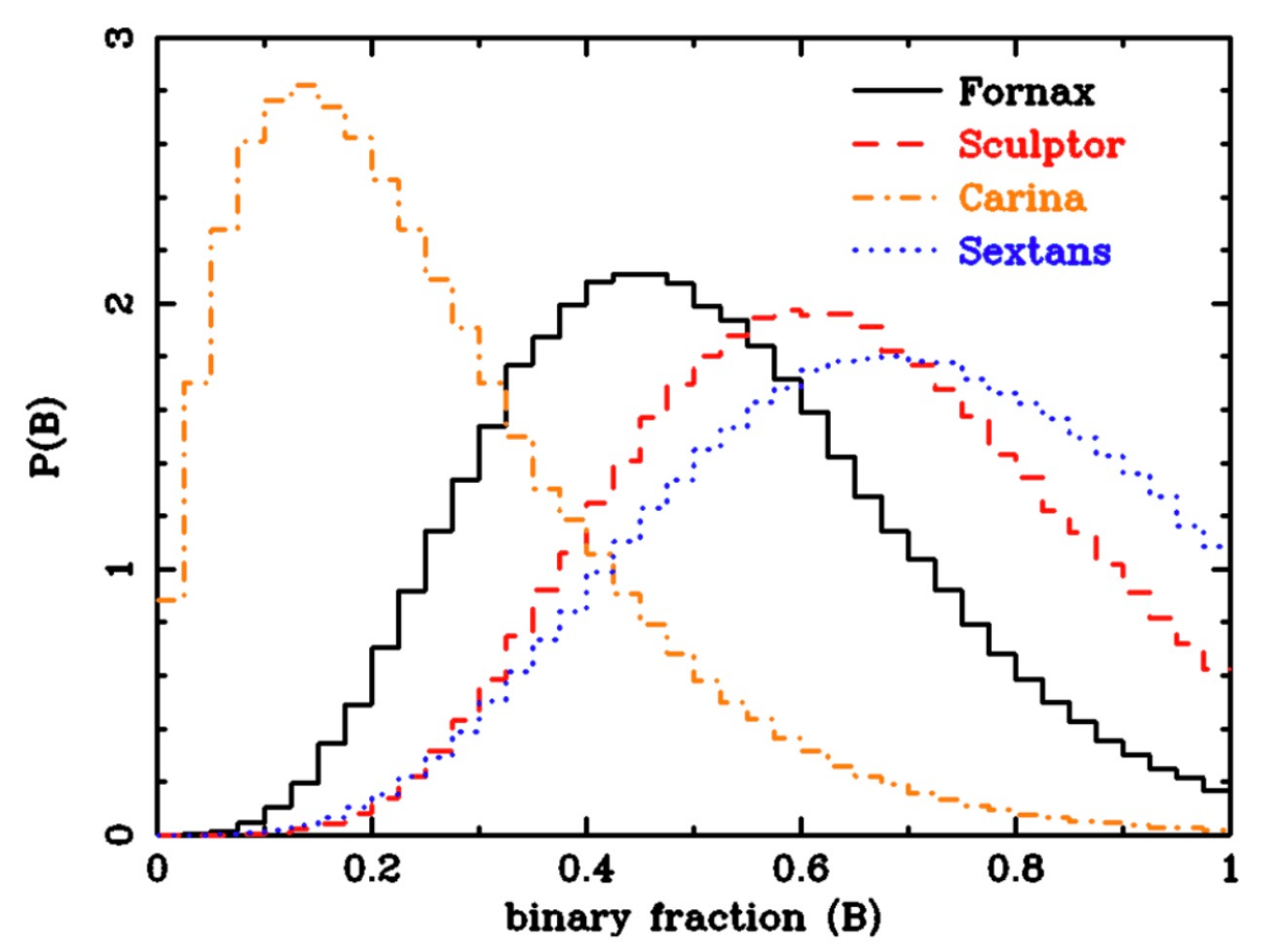}}
\caption{
Probability distributions for binary fractions of stars in four dwarf galaxies, based on radial velocity measurements with 1-3\,km~$s^{-1}$ precision. From \citet{minor2013}. 
\label{fig:mwsat_5}
}
\end{figure}

More than half of the stars in the Galactic field are in binary systems \citep{duquennoy1991}. These stellar systems have an important role in stellar astrophysics. In fact, basic principles of classical mechanics allow to link the evolution of the observables (velocity and magnitude) to intrinsic characteristics of the two components like their masses, their separation and, in some cases, their luminosities and radii.

These objects have therefore been used to derive fundamental relations like the mass-radius and mass-luminosity relations which have been used as a benchmark to test stellar structure models \citep{kuiper1938,huang1956}. Moreover, eclipsing binaries are among the few classes of objects which can be used as standard candles to determine the distance of nearby stellar systems. In collisional stellar systems, like in open and globular clusters, they also provide the energy budget necessary to halt and eventually reverse core-collapse \citep{hut1992}.

Furthermore, the interplay between stellar evolution and hydrodynamics in binaries causes gas transfer between the components and is responsible for the formation of peculiar objects like, for example, novae, cataclysmic variables, millisecond pulsars, low-mass X-ray binaries, SNe Ia and blue straggler stars. 

The fraction and the characteristics (distribution of periods, mass-ratios, eccentricities) of a binary population depend on the environment where they live. Indeed, in high-density stellar systems frequent interactions with singles and other binary stars ionise the wide binaries and produce exchanges among the stars involved in the interaction thus altering the distribution of mass-ratios. Of particular interest is exploring the effect of metallicity on binaries, in contrast with the solar neighbourhood, $\rm[Fe/H]=0$. This is ultimately expected to have an effect on SN rates (in particular SN type Ia), and consequently will affect the chemical evolution of the system. What is currently completely unknown is whether these binary fractions or period distributions change with effects such as metallicity and distance from galaxy centre.

Finally, undetected binaries will affect the measured velocity dispersion of the smallest galaxies, like the UFD galaxies, for which the intrinsic velocity dispersion of the stellar component is of the order of just a few km\,$s^{-1}$ \citep{minor2013, McConnachie2010}. For a binary fraction equal to 1 while assuming the binary properties of the solar neighbourhood, an intrinsic line-of-sight velocity dispersion $\sigma_{los}=8$\,km\,$s^{-1}$ would inflate to 9\,km\,$s^{-1}$ \citep{spencer2017}. However, for galaxies with low intrinsic dispersion, $\lesssim2$\,km\,$s^{-1}$, the same study shows that the measured $\sigma_{los}$ can be $1.5-4$ times higher, 
even with a modest binary fraction of 0.3.  Therefore, to correctly understand dark matter halo properties at the smallest galactic scales, it is important to quantify the effect of binary motion on the velocity distribution \citep[e.g.][]{koposov2011}.

It is therefore of crucial importance to study the relative frequency and the characteristics of binary stars in different environments. However, to date, a thorough census of the properties of binary systems is available only in the Solar neighbourhood \citep[e.g.][]{raghavan2010,moe2017,pricewhelan2020,el-badry2021,penoyre2022}. Sparse studies, performed using low- and intermediate-resolution radial velocity surveys on relatively small numbers ($<500$) of stars have been performed on stellar systems like star clusters and dwarf galaxies, see Fig.~\ref{fig:mwsat_5} \citep{minor2013,spencer2018}.

Estimates of the binary fraction in star clusters can be made on the basis of the colour-magnitude distribution of main sequence stars \citep{sollima2010,milone2012}. This technique, while allowing to detect hundreds of binaries with high mass-ratios ($q>0.5$), requires very accurate photometry and it is blind to low-mass ratio binaries. Observations at random cadence based on spectra at intermediate resolution ($R\sim20\,000$) in GCs have revealed that binaries are less common in high-density systems, compared to those of lower density \citep[e.g.][]{lucatello2015}, as expected due to more frequent stellar dynamical interaction. Given the typical uncertainty of individual radial velocities ($\sim1$\,km\,$s^{-1}$) and the non-optimal time series, the efficiency of binary detection in this study is generally $<50$\,\%. Interestingly, the fraction of binaries significantly changes in the cluster sub-populations, with the Na-rich/O-poor populations hosting a significantly lower frequency of binaries. This evidence has a deep relevance for the origin of the multiple populations in GCs since it provides a strong constraint to the initial density of the various sub-populations \citep[see][]{gratton2020}.

\begin{figure}
\centering
\resizebox{0.90\hsize}{!}{
\includegraphics{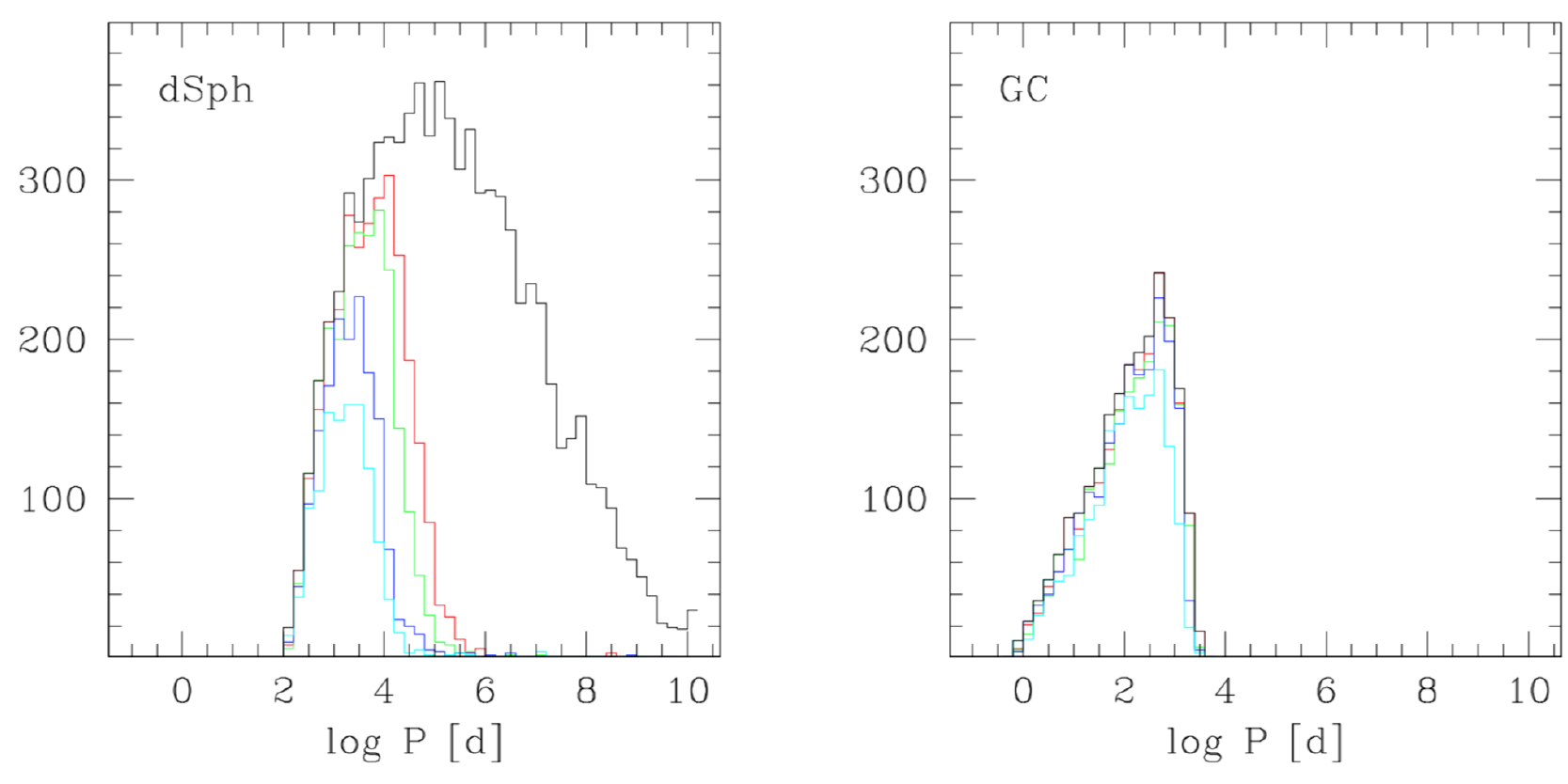}}
\caption{
The number of detected binaries with period, P, during simulated 5-year mock observations with 5 epochs for a typical dSph galaxy (Sculptor) on the left, and 4 epochs for a typical globular cluster on the right. Black lines show the assumed underlying distributions while coloured lines show different assumption of the radial velocity precision, $\sigma_\circ=0.02$ (red), 0.1 (green), 0.5 (blue) and 1\,km\,$s^{-1}$ (cyan). Credit: A. Sollima.  
\label{fig:mwsat_6}
}
\end{figure}

Dwarf galaxies provide an excellent environment for binary studies, as they are large and diffuse enough that their binary systems should remain mostly unaltered. These galaxies can thus be used as test-beds to understand star formation, and their binary fractions and orbital characters also provide information about the initial conditions under which the stars in the galaxy formed. Fig.~\ref{fig:mwsat_6} shows how precision of line-of-sight velocities affect the detectable part of the binary period distributions. 

Currently, not much is known about the fraction and period distribution of binary star systems in dwarf galaxies. The most comprehensive studies were carried out by \cite{minor2013, spencer2018} and \citet[][]{arroyo-polonio2023} on between 4 and 7 of the "classical" dSph galaxies satellites of the Milky Way. These studies were based on multi-epoch data with radial velocity precision of $1-3$\,km\,$s^{-1}$, thus limited to binaries with periods $P<10$ years, and in general consisted of samples obtained with heterogeneous facilities and/or small number statistics. 

While \cite{spencer2018} found that the dwarf galaxies analysed exhibited a wide range of binary fractions and rejected the possibility of a common binary fraction with a significance level of 99\,\%, \citet{arroyo-polonio2023} implemented a more robust determination of uncertainties and found that these same dwarf galaxies could share the same binary fraction. This finding reopens the possibility of using the binary stellar population of dSphs as templates for those in UFDs, for which determining the properties is even more challenging. Hence, there remains significant work to be done in this field, but it necessitates better datasets in terms of repeated observations and velocity uncertainties. 

In the magnitude regime for which target stars in Milky Way dwarf galaxies are going to be accessible to HRMOS, which corresponds to $1-2$ magnitudes below the tip of the red giant branch in some cases, it might be important to consider the effect of jitter. Jitter refers to the random motions in stars that arise from intrinsic factors such as material movement on the star's surface. For low surface gravity red giant stars, these movements can reach velocity amplitudes of up to 1\,km\,$s^{-1}$ \citep{Hekker2008} and could therefore be dominating the uncertainties in the velocity measurements for the velocity precision of HRMOS ($<1$\,km\,$s^{-1}$). We therefore computed what would be the reach of HRMOS, in the absence or presence of jitter, using a refined version of the methodology of \citet{arroyo-polonio2023}. 

To accomplish this we generated mock data, assuming a binary fraction of 0.4 and $\overline{logP\left[\text{d}\right]} = 4.8$. We considered a sample of the 500 brightest red giant branch stars with 5 repeated observations over a span of 5 years, evenly spaced in time, for a dwarf galaxy at 80\,kpc. In this case, the average jitter is of the order of 0.35\,km\,$s^{-1}$. While it remains challenging to place simultaneous tight constrains on both the binary fraction and the mean of the period distribution $\overline{logP\left[\text{d}\right]}$, we find that, when fixing the period distribution, HRMOS can deliver exquisite precision on the binary fraction with only 5 observations of 500 stars spaced over only 5 years, both when including or neglecting the effects of jitter ($\pm0.04-0.06$ at a $1-\sigma$ confidence level), already with a velocity uncertainty of $\pm0.2-0.5$\,km\,$s^{-1}$.

\subsection{Hierarchical Galaxy formation}

The {\sl Gaia} mission opened up a window to investigate many of the previous merger events that had occurred in the history of the Milky Way \citep[][]{GDR2_helmi2018,ibata2019,ibata2021}. With large spectroscopic surveys such as 4MOST and WEAVE, this is expected to reach down to smaller systems, where past merger events of the Magellanic Clouds and the dSph galaxies will almost certainly be discovered. Currently, very little is known observationally about hierarchical galaxy formation on smaller scales. How frequent are such mergers? How chemically distinct were they from the host galaxy? How did they influence the evolution of the host galaxy? 

Once these substructures are identified, it will become crucial to follow them up with high precision radial velocities and chemical abundances to fully characterise such events, and quantify the likely small chemical differences with the main host \citep[e.g.][]{cicuendez2018}. Furthermore, large spectroscopic surveys are expected to discover various abundance gradients in the Milky Way satellite galaxies along with possible intrinsic chemical abundance dispersion \citep[e.g.][]{alexander2023}, which will need a follow-up for an accurate quantification of these effects, so they can be contrasted against simulations of these systems.

\subsubsection{Requirements for the study of stellar population in Local Group dwarf galaxies}

Only by applying an instrument such as HRMOS to a wide range of galaxies will we be able to directly observe and quantify the imprints of the first zero-metallicity stars, and the impact of various nucleosynthetic sites, that is understanding their dependence on environment, their nucleosynthetic yields, and constraining their time-delay distribution functions. In addition, such an instrument allows us to measure stellar binary fractions and quantify substructure and previous merger events in galaxies of different sizes, and constrain the properties of the various building blocks of the Milky Way. Thus directly disentangling the detailed hierarchical structure formation of our Galaxy.
\clearpage\pagebreak


\frame{\vspace{-7mm}
\paragraph{Uniqueness}
The HRMOS instrument is visioned on the VLT in around year 2030. By that time, 4MOST (also at Paranal) will have run a dedicated survey of the LMC and SMC, providing chemical abundances of various degrees for $\sim400\,000$ giant stars \citep{cioni2019}. The large dSph galaxies Sagittarius, Fornax and Sculptor and all smaller dwarf galaxies in the 4MOST footprint, will be targeted by 4DWARFS, the 4MOST survey of dwarf galaxies and the stellar streams \citep{skuladottir2023dwarfs}. 4DWARFS will provide line-of-sight velocities and chemical abundances for $\sim130\,000$ stars in close to 50 individual dwarf galaxies, and dozens of stellar streams.

Importantly, all 4MOST targets will be observable with HRMOS in Paranal. 4MOST will thus provide excellent, high-quality catalogues for more detailed follow-up studies, with more specific scientific goals in mind. HRMOS will provide the facility to do so efficiently.
In particular, the combination of the large collecting area of the VLT, high accuracy in line-of-sight velocities and high precision in chemical abundances will make HRMOS a unique facility to investigate mergers and substructures within the Milky Way satellite galaxies. This combination of high multiplicity and high precision in line-of-sight velocities is particularly important in the case of the binary stars which (i) constitute a relatively small fraction of the whole population; (ii) can be detected with a limited efficiency; and (iii) where the period distribution goes beyond $\log~ P$ [days]$>$3. There is currently no available or planned instrument able to achieve the science goals presented here. The wavelength coverage of MOONS is limited to $>650$\,nm, and the resolution is significantly lower ($R<10\,000$ at $<1500$\,nm). This significantly limits the precision of line-of-sight velocities, and  drastically decreases the abundance information obtainable with its spectra, especially in relation to metal-poor stars and the heavy elements in general. The WST which is being investigated, in its high resolution mode, will only have a resolving power in the range $R=20\,0000-40\,000$. Last but not least, we point out that the five largest near-by Milky Way satellites are in the Southern hemisphere, and instruments in the Northern hemisphere such as MSE will therefore not be able to reach these systems.  HRMOS is therefore fundamental for the science of Milky Way satellite galaxies. 
}

\begin{table*}[t]%
\centering
\caption{Instrument requirement summary (Milky Way satellite galaxies)}%
\footnotesize
\begin{tabularx}{\textwidth}{lXX}
\hline
\textbf{Parameter} & \textbf{Value}   & \textbf{Justification}  \\
\hline
Resolution ($R$)    & 60\,000     & Higher resolution will not result in significantly more accurate abundances, but will significantly reduce the number of observable targets (see Fig.~\ref{fig:mwsat_1}, \ref{fig:mwsat_3} and~\ref{fig:mwsat_4}). It will almost exclude the (accurate and precise) use of the blue wavelength range ($<420$\,nm), and negatively affect the efficiency of any chemical abundance survey of satellite galaxies. Binarity: In order to be able to reach faint targets/smaller systems, 60\,000 is preferred. But for the largest systems, LMC, SMC and SGR, and nearby clusters $R=80\,000$ is also suitable.   \\
Spectral range & 402\,nm [Th], 406\,nm [Pb], 481\,nm [Zn]. Other key elements have some flexibility (of various degree), see Fig.~\ref{fig:mwsat_3}.  & Key elements: Fe, (C, N, O), Al, Mg, Mn, Ni, Cu, Zn, 1st peak (Sr, Y and/or Zr), Ba, Eu, Tb/Dy/Ho, Th, Pb   \\
Multiplexing   & 100      & Minimum limit: 40 fibres/field Maximum useful limit: 500 fibres/field. Target density differs significantly between galaxies, and is shown for the largest galaxies in Fig.~\ref{fig:mwsat_1} Target density goes from $>2000$ stars/ field in the LMC to $<40$ in the UFDs.    \\
Stability      & $\lesssim200$\,m\,s$^{-1}$         & Needed to unambiguously identify member stars of substructure within the galaxies. To resolve degenerancies in the period distribution parameters, a precision of $\lesssim200$\,m\,s$^{-1}$ is required (including measurement errors because of noise).   \\
Fibre spacing  & Similar to FLAMES        & The FLAMES fibre spacing and multiplicity has proven very efficient in numerous existing studies of satellite galaxies.     \\
Efficiency          &     For RGB star of $V=16.5$ to reach $\rm SNR\approx50$ per resolution element at 520\,nm in 1\,hour. For an RGB star of $B=17.5$ to reach $\rm SNR\approx50$ per resolution element at 390\,nm in 5\,hours. The ability to reach $\rm SNR\approx20$ per resolution element, for a $V=18$ giant at $\sim520$\,nm in 1\,hour exposure        &   High efficiency in the blue is essential for our scientific goals. \\
\hline
\end{tabularx}
\end{table*}

%% file: acronyms.tex
\section*{List of acronyms} \label{sec:acronyms}

4MOST - 4-metre Multi-Object Spectroscopic Telescope \\
AAT - Anglo-Australian Telescope \\
ADC -  Atmospheric dispersion corrector\\
AGB - Asymptotic giant branch \\
ANDES - ArmazoNes high Dispersion Echelle Spectrograph\\
APOGEE - Apache Point Observatory Galactic Evolution Experiment \\
CCD -  Charge-coupled device\\
CEMP - Carbon-enhanced metal-poor \\
CHEOPS - CHaracterising ExOPlanet Satellite\\
DESI -  Dark Energy Spectroscopic Instrument\\
DSPH - Dwarf spheroidal \\
ELT -  Extremely Large Telescope\\
EMP -  Extremely metal-poor \\
EP - Excitation potential \\
ESA - European Space Agency\\
ESO - European Southern Observatory\\
ESPRESSO - Echelle SPectrograph for Rocky Exoplanet and Stable Spectroscopic Observations\\
EW - Equivalent width \\
FLAMES - Fibre Large Array Multi Element Spectrograph\\
FNX - Fornax \\
FOV - Field-of-view \\
FUV - Far ultraviolet \\
FWHM - Full width at half maximum \\
GALAH - GALactic Archaeology with HERMES\\
GC - Globular cluster \\
HARPS - The High Accuracy Radial velocity Planet Searcher\\
HAYDN - High-precision AsteroseismologY of DeNse stellar fields\\
HERMES -  The high efficiency and resolution Mercator echelle spectrograph\\
HR - High-resolution \\
HRMOS - High-resolution multi-object spectrograph \\
HST - Hubble Space Telescope \\
LAMOST - The Large Sky Area Multi-Object Fibre Spectroscopic Telescope \\
LBT - The Large Binocular Telescope\\
LIM - Low- and intermediate-mass \\
LMC -  Large Magellanic Cloud
LR - Low-resolution \\
LSST - Large Synoptic Survey Telescope \\
MHD - Magneto-hydrodynamics \\
MOONS - Multi-Object Optical and Near-infrared Spectrograph \\
MOS - Multi-object spectrograph \\
MS - Main sequence \\
MSE - Maunakea Spectroscopic Explorer \\
MW - Milky Way \\
NIR - Near infrared \\
NSM - Neutron star merger \\
OC - Open cluster \\
PEPSI - Potsdam Echelle Polarimetric and Spectroscopic Instrument \\
PISN - Pair-instability supernova \\
PLATO - PLAnetary Transits and Oscillations of stars \\
PMS - Pre-main sequence \\
R - Resolving power \\
RAVE - The Radial Velocity Experiment \\
RC - Red clump \\
RV - Radial velocity \\
SDSS -  Sloan Digital Sky Survey \\
SEGUE - Sloan Extension for Galactic Understanding and Exploration \\
SGR -  Sagittarius \\
SMC -  Small Magellanic Cloud \\
SN - Supernova \\
SNR - Signal-to-noise ratio \\
STC - Science and Technical Committee \\
TESS - Transiting Exoplanet Survey Satellite \\
TNG - Telescopio Nazionale Galileo \\
UES - The Utrecht Echelle Spectrograph \\
UCLES - University College London Echelle Spectrograph \\
UFD - Ultra-faint dwarf \\
UVES - The UltraViolet-Visual Echelle Spectrograph\\
VLT - Very Large Telescope \\
WEAVE -  WHT Enhanced Area Velocity Explorer\\
WFIRST - The Wide Field Infrared Survey Telescope\\
WHT - William Herschel Telescope \\
WST - The Wide-Field Spectroscopic telescope \\
ZAMS - Zero-age main sequence \\